\documentstyle[12pt]{article}
\catcode`\@=11
\@addtoreset{equation}{section}

\def\theequation{\thesection.\arabic{equation}}

\newtoks\@stequation

\def\subequations{\refstepcounter{equation}%
  \edef\@savedequation{\the\c@equation}%
  \@stequation=\expandafter{\theequation}
  \edef\@savedtheequation{\the\@stequation}
  \edef\oldtheequation{\theequation}%
  \setcounter{equation}{0}%
  \def\theequation{\oldtheequation\alph{equation}}}

\def\endsubequations{\setcounter{equation}{\@savedequation}%
  \@stequation=\expandafter{\@savedtheequation}%
  \edef\theequation{\the\@stequation}\global\@ignoretrue
  \vspace*{-12pt} \\}

\def\hybrid{\topmargin -20pt	\oddsidemargin 0pt
	\headheight 0pt	\headsep 0pt
	\textwidth 6.25in	
	\textheight 9.5in	
	\marginparwidth .875in
	\parskip 5pt plus 1pt	\jot = 1.5ex}

\def\baselinestretch{1.2}

\catcode`\@=11

\def\marginnote#1{}
\newcount\hour
\newcount\minute
\newtoks\amorpm
\hour=\time\divide\hour by60
\minute=\time{\multiply\hour by60 \global\advance\minute by-\hour}
\edef\standardtime{{\ifnum\hour<12 \global\amorpm={am}%
	\else\global\amorpm={pm}\advance\hour by-12 \fi
	\ifnum\hour=0 \hour=12 \fi
	\number\hour:\ifnum\minute<10 0\fi\number\minute\the\amorpm}}
\edef\militarytime{\number\hour:\ifnum\minute<10 0\fi\number\minute}
\def\draftlabel#1{{\@bsphack\if@filesw {\let\thepage\relax
   \xdef\@gtempa{\write\@auxout{\string
      \newlabel{#1}{{\@currentlabel}{\thepage}}}}}\@gtempa
   \if@nobreak \ifvmode\nobreak\fi\fi\fi\@esphack}
	\gdef\@eqnlabel{#1}}
\def\@eqnlabel{}
\def\@vacuum{}
\def\draftmarginnote#1{\marginpar{\raggedright\scriptsize\tt#1}}

\def\draft{\oddsidemargin -.2truein
	\def\@oddfoot{\sl preliminary draft \hfil
	\rm\thepage\hfil\sl\today\quad\militarytime}
	\let\@evenfoot\@oddfoot	\overfullrule 3pt
	\let\label=\draftlabel
	\let\marginnote=\draftmarginnote
   \def\@eqnnum{(\theequation)\rlap{\kern\marginparsep\tt\@eqnlabel}%
\global\let\@eqnlabel\@vacuum}  }

\def\preprint{\twocolumn\sloppy\flushbottom\parindent 2em
	\leftmargini 2em\leftmarginv .5em\leftmarginvi .5em
	\oddsidemargin -.5in	\evensidemargin -.5in
	\columnsep .4in	\footheight 0pt
	\textwidth 10.in	\topmargin  -.4in
	\headheight 12pt \topskip .4in
	\textheight 6.9in \footskip 0pt
	\def\@oddhead{\thepage\hfil\addtocounter{page}{1}\thepage}
	\let\@evenhead\@oddhead	\def\@oddfoot{}	\def\@evenfoot{} }

\def\titlepage{\@restonecolfalse\if@twocolumn\@restonecoltrue\onecolumn
     \else \newpage \fi \thispagestyle{empty}\c@page\z@
	\def\thefootnote{\fnsymbol{footnote}} }

\def\endtitlepage{\if@restonecol\twocolumn \else \newpage \fi
	\def\thefootnote{\arabic{footnote}}
	\setcounter{footnote}{0}}  

\catcode`@=12
\relax

\def\figcap{\section*{Figure Captions\markboth
	{FIGURECAPTIONS}{FIGURECAPTIONS}}\list
	{Figure \arabic{enumi}:\hfill}{\settowidth\labelwidth{Figure
999:}
	\leftmargin\labelwidth
	\advance\leftmargin\labelsep\usecounter{enumi}}}
 \relax
\def\tablecap{\section*{Table Captions\markboth
	{TABLECAPTIONS}{TABLECAPTIONS}}\list
	{Table \arabic{enumi}:\hfill}{\settowidth\labelwidth{Table
999:}
	\leftmargin\labelwidth
	\advance\leftmargin\labelsep\usecounter{enumi}}}
 \relax
\def\reflist{\section*{References\markboth
	{REFLIST}{REFLIST}}\list
	{[\arabic{enumi}]\hfill}{\settowidth\labelwidth{[999]}
	\leftmargin\labelwidth
	\advance\leftmargin\labelsep\usecounter{enumi}}}
 \relax
\def\R{{\rm I\!R}}

\def\one{{\mathchoice {\rm 1\mskip-4mu l} {\rm 1\mskip-4mu}
{\rm 1\mskip-4.5mu l} {\rm 1\mskip-5mu l}}}
\def\Q{{\mathchoice
{\setbox0=\hbox{$\displaystyle\rm Q$}\hbox{\raise 0.15\ht0\hbox to0pt
{\kern0.4\wd0\vrule height0.8\ht0\hss}\box0}}
{\setbox0=\hbox{$\textstyle\rm Q$}\hbox{\raise 0.15\ht0\hbox to0pt
{\kern0.4\wd0\vrule height0.8\ht0\hss}\box0}}
{\setbox0=\hbox{$\scriptstyle\rm Q$}\hbox{\raise 0.15\ht0\hbox to0pt
{\kern0.4\wd0\vrule height0.7\ht0\hss}\box0}}
{\setbox0=\hbox{$\scriptscriptstyle\rm Q$}\hbox{\raise 0.15\ht0\hbox to0pt
{\kern0.4\wd0\vrule height0.7\ht0\hss}\box0}}}}
\def\C{{\mathchoice
{\setbox0=\hbox{$\displaystyle\rm C$}\hbox{\hbox to0pt
{\kern0.4\wd0\vrule height0.9\ht0\hss}\box0}}
{\setbox0=\hbox{$\textstyle\rm C$}\hbox{\hbox to0pt
{\kern0.4\wd0\vrule height0.9\ht0\hss}\box0}}
{\setbox0=\hbox{$\scriptstyle\rm C$}\hbox{\hbox to0pt
{\kern0.4\wd0\vrule height0.9\ht0\hss}\box0}}
{\setbox0=\hbox{$\scriptscriptstyle\rm C$}\hbox{\hbox to0pt
{\kern0.4\wd0\vrule height0.9\ht0\hss}\box0}}}}

\font\fivesans=cmss10 at 4.61pt
\font\sevensans=cmss10 at 6.81pt
\font\tensans=cmss10
\newfam\sansfam
\textfont\sansfam=\tensans\scriptfont\sansfam=\sevensans\scriptscriptfont
\sansfam=\fivesans
\def\sans{\fam\sansfam\tensans}
\def\Z{{\mathchoice
{\hbox{$\sans\textstyle Z\kern-0.4em Z$}}
{\hbox{$\sans\textstyle Z\kern-0.4em Z$}}
{\hbox{$\sans\scriptstyle Z\kern-0.3em Z$}}
{\hbox{$\sans\scriptscriptstyle Z\kern-0.2em Z$}}}}

\mathchardef\endbar="375

\makeatletter
\newcounter{pubctr}
\def\publist{\@ifnextchar[{\@publist}{\@@publist}}
\def\@publist[#1]{\list
	{[\arabic{pubctr}]\hfill}{\settowidth\labelwidth{[999]}
	\leftmargin\labelwidth
	\advance\leftmargin\labelsep
	\@nmbrlisttrue\def\@listctr{pubctr}
	\setcounter{pubctr}{#1}\addtocounter{pubctr}{-1}}}
\def\@@publist{\list
	{[\arabic{pubctr}]\hfill}{\settowidth\labelwidth{[999]}
	\leftmargin\labelwidth
	\advance\leftmargin\labelsep
	\@nmbrlisttrue\def\@listctr{pubctr}}}
 \relax
\makeatother
\newskip\humongous \humongous=0pt plus 1000pt minus 1000pt
\def\caja{\mathsurround=0pt}
\def\eqalign#1{\,\vcenter{\openup1\jot \caja
	\ialign{\strut \hfil$\displaystyle{##}$&$
	\displaystyle{{}##}$\hfil\crcr#1\crcr}}\,}
\newif\ifdtup
\def\panorama{\global\dtuptrue \openup1\jot \caja
	\everycr{\noalign{\ifdtup \global\dtupfalse
	\vskip-\lineskiplimit \vskip\normallineskiplimit
	\else \penalty\interdisplaylinepenalty \fi}}}
\def\eqalignno#1{\panorama \tabskip=\humongous
	\halign to\displaywidth{\hfil$\displaystyle{##}$
	\tabskip=0pt&$\displaystyle{{}##}$\hfil
	\tabskip=\humongous&\llap{$##$}\tabskip=0pt
	\crcr#1\crcr}}
\relax
\hybrid
\def\a{\alpha}
\def\b{\beta}
\def\g{\gamma}  
\def\d{\delta}

\def\m{\mu}

\def\t{\tau}
\def\p{\pi}
\def\ps{\psi}
\def\r{\rho}
\def\th{\theta}
\def\s{\sigma}
\def\l{\lambda}
\def\o{\omega}

\def\et{\eta}
\def\L{\Lambda}

\def\Ga{\Gamma}

\def\O{\Omega}
\def\D{\Delta}

\def\bw{\bar{w}}
\def\bv{\bar{v}}

\def\br{\bar{r}}
\def\bs{\bar{s}}

\def\bps{\bar{\psi}}

\def\tL{\tilde{L}}
\def\tu{\tilde{u}}

\def\F{{\cal F}}
\def\Fh{\hat{{\cal F}}}
\def\Bh{\hat{{\cal B}}_H}
\def\Ch{\hat{{\cal C}}_{g/h}}

\def\P{{\cal P}}

\def\G{{\cal G}}

\def\T{{\cal T}}
\def\C{{\cal C}_{g/h}}
\def\cO{{\cal O}}

\def\B{{\cal B}_H}
\def\Bo{{\cal B}_O}

\def\ve{\vert}

\def\ra{\rightarrow}
\def\lra{\leftrightarrow}
\def\ti{\times}
\def\oti{\otimes}

\def\xx{\hbox{ }^*_*}
 
\def\on{\cdot 1}

\def\sux{(SU(3)_x)_M^\#}
\def\sumh{SU(3)_M^\#}
\def\suc{SU(3)/SU(2)_{\rm irr}}
\def\corr{\one +  
2 L^{ab}_{\infty} (\T_a^1 \T_b^2 \ln  y + \T_a^1 \T_b^3 \ln(1-y) ) }
\def\corrH{\one +  
2 L^{ab}_{H,\infty} (\T_a^1 \T_b^2 \ln  y + \T_a^1 \T_b^3 \ln(1-y) ) }

\def\corrHs{\one  +
2 L^{ab}_{H,\infty} (\T_a^1 \T_b^2 \ln  y^* +\T_a^1 \T_b^3 \ln(1-y^*) ) }
\def\corrg{\one  +
2 L^{ab}_{g,\infty} (\T_a^1 \T_b^2 \ln  y +\T_a^1 \T_b^3 \ln(1-y) ) }
\def\corrgt{\one  +
2 L^{ab}_{g,\infty} (\T_a^1 [\T_b^2 + \T_b^3]  \ln  (-y) 
+\T_a^1 \T_b^3 \ln\left(1-\frac{1}{y}\right) ) }
\def\corrgs{\one  +
2 L^{ab}_{g,\infty} (\T_a^1 \T_b^2 \ln  y^* + \T_a^1 \T_b^3 \ln(1-y^*) ) }
\def\corrh{\one  -
2 L^{ab}_{h,\infty} (\T_a^1 \T_b^2 \ln  y + \T_a^1 \T_b^3 \ln(1-y) ) }

\def\corrgh{\one  +
2 L^{ab}_{g/h,\infty} (\T_a^1 \T_b^2 \ln  y + \T_a^1 \T_b^3 \ln(1-y) ) }
\def\corrghs{\one  +
2 L^{ab}_{g/h,\infty} (\T_a^1 \T_b^2 \ln  y^* + \T_a^1 \T_b^3 \ln(1-y^*) ) }
\def\corrhp{\one  +
2 L^{ab}_{h,\infty} (\T_a^1 \T_b^2 \ln  y + \T_a^1 \T_b^3 \ln(1-y)) }
\def\ok{\cO (k^{-2} )}
\def\oko{\cO (k^{-1} )}
\def\ox{\cO (x^{-2} )}
\def\oxo{\cO (x^{-1} )}
\def\tN{\tilde{N}}
\def\tR{\tilde{R}}
\def\tM{\tilde{M}}

\def\nl{\newline}
\def\limit#1#2{\smash { \mathop{#1} \limits_{#2} }  }
\def\thefootnote{\fnsymbol{footnote}}
\def\be{\begin{equation}}
\def\ee{\end{equation}}
\def\bs{\begin{subequations}}
\def\es{\end{subequations}}
\def\ben{\begin{enumerate}}
\def\een{\end{enumerate}}

\def\G{{\cal G}}

\def\vs{\vskip}
\def\arg{{\rm arg}}
\def\sign{{\rm sign}}

\def\nl{\newline}
\def\ed{\end{document}}

\def\bibtem#1{\bibitem{#1} }
\def\cit#1{\cite{#1}}

\def\sp{\quad, \quad}
\def\pe{\quad . }
\def\ssp{, \;\;}

\begin{document}
\begin{titlepage}
\begin{center}

\hfill CERN-TH/96-03 \\
\hfill CPTH-S428.0196 \\
\hfill UCB-PTH-96/02 \\ 
\hfill LBL-38128 \\  
\hfill hep-th/9601070  \\

\vskip .2in

{\large \bf  
Semi-Classical Blocks and Correlators in \\  
Rational and Irrational Conformal Field Theory }
\vskip .4in
{\bf M.B. Halpern}
\footnote{On leave from the Department of Physics, University of California,
 Berkeley, CA 94720, USA.} 
\footnote{e-mail address: mbhalpern@theor3.lbl.gov, 
halpern@vxcern.cern.ch}
\\
\vskip .15in
{\em Theory Division, CERN, CH-1211\\
Geneva 23, SWITZERLAND} 
\\
\vskip .2in
and
\vskip .15in
{\bf N.A. Obers}
\footnote{e-mail address: obers@orphee.polytechnique.fr}
\vskip .1in
{\em Centre de Physique Th\'eorique
\footnote{Laboratoire Propre du CNRS UPR A.0014}  \\
Ecole Polytechnique \\
F-91128 Palaiseau,
FRANCE } 
\end{center}

\vskip .3in

\begin{center} {\bf ABSTRACT } \end{center}
\begin{quotation}\noindent
The generalized Knizhnik-Zamolodchikov equations of irrational conformal 
field theory provide a uniform description of rational and irrational conformal
field theory. Starting from the known high-level solution of these equations, 
we first construct the high-level conformal blocks and correlators of all the 
affine-Sugawara and coset constructions on simple $g$. 
Using intuition gained from these cases, we then identify a simple class
of irrational processes  whose high-level
blocks and correlators we are also able to construct. 
\end{quotation}
 \vskip 0.7cm
CERN-TH/96-03\\
January 1996 
  \\ Revised version 
 \end{titlepage}
\vfill
\eject
\def\baselinestretch{1.2}
\baselineskip 16 pt
\noindent
\setcounter{equation}{0}
\tableofcontents
\section{Introduction}
In recent years we have learned that the generic conformal field theory
has irrational central charge, even when the theory is unitary. The study
of this subject is called irrational conformal field theory (ICFT), which
properly includes rational conformal field theory (RCFT) as a small subspace,
\be
  {\rm ICFT} \supset \supset {\rm RCFT}
\ee
where RCFT is understood here as the affine-Sugawara [1-6] and coset 
constructions [1,2,7,8]. A comprehensive review of ICFT is found in Ref.\cit{rev}.
     
The foundation of ICFT is affine Lie algebra [10,1] and the general 
affine-Virasoro construction [11,12],
\be
              T= L^{ab} \xx J_a J_b \xx
\label{avc} \ee
on the currents $J_a$, $a=1 \ldots {\rm dim}\,g$ of the general affine algebra.
The construction (\ref{avc}) is summarized by the Virasoro master equation
[11,12] for the inverse inertia tensor $L^{ab}$, and the system may be
  understood 
as a conformal spinning top. 

The solutions of the master equation show a symmetry hierarchy \cit{lie} in 
ICFT,  
\be
\mbox{ICFT} \supset \supset H\mbox{-invariant CFTs} \supset \supset 
\mbox{Lie}\;\,h\mbox{-invariant CFTs}
\supset \supset \mbox{RCFT}
\ee
where the $H$-invariant CFTs, which are also generically irrational, include
all theories with a symmetry $H$, where $H$ may be a finite group or a Lie
group.  
In this hierarchy, the RCFTs are understood as special cases of exceptionally
high symmetry, with ever-increasing symmetry breakdown to the left.
The generic ICFT is completely asymmetric.
    
The central computational tools of the subject are the generalized 
Knizhnik-Zamolod- \linebreak chikov (KZ) equations of ICFT \cit{fc}, 
which provide a unified description of rational and irrational conformal field
theory, including powerful new tools for RCFT. In particular, the recent 
solution of these equations for the general coset correlators [15,16,14] 
appears 
to be inaccessible by other methods.

Moreover, the semi-classical or high-level solution of the generalized
KZ equations has been known for some time, providing  a uniform and apparently
simple description of all ICFT $\supset\supset$ RCFT on simple $g$.
The high-level solution is deceptively simple, however, because it is expressed
in a Lie algebra basis, which is not the block basis in which conformal blocks 
are conventionally expressed, and it is only in solving the general problem,
$$    
\bullet \;\;\,\mbox{Lie algebra basis }\;\, \ra \;\,  \mbox{block basis} 
$$
that one confronts the full complexity of the ever-increasing symmetry 
breakdown of ICFT.

In this paper we begin the study of the known high-level solutions, obtaining
the high-level conformal blocks and non-chiral correlators of the simplest
and most symmetric cases. 

In particular, we will first find closed-form expressions for the high-level 
conformal blocks and correlators of all the affine-Sugawara and coset 
constructions. Both results are new.
    
Using intuition gained in this analysis, we then identify what we believe to be
the simplest and most symmetric class of correlators in ICFT, which we call
$$
\bullet \;\;\,  \mbox{the}\; L(g;H)\mbox{-degenerate processes in the }
H\mbox{-invariant CFTs.} 
$$
This is the set of correlators each of whose external states has completely
degenerate conformal weights. The set includes all the affine-Sugawara
correlators, a highly-symmetric set of coset correlators and a presumably
large set of irrational correlators, examples of which are known. For this
class of processes, we are also able to find general  expressions for the
high-level conformal blocks and non-chiral correlators, and we discuss 
an irrational example with octohedral symmetry in some detail.

\section{The High-Level Chiral Correlators of ICFT}

Our starting point is the set of high-level four-point chiral
correlators of ICFT,
\bs
\be
Y_L^\a (y) = \bv_g^\b [ \corr ]_{\b}{}^{\a} + \ok
\label{yle} \ee
\be
L^{ab}_{\infty} ={ P^{ab} \over 2k } 
\;\;\;\;, \;\;\;\; a,b = 1\ldots {\rm dim}\, g
\label{hll}\ee
\label{cor} \es
on simple${}^{\rm a}$\footnotetext{${}^{\rm a}$The chiral correlators 
(\ref{cor}) and the results of this paper also apply to ICFT on semisimple
compact $g= \oplus_I g_I$ with $k_I=k,\forall \; I $.} 
compact $G$, where $G$ is the Lie group whose algebra is $g$, and $k$
is the level of affine $g$. These correlators were conjectured in Ref.\cit{wi},
derived in Ref.\cit{wi2}, and were also obtained as solutions of the 
generalized Knizhnik-Zamolodchikov (KZ) equations of ICFT in Ref.\cit{fc}. 

In what follows, we discuss the conventions, notation and concepts involved in 
the
result (\ref{cor}).
\nl
A. Logarithms. The variable $y$ in (\ref{yle}) is complex,  
and  the logarithms in (\ref{yle}) are defined with natural cuts: $\ln y$ is
defined for $|\arg(y)| <\p$, with its cut to the left, and $\ln(1-y)$ is
defined for $|\arg(1-y)| < \p $ (or equivalently $| \arg(-y)| < \p$), 
with its cut to the right for $y>1$. 
\nl 
B. Inverse inertia tensors.
  The symmetric matrix $L^{ab}_{\infty}$ in (\ref{yle},b)
is the high-level form $L \ra L_\infty$ of the inverse inertia tensor of any
high-level smooth solution of the Virasoro master equation.  
The matrix $P^{ab}$, which must solve the relation [17,18] 
\be
P^{ac} \eta_{cd} P^{db} = P^{ab}
\ee
is the high-level projector of the $L$ theory and $\eta_{ab}$ is the Killing
metric of $g$.
The high-level central charge of the theory is $c(L_{\infty}) ={\rm rank}\,P$.

The chiral correlators (\ref{cor}) provide a uniform high-level description of 
the rational and irrational conformal field theories on $g$, including
\be
P^{ab}_g=\eta^{ab}\sp P^{ab}_{g/h}=P_g^{ab}-P_h^{ab}
\ee
for the affine-Sugawara and coset constructions respectively,
where $\eta^{ab}$ is the inverse Killing metric of $g$. More generally, 
the projectors $P$ are closely related to the adjacency matrices of 
graph theory \cit{gt} and generalized graph theory \cit{ggt} in the partial 
classification of ICFT.
For example, one has \cit{gt}
\be
P_{ij,kl} = \th_{ik} (\G_n) \d_{ij,kl} \sp 
1 \leq i < j \leq n \sp   
1 \leq k < l \leq n
\label{gtp} \ee
in the graph theory ansatz on $SO(n)$, where 
 $a=(ij)$ is the adjoint index and  
$\th (\G_n)$ is the adjacency 
matrix of any graph $\G_n$ of order $n$. The level-families classified by the 
graphs and generalized graphs
 are generically unitary and irrational on non-negative
 integer levels of the affine
algebras.   
\nl
C. Matrix irreps. The matrices
\be 
( \T_a^i)_{\a_i}{}^{\b_i}  
\sp \a_i , \b_i   =1 \ldots {\rm dim}\, \T^ i \sp i =1 \ldots 4
\ee
are irreducible matrix representations (irreps) of $g$, which satisfy 
\be
[\T_a , \T_b ] = i f_{ab}{}^c \T_c \sp a,b,c = 1 \ldots {\rm dim}\,g 
\ee 
where $f_{ab}{}^c$ are the structure constants of $g$.
The labels  $a,\b,\ldots$ are composite indices, e.g. 
$\a=(\a_1 \a_2 \a_3 \a_4)$, 
and  multiplication of irreps is by tensor product, so that
 \bs
\be
(\one )_{\a}^\b = \d_{\a}^\b \equiv 
\d_{\a_1}^{\b_1} \d_{\a_2}^{\b_2} \d_{\a_3}^{\b_3} 
\d_{\a_4}^{\b_4}
\ee
\be
(\T_a^1)_{\a}{}^\b \equiv
 (\T_a^1)_{\a_1}{}^{\b_1} \d_{\a_2}^{\b_2} 
\d_{\a_3}^{\b_3} 
\d_{\a_4}^{\b_4}
\ee
\be
(\T_a^1 \T_b^2 )_{\a}{}^\b \equiv
 (\T_a^1)_{\a_1}{}^{\b_1} (\T_b^2)_{\a_2}{}^{\b_2} 
\d_{\a_3}^{\b_3} 
\d_{\a_4}^{\b_4}
\pe
\ee
\es 
\nl 
D. Broken affine primary fields. The chiral 
 correlators (\ref{cor}) may be understood  
schematically as the high-level form of the averages 
\be
Y_L^\a \sim \langle R_L^{\a_1} (\T^1) 
R_L^{\a_2} (\T^2) R_L^{\a_3} (\T^3) R_L^{\a_4} (\T^4)\rangle
\ee
where $R_L^\a (\T) $, $\a =1 \ldots {\rm dim}\, \T$ is the broken affine
primary field of the $L$ theory corresponding to 
irrep $\T$ of $g$.
 The correlators are written assuming an $L$-basis
\cit{wi} for each $\T^i$, where the conformal weight matrix of the 
broken affine primary field $R_L^{\a_i} (\T^i)$ is diagonal,  
\be
(L^{ab} \T_a^i \T_b^i )_{\a_i}{}^{\b_i} = \D_{\a_i} (\T^i) 
\d_{\a_i}^{\b_i} 
\sp \D_{\a_i} (\T^i)= \cO (k^{-1})
\pe 
\label{cwm} \ee
Fig.1  shows our conventions for the s and t-channels
of the correlators, and  the 13 channel is the u-channel.
\begin{center}
\begin{picture}(200,100)
\put (10,10){\line(4,3){40}}
\put (10,70){\line(4,-3){40}}
\put (50,40){\line(1,0){90}}
\put (140,40){\line(4,3){40}} 
\put (140,40){\line(4,-3){40}} 
\put(2,5){1}
\put(2,70){2}
\put(185,70){3}
\put(185,5){4}
\put(0,40){\vector(1,0){30}}
\put(95,80){\vector(0,-1){30}}
\put(15,45){s}
\put(100,65){t}
\put(40,-10){Fig. 1. The correlators.}
\end{picture}
\end{center}
\vskip 1cm

In ICFT, broken affine primary states are the only states whose conformal
weights are $\cO( k^{-1})$. In the affine-Sugawara constructions, conformal
weights of the form integer plus $\cO( k^{-1})$ are integer descendants of
affine primary states, but this is not necessarily true for the coset
constructions${}^{\rm b}$\footnotetext{${}^{\rm b}$See for example the
conformal weights of Ref.\cit{wi} under the coset construction
$(SU(n)_{k_1} \ti SU(n)_{k_2} )/SU(n)_{k_1 + k_2}$ when $k_1=k_2=k$.}
and beyond, where we know only that the corresponding states are broken
affine secondary. 
\nl
E. Global Ward identity.
 The objects $\bv_g^\a$ are  arbitrary linear combinations of
$g$-invariant tensors of $\T^1 \oti \cdots \oti \T^4$, which satisfy the
$g$-global Ward identity,
\be
\bv_g^\b ( \sum_{i=1}^4 \T_a^i)_{\b} {}^\a =0 \;\;\;\;, \;\;\;\;
a= 1 \ldots {\rm dim}\, g 
\pe
\ee
\nl
F. Hermiticity.  The matrix irreps $\T_a$ satisfy  the
hermiticity condition,  
\bs
\be
\T_a^{\dagger} = \r_a{}^b \T_b
\ee
\be
(\T_a^{\dagger})_\a{}^\b \equiv \et_{\a \r} \et^{\b \s} 
(\T_a)_\s{}^\r {}^*
\ee
\es
 where star is complex conjugation and $\et_{\a\b}= \eta_{\a \b}^*$ is the
carrier space metric of irrep $\T$. Moreover, we will consider only
unitary theories (non-negative integer level of the affine algebra and 
$L^\dagger (m) =L(-m)$), 
for which the inverse inertia tensor satisfies
\be
L^{ab} {}^* = L^{cd} (\r^{-1} ) _c{}^a(\r^{-1} ) _d{}^b 
\ee
and similarly for $L_{\infty}$. It follows that all the matrices  
 in (\ref{cor}) are hermitean, e.g.
\be
(2 L^{ab}_{\infty} \T_a^1 \T_b^2 )^{\dagger} = 2  L^{ab}_{\infty}
\T_a^1 \T_b^2 
\ee
with orthonormal, complete sets of eigenvectors and real eigenvalues. \nl
G. $SL(2,\R)$ gauge. The chiral 
 correlators (\ref{cor}) are given in the 2-3 symmetric KZ gauge \cit{kz},
\bs
\be
Y^\a (y)= (\prod_{i< j}^4 z_{ij}^{\g_{ij}}) A^\a (z_1,z_2,z_3,z_4)
\sp 
y = {z_{12} z_{34} \over z_{14} z_{32}} 
\ee
\be
\g_{12} = \g_{13} =0 \;\;\;\;, \;\;\;\; \g_{14} = 2  \D_{\a_1}
\;\;\;\;, \;\;\;\;
\g_{23} = \D_{\a_1}+\D_{\a_2}+\D_{\a_3}-\D_{\a_4}
\ee
\be
\g_{24} =- \D_{\a_1}+\D_{\a_2}-\D_{\a_3}+\D_{\a_4} \;\;\;\;, \;\;\;\;
\g_{34} = -\D_{\a_1}-\D_{\a_2}+\D_{\a_3}+\D_{\a_4}
\ee
\es
where $A^\a(z)$ are the non-invariant chiral four-point correlators. \nl
H. Limiting behavior.  For any conformal field theory in the KZ gauge, 
the conformal weights $\D_{({\rm s})}$, $\D_{({\rm u})}$ and $\D_{({\rm t})}$ of the s, u and 
t-channel intermediate states appear in the limiting behavior,
\be
Y^\a (y) \sim \left\{ \matrix{ y^{\D_{({\rm s})} -  \D_{\a_1}(\T^1)-\D_{\a_2}(\T^2)
} & ,\;\;y \ra 0 \cr
(1-y)^{\D_{({\rm u})} -  \D_{\a_1}(\T^1)-\D_{\a_3}(\T^3)
} & ,\;\;y \ra 1 \cr
\left( \frac{1}{y} \right)^{\D_{({\rm t})} +\D_{\a_1}(\T^1)- \D_{\a_4}(\T^4)
} & ,\;\;y \ra \infty \cr} \right.
\pe
\label{lyb2} \ee
Here, we will use these facts in the high-level form
\be
y^{ -  \D_{\a_1}(\T^1)-\D_{\a_2}(\T^2)} =
1  - [ \D_{\a_1}(\T^1) + \D_{\a_2}(\T^2) ] \ln y + \cO (k^{-2}) 
\label{lyb} \ee
where we have recalled that the conformal weights of the broken affine 
primary fields are  $\cO (k^{-1})$. \nl
I. High-level OPEs. 
 In Ref.[16,14], it was shown that the high-level chiral correlators 
(\ref{cor}) 
have physical singularities in all channels, and that the high-level
fusion rules among the 
broken affine primaries follow the Clebsch-Gordan coefficients of their
corresponding matrix irreps. In further detail, the high-level 
OPEs of the broken affine primaries can be written schematically as
\be
\label{ope}
\eqalign{
R(\T^1,z)^{\a_1}  R(\T^2,w)^{\a_2}  \; =  
 \sum_{i, \a_i}  & { [ C(T^1,T^2,T^i) + \oko]^{\a_1 \a_2}{}_{\a_i} 
   R(\T^i,w)^{\a_i} \over (z-w)^{\D_{\a_1}(\T^1)  + 
\D_{\a_2}(\T^2)  -  \D_{\a_i}(\T^i)} }  \cr
& \;\;\;\;\;\;\; +  \oko \cdot ( \mbox{broken affine secondaries}) \cr}  
\ee
where the level-independent tensor $C(\T^1,\T^2,\T^i)$ is proportional to
 the Clebsch-Gordan coefficients and the broken affine 
secondaries enter  only at the next order of the high-level expansion.

\vs .4cm
\noindent \underline{Symmetry hierarchy in ICFT}
\vs .3cm 

The high-level correlators (\ref{cor}) provide a uniform description of all
ICFT on simple $g$, which is a bewildering variety \cit{rev} of theories
and correlators. In this
paper we make the first attempt to identify simpler, more symmetric
correlators among these varieties. 

Towards this end, we remind the reader of the symmetry hierarchy \cit{lie}
in ICFT,
\be
\mbox{ICFT} \supset \supset H\mbox{-invariant CFTs} \supset \supset 
\mbox{Lie}\;\,h\mbox{-invariant CFTs}
\supset \supset \mbox{RCFT}
\ee
which organizes the space of ICFTs on $G$ 
according to the residual symmetry group $H \subset G$ of the theory. 
As seen in this hierarchy, the generic ICFT has no residual symmetry 
group${}^{\rm c}$\footnotetext{${}^{\rm c}$In the graph theory ansatz \cit{gt} 
on $SO(n)$, 
whose high-level projectors are given in (\ref{gtp}), this corresponds to
the fact that the generic graph has no symmetry.}, 
and these generic theories are expected to be the most complex.
Consequently, we focus here on the theories with a symmetry, which are also
generically irrational.

The set of all ICFTs with a non-trivial symmetry group $H$ (which
may be a discrete subgroup of $G$ or a Lie subgroup) is called the set
of $H$-invariant CFTs. Among the $H$-invariant CFTs, the subspace of 
theories with a Lie symmetry
 is called the set of Lie $h$-invariant CFTs, where $h \subset g$.
This subspace includes the affine-Sugawara and coset constructions 
as a much smaller subspace.

When a theory $L$ is an $H$-invariant CFT, the correlators (\ref{cor})
also satisfy the 
 global $H$-invariance condition, 
\be
Y_H \O (H) =Y_H  \;\;\;\;, \;\;\;\; \O (H) \in G \;\;\;\;, \;\;\;\;
\O (H)_\a {}^\b = \prod_{i=1}^4 \O(H, \T^i)_{\a_i}{}^{\b_i} 
\label{lh} \ee
 where $\O(H,\T^i)_{\a_i}{}^{\b_i} $ is the
 subgroup  $H$ in matrix irrep $\T^i$. When the theory is a  Lie
$h$-invariant CFT, the condition (\ref{lh}) reduces to the $h$-global Ward 
identity
\be
Y_{{\rm Lie}\,h} 
\sum_{i=1}^4 \T_a^i =0 \;\;\;\;, \;\;\;\; a =1\ldots {\rm dim}\, h
\label{lh1} \ee
which applies for example in the cases of the affine-Sugawara construction
(with $h=g$) and the
 $g/h$ coset constructions.

For the affine-Sugawara and $g/h$ coset constructions, it is known 
[4,15] that the resolution of chiral correlators into conformal blocks
is a basis change from the Lie algebra basis to the block basis, using the
$h$-invariant tensors defined by (\ref{lh1}). More generally, one expects 
that the $H$-invariant tensors defined by (\ref{lh}) will play an analogous
role in finding the block bases of the $H$-invariant CFTs. 

\section{The Affine-Sugawara Constructions \label{sec3}}
\subsection{The affine-Sugawara blocks}
The simplest and most symmetric conformal field theories are the
affine-Sugawara constructions [1-6] on $G$, whose 
high-level correlators are described by (\ref{cor}) with
\bs
\be
 L^{ab}_{g,\infty} ={ \et^{ab} \over 2k }
\ee
\be
Y_g(y) \sum_{i=1}^4 \T_a^i =0 \;\;\;\;, \;\;\;\; a= 1 \ldots {\rm dim}
\,g 
\ee
\label{as} \es
where $P_g^{ab}= \et^{ab}$ is the inverse Killing metric of $g$.
In this case, the correlators (\ref{cor}) are the high-level solutions of the
KZ equations [3,4] for any correlator on simple $g$. 

We begin by  defining the s-channel block basis  of $g$-invariants
$v({\rm s},g)^m$ as the solutions of  the simultaneous eigenvalue problem and
$g$-global condition 
\bs
\be
(2 L^{ab}_{g,\infty} \T_a^1 \T_b^2)_\a {}^\b v({\rm s},g)_\b^m=
(\D_{({\rm s})}^g (m) -\D^g (\T^1) -\D^g (\T^2) ) v({\rm s},g)_\a^m
\label{asev} \ee
\be
\sum_{i=1}^4 (\T_a^i )_\a {}^\b v({\rm s},g)_\b^m =0 \;\;\;\;, \;\;\;\; 
a= 1 \ldots {\rm dim}\,g 
\pe
\label{gwi} \ee 
\label{asev2} \es
The $g$-global condition (\ref{gwi})
is compatible with the eigenvalue problem because the
generators  $\sum_{i=1}^4 \T_a^i$ commute with 
$L_{g}^{ab} \T_a^1 
\T_b^2$. Here $\D_g(\T^i)$, $i=1,2$ are the high-level forms of the
 conformal weights of external affine primary states,  
\bs
\be
\D_{\a_i} (\T^i) \ve_{L=L_g} = \D^g (\T^i) = {I(\T^i) \over x + \tilde{h} }
= {I(\T^i) \over x } + \ox 
\ee
\be
x = {2k \over \ps_g^2} 
\ee
\es
where $\ps_g$, $\tilde{h}$, $I(\T)$ and $x$ are respectively the highest root
and dual Coxeter number of $g$, the invariant Casimir of irrep $\T$ and the
invariant level of the affine algebra. The high-level form of the relation  
\be
2 L_{g}^{ab}  \T^1_a \T^2_b = 
 L_{g}^{ab}  (\T^1_a + \T^2_a) (\T^1_b + \T^2_b)   
- (\D^g (\T^1) + \D^g (\T^2)) \one 
\ee
tells us that the quantities  in (\ref{asev}) 
\be
\D_{({\rm s})}^g(m) \equiv  \D^g(\T^m)
\ee
are the high-level conformal weights of an irrep $\T^m$ in
$\T^1 \oti \T^2$, hence the conformal weights of affine primary states 
exchanged in the s-channel. 
The dual eigenvalue problem is 
\bs
\be
\bv({\rm s},g)^\b_m (2 L^{ab}_{g,\infty} \T_a^1 \T_b^2)_\b {}^\a =
\bv({\rm s},g)^\a_m (\D_{({\rm s})}^g (m) -\D^g (\T^1) -\D^g (\T^2) )
\label{dep} \ee
\be
\bv({\rm s},g)^\b_m \sum_{i=1}^4 (\T_a^i )_\b {}^\a =0 \;\;\;\;, \;\;\;\; 
a= 1 \ldots {\rm dim}
\,g 
\label{gvw} \ee 
\es
where $\bv({\rm s},g)_m^\a = v({\rm s},g)_\b^m {}^* \et^{\b \a} $ and 
$\et_{\a \b}= \prod_{i=1}^4 \et_{\a_i \b_i} $ is the product of the carrier
space metrics. 

Because $ 2L_{g,\infty}^{ab} \T_a^1 \T_b^2$ is hermitean we  know
that the eigenvectors are  orthonormal and complete,  
\be
\bv({\rm s},g)_m v({\rm s},g)^n =\d_m^n \;\;\;\; , \;\;\;\;
 v({\rm s},g)^m_{\a} \bv({\rm s},g)_m^\b = (I_g)_\a^\b
\label{com} \ee
where $I_g$ is the projector onto the $G$-invariant subspace of 
$\T^1 \oti \cdots \oti \T^4$. The relation
\be
[ L_{g,\infty}^{ab} \T_a^i \T_b^j, I_g] = 0\sp 1 \leq i,j \leq 4 
\label{lic} \ee
also holds on the $G$-invariant subspace defined by (\ref{asev2}). 
An explicit solution to the eigenvalue problem and global condition in
(\ref{asev2}) is known \cit{wi2}
\bs
\be
\bv({\rm s},g)_m^\a = \sum_{\a_r\a_{\br}}\bw_{\rm s}(r,\xi)^{\a_1 \a_2 \a_{r} } 
\bw_{\rm s}(\br,\xi ') ^{\a_3 \a_4 \a_{\br} }  \et_{\a_r \a_{\br} }
\;\;\;\;, \;\;\;\; m=(r,\xi,\xi ') 
\label{cl} \ee
\be
\bw_{\rm s}(r,\xi)^{\b_1 \b_2 \b_r }(T_a^1 + \T_a^2 + \T_a^r) _{\b_1 \b_2 \b_r}
{}^{\a_1 \a_2 \a_r} =0 \;\;\;\;, \;\;\;\; 
a= 1 \ldots {\rm dim}\,g 
\ee
\be
\bw_{\rm s}(r,\xi)^{\b_1\b_2 \a_r }[2L_{g,\infty}^{ab}T_a^1 \T_b^2 ]_{\b_1 \b_2}
{}^{\a_1 \a_2 } =\bw_{\rm s} (r,\xi)^{\a_1 \a_2 \a_r } [ \D^g(\T^r) 
- \D^g(\T^1 ) - \D^g(\T^2 )]  
\ee
\label{vt} \es
where $\bw_{\rm s}(r,\xi)^{\a_i\a_j\a_r}$ are the Clebsch-Gordan coefficients of
$\T^i \oti \T^j $ into irrep $\T^r$, $\xi$ labels copies of the same 
irrep $\T^r$ and $\br$ is the conjugate representation of $r$. Using 
(\ref{vt}), it is easy to check directly that $\D^g_{({\rm s})}(m) = $
$\D^g(\T^m)$  
in (\ref{asev}) is the conformal weight  of irrep $m$ under the affine-Sugawara 
construction.

As an explicit example, one finds for 
 $n \bar{n} \bar{n} n$
correlators on $SU(n)$ that the invariant tensors (\ref{vt}) are
\bs
\be
\bv({\rm s},SU(n))_V^\a = v({\rm s},SU(n))_\a^V = \frac{1}{n}\d_{\a_1 \a_2}\d_{\a_3 \a_4 } 
\ee
\be
\bv({\rm s},SU(n))_A^\a = v({\rm s},SU(n))_\a^A = \frac{1}{\sqrt{n^2-1}}[
\d_{\a_1 \a_3}\d_{\a_2 \a_4} 
- \frac{1}{n} \d_{\a_1 \a_2 }\d_{\a_3  \a_4} ]
\ee
\label{vts} \es
where $V$ and $A$ are vacuum and adjoint. This is the original example \cit{kz} 
considered by Knizhnik and Zamolodchikov, although our Clebsch basis
(\ref{vt}), (\ref{vts}) is slightly different than theirs (see Appendix B).

{} From  (\ref{cor}),(\ref{as}) and the completeness relation (\ref{com}), we 
use eigenvector expansions to define
the s-channel conformal 
blocks $\F_g^{({\rm s})}(y)$ of the affine-Sugawara construction
\bs
\be
\bv_g^\a = \sum_m d({\rm s})^m \bv({\rm s},g)_m^\a
\ee
\be
Y_g^\a(y)= \sum_{m,n} d({\rm s})^m \F_g^{({\rm s})}(y)_m{}^n \bv({\rm s},g)_n^\a
\ee
\be
\F_g^{({\rm s})}(y)_m {}^n 
= \bv({\rm s},g)_m [ \corrg ] v({\rm s},g)^n + \ok 
\label{asbl} \ee
\label{asb1} \es
as the coefficients of the chiral correlators expanded in the block basis. 
Here, $d({\rm s})^m$ are a set of undetermined constants.

To study the small $y$ behavior of the s-channel blocks, we rearrange 
(\ref{asbl}) as follows, 
\bs
\be
 \F_g^{({\rm s})}(y)_m {}^n 
= \bv({\rm s},g)_m [\one + 2  L^{ab}_{g,\infty} \T_a^1 \T_b^2 \ln y ]  
[\one +  2 L^{ab}_{g,\infty} \T_a^1 \T_b^3 \ln (1-y) ]  
 v({\rm s},g)^n + \ok 
\ee
\be  
\label{lbp} 
\;\;\;\;\;\;\;\;\;\;\;\;\;\;  \eqalign{
= [ 1 + (\D^g_{({\rm s})}(m) - & \D^g(\T^1) - \D^g(\T^2) ) \ln y ] \cr 
 & \ti \bv({\rm s},g)_m [ \one + 2 L^{ab}_{g,\infty} \T_a^1 \T_b^3 \ln (1-y) ]  
v({\rm s},g)^n + \ok \cr}  
\ee
\be
=  y^{\D^g_{({\rm s})}(m) - \D^g(\T^1) - \D^g(\T^2) } 
\left[ \d_m^n - c({\rm s},g)_m{}^n \sum_{p=1}^{\infty} {y^p \over p} \right] + \ok   
\label{asb} \ee  
\be
c({\rm s},g)_m{}^n = \bv({\rm s},g)_m 2 L^{ab}_{g,\infty} \T_a^1 \T_b^3 v({\rm s},g)^n
\label{ascm} \ee
\label{iss} \es 
where we have used the dual eigenvalue problem (\ref{dep}) to obtain
(\ref{lbp}) and the high-level relation (\ref{lyb}) to obtain (\ref{asb}).
We note in particular that the eigenvector resolution 
correctly guarantees that each
block has a unique leading singularity, 
\bs
\be 
 \F_g^{({\rm s})}(y)_m {}^n 
\mathrel{\mathop\sim_{y \ra 0} }
\Ga_g^{({\rm s})}(m,n) y^{\D^g_{({\rm s})}(m,n) - \D^g(\T^1) - \D^g(\T^2) } + \ok 
\ee
\be
\D^g_{({\rm s})}(m,n) =  
\left\{ \matrix{ 
   \D^g_{({\rm s})}(m) + \ok  & ,\;\;n =m  \cr 
 1 + \cO (k^{-1} )  & , \;\;  n \neq m \cr}
\right.  
\ee
\be
\Ga_g^{({\rm s})} (m,n) =  
\left\{ \matrix{ 1  + \ok & ,\;\; n =m \cr 
 -c({\rm s},g)_m{}^n + \ok  & ,\;\; n \neq m \cr}
\right.  
\label{asr} \ee
\label{asb3} \es 
labelled by $m$ and $n$, which is followed by integer-spaced
secondaries from $\ln(1-y)$.  
According to 
eqs.(\ref{lyb2}) and (\ref{asb3}), 
 the leading singularities of the $n=m$ blocks correspond
to the s-channel exchange of affine primary states, with residue
$\Ga_g(m,m) = \cO (k^0)$, while the leading 
singularities of the $n \neq m$ blocks are affine secondaries, with
$\Ga_g(m,n\neq m) = \cO (k^{-1})$. This pattern is in agreement with the 
general OPE (\ref{ope}). Beyond the leading residues, diagonal blocks
begin at $\cO (k^0) $ and off-diagonal blocks begin at $\oko$.  

If $c({\rm s},g)_m{}^n =0 $ for some $n\neq m$, then this block begins 
at $\ok$, and we obtain no information beyond this fact in our approximation.
Although we are not aware of any examples of this
phenomenon among the affine-Sugawara blocks, examples do occur in
the coset constructions and irrational processes (see Appendix D and
Section \ref{sec6}). 

We also note that,  although we have solved the generalized KZ
equations through  $\cO (k^{-1})$, we are not able to determine the
$\cO (k^{-1})$ part of the $n\neq m$ conformal weights in this approximation.
Of course, under the affine-Sugawara constructions all conformal weights have
the form $\D^g(\T)$+integer, so we can guess the exact result 
\be
\D_{({\rm s})}^g(m,n) = \D^g_{({\rm s})} (m) + 1 - \d_{m,n} \sp \forall \;m,n 
\label{exr} \ee
for the conformal weights of the blocks, which we believe to be correct 
(see Appendix B). 

To define block bases for the other channels,  we also
introduce the u and t-channel $g$-invariants as solutions to their
corresponding eigenvalue problems,
\bs
\be
2 L^{ab}_{g,\infty}
 \T_a^1 \T_b^3 v({\rm u},g)^m = (\D^g_{({\rm u})}(m) -\D^g(\T^1) -\D^g(\T^3))
v({\rm u},g)^m
\label{asevu} \ee
\be
2 L^{ab}_{g,\infty}
 \T_a^2 \T_b^3 v({\rm t},g)^m = (\D^g_{({\rm t})}(m) -\D^g(\T^2) -\D^g(\T^3))
v({\rm t},g)^m
\label{asevt} \ee 
\be
(\sum_{i=1}^4 \T_a^i ) v({\rm u},g)^m = 
(\sum_{i=1}^4 \T_a^i ) v({\rm t},g)^m = 0
\;\;\;\;, \;\;\;\;\ a= 1\ldots  {\rm dim}\,g
\label{ascut} \ee
\be
\bv({\rm u},g)_m v({\rm u},g)^n =
\bv({\rm t},g)_m v({\rm t},g)^n =\d_m^n 
\ee
\be 
\;\;\;\; v({\rm u},g)^m \bv({\rm u},g)_m =
 v({\rm t},g)^m \bv({\rm t},g)_m = I_g
\pe \label{ocev2} \ee
\label{ocev} \es
Here 
\be
\D_{({\rm u})}^g(m) \equiv  \D^g(\T^m) \sp 
\D_{({\rm t})}^g(m') \equiv \D^g(\T^{m'})
\ee
are the high-level (affine-primary) 
conformal weights under the affine-Sugawara construction of irreps $\T^m$ and
$\T^{m '}$
in $\T^1 \oti \T^3$ and $\T^2 \oti \T^3$ respectively. Explicit forms
of the u and t-channel invariants are obtained formally by a $2\lra 3$
and a $2 \lra 4$ interchange respectively in eq.(\ref{vt}). 

In analogy to the s-channel  blocks $\F^{({\rm s})}_g$ in eq.(\ref{asb1}),  
we define the u-channel blocks $\F^{({\rm u})}_g$ using the 
corresponding u-channel invariants,  
\bs
\be Y_g(y) = \sum_{m,n} d({\rm u})^m \F^{({\rm u})}_g(y)_m{}^n  \bv({\rm u},g)_n
\ee
\be
\F^{({\rm u })}_g(y)_m{}^n = \bv({\rm u},g)_m [\corrg ] v({\rm u},g)^n + \ok
\label{asbu} \ee
\be
\;\;\; \;\; \;\;\;\;\;\;\;\;\;
 =  (1-y)^{\D^g_{({\rm u})}(m) - \D^g(\T^1) - \D^g(\T^3) } 
\left[ \d_m^n - c({\rm u},g)_m{}^n \sum_{p=1}^{\infty} {(1-y)^p \over p} \right] + \ok   
\label{asbu2} \ee
\be
\label{ascmu}
c({\rm u},g)_m{}^n = \bv({\rm u},g)_m2  L^{ab}_{g,\infty} \T_a^1 \T_b^2 
v({\rm u},g)^n
\pe \ee
\label{asbu3} \es
The expansion (\ref{asbu2}) is obtained from (\ref{asbu}) following steps
analogous to those in (\ref{iss}). The 
limiting behavior of the u-channel blocks   
\bs
\be
\F^{({\rm u})}_g(y)_m{}^n 
\mathrel{\mathop \sim_{y \ra 1} } 
\Ga_g^{({\rm u})}(m,n)(1-y)^{\D_{({\rm u})}^g(m,n)-\D^g(\T^1) -\D^g(\T^3) } + \ok 
\label{asu} \ee
\be
\D^g_{({\rm u})}(m,n) =  
\left\{ \matrix{ 
   \D^g_{({\rm u})}(m) + \ok  & \ssp n =m \cr 
 1 + \cO (k^{-1} )  & \ssp n \neq m \cr}
\right.  
\ee  
\be
\Ga_g^{({\rm u})} (m,n) =   
\left\{ \matrix{ 1 +\ok  & \ssp n =m \cr 
 -c({\rm u},g)_m{}^n + \ok   & \ssp n \neq m \cr}
\right.  
\label{asru} \ee
\label{ucr} \es 
(followed by integer-spaced secondaries) is  easily read from (\ref{asbu2}).  
As seen above for the s-channel blocks, the diagonal u-channel
 blocks show affine primary conformal weights 
with residue $\cO (k^0)$, while the off-diagonal blocks show integer
descendants of affine primaries,  
and we are again unable to determine the $\cO(k^{-1})$
part of the off-diagonal conformal weights. 

\vs .4cm
\noindent \underline{Analytic blocks}
\vs .3cm

It is clear from the discussion above that the s and u-channel blocks 
$\F_g^{({\rm s})}$ and $\F_g^{({\rm u})}$ are high-level forms of analytic
blocks, but identification of the analytic t-channel blocks is more subtle.
We begin by defining the preliminary t-channel blocks $\F^{({\rm t})}_g$
as the coefficients in the t-channel eigenbasis  
\bs
\be Y_g(y) 
= \sum_{m,n} d({\rm t})^m \F^{({\rm t})}_g(y)_m{}^n  \bv({\rm t},g)_n
\ee
\be
\F^{({\rm t})}_g(y)_m{}^n = \bv({\rm t},g)_m [\corrg ] v({\rm t},g)^n + \ok 
\label{asbt} \ee
\es
in parallel with our expansions above for the s and u-channels.

We must next consider continuation of the logarithms to the t-channel,
for which we use the following two rules
\bs
\be
\ln (1-y) = \ln (-y) + \ln \left( 1 - {1\over y} \right) 
\sp | \arg (-y)| < \p 
\label{ru1} \ee
\be
\ln y = \ln (-y) - i \p \sign (\arg(-y)) 
 \label{ru2} \ee
\label{rul} \es
throughout this paper.
The left side of (\ref{ru2}) is defined for  $| \arg (y)| < \p$ (so that 
$\ln y$ has its cut
is to the left), while the right side of (\ref{ru2}) is defined for
 $ | \arg (-y)| < \p$ (so that $\ln(-y)$ has its  cut to the right). 

In the finite-level 
example of Appendix B, the relation (\ref{ru2}) is used in the 
equivalent form
\be
y^{\nu} = (-y)^{\nu} \exp[-i\pi \nu \sign (\arg(-y))] 
\label{ru2b} \ee
to continue singular  s-channel factors to the t-channel. The non-analytic
phase in (\ref{ru2}) and (\ref{ru2b}) is therefore associated to operator
ordering in the four-point Green function, as discussed in Ref.\cit{ms}.
The continuation (\ref{ru1}) is also  seen in the  example of Appendix B
 as the 
high-level limit of well-known continuation formulae for hypergeometric 
functions.

We must therefore factor the non-analytic phase out of the preliminary
t-channel blocks to obtain the analytic t-channel blocks. 
 More precisely, we define
\bs
\be
\F_g^{({\rm t})}(y)_m{}^n =  \Fh_g^{({\rm t})}(y)_m{}^p  U_g (y)_p {}^n  
\sp
\Fh_g^{({\rm t})}(y)_m{}^n =  \F_g^{({\rm t})}(y)_m{}^p  (U_g (y)^{-1})_p {}^n  
\label{aast} \ee
\be
\label{pm} 
\eqalign{
U_g(y)_p {}^n & = 
\bv({\rm t},g)_p  
\exp [- 2 \p i L_{g,\infty}^{ab} \T_a^1 \T_b^2 \sign (\arg (-y)) ] 
v({\rm t},g)^n + \ok \cr
& = \d_p^n - i \p c'({\rm t},g)_p{}^n   \sign (\arg (-y)) + \ok  \cr}  
\ee
\be
\eqalign{
(U_g(y)^{-1})_p {}^n  & = 
\bv({\rm t},g)_p   
\exp [ 2 \p i L_{g,\infty}^{ab} \T_a^1 \T_b^2 \sign (\arg (-y)) ] 
v({\rm t},g)^n  + \ok \cr
& = \d_p^n  +   i \p    c'({\rm t},g)_p{}^n   \sign (\arg (-y)) + \ok  \cr}  
\ee
\be
c'({\rm t},g)_p {}^n = \bv({\rm t},g)_p  
2 L^{ab}_{g,\infty}  \T_a^1 \T_b^2 v({\rm t},g)^n  
\ee
\be
U_g (y^*) = U_g(y)^{-1} \sp  U_g(y)^{\dagger} = U_g(y)^{-1}  
\label{ppm} \ee
\es
where $\Fh_g^{({\rm t})}(y)$ are the analytic t-channel
blocks and $U_g(y)_m{}^n$ is the  
non-analytic unitary phase matrix of the affine-Sugawara constructions
(unitary because the sign function is real). 
Then we find the explicit form of the analytic 
t-channel blocks, 
\bs
\be
\Fh_g^{({\rm t})}(y)_m{}^n    
= \bv({\rm t},g)_m [\corrgt ]  v({\rm t},g)^n + \ok 
\label{atcb} \ee
\be 
\label{fta} 
\eqalign{
\;\;\;\;\;\;\;\;\;\; 
 = \bv({\rm t},g)_m  
[\one  -  (2  L^{ab}_{g,\infty} & \T_a^2 \T_b^3 + \sum_{i=1}^3 
\D^g(\T^i) - \D^g(\T^4) ) \ln (-y) ]  \cr
  \ti & [\one  +
 2  L^{ab}_{g,\infty} \T_a^1 \T_b^3 
\ln \left(1- \frac{1}{y}  \right) ]  v({\rm t},g)^n + \ok \cr}  
\ee
\be  
\;\;\;\;\;\;\;\;\; =  (-y)^{-\D^g_{({\rm t})}(m) - \D^g(\T^1) + \D^g(\T^4) } 
\left[ \d_m^n   - c({\rm t},g)_m{}^n  
 \sum_{p=1}^{\infty} 
\left(\frac{1} {y}\right)^p { 1 \over p} \right] + \ok  
\label{ftad} \ee  
\be
\label{ascmt}
c({\rm t},g)_m{}^n = \bv({\rm t},g)_m 2 L^{ab}_{g,\infty} \T_a^1 \T_b^3   
v({\rm t},g)^n  
\pe \ee
\label{asb2b} \es
To obtain (\ref{fta}), we have used the  
identity 
\bs
\be
\bv({\rm t},g)_m [2 L^{ab}_{g,\infty} ( 
\T_a^1 \T_b^2 +  
\T_a^2 \T_b^3 + \T_a^3\T_b^1) - \g_g \one ] =0 
\ee
\be
\g_g = \D^g( \T^4)-\D^g( \T^1)-\D^g( \T^2)- \D^g( \T^3)
\ee
\label{ggwi} \es
which follows from the $g$-global Ward identity (\ref{ascut}).

{} From (\ref{ftad}) we read the limiting behavior of the analytic t-channel 
blocks 
\bs
\be
\Fh^{({\rm t})}_g(y)_m{}^n 
\mathrel{\mathop \sim_{y \ra \infty} }
\Ga_g^{({\rm t})} (m,n)   
 ( -y)^{-\D_{({\rm t})}^g(m,n)-\D^g(\T^1) +\D^g(\T^4) } + \ok
\label{tlim} \ee
\be
\D^g_{({\rm t})}(m,n) =  
\left\{ \matrix{ 
   \D^g_{({\rm t})}(m) + \ok  & \ssp n =m \cr 
 1 + \cO (k^{-1} )  & \ssp n \neq m \cr}
\right.  
\ee
\be  
\Ga_g^{({\rm t})} (m,n) =   
\left\{ \matrix{ 1 +\ok  & \ssp n =m \cr 
 -c({\rm t},g)_m{}^n  + \ok     & \ssp n \neq m \cr}
\right.  
\label{asrt} \ee
\label{usl} \es
and the remarks below (\ref{asb3}) apply in this case as well. 
In particular, one might guess the exact u and t-channel  results 
\bs
\be
\D_{({\rm u})}^g(m,n) = \D^g_{({\rm u})} (m) + 1 - \d_{m,n} \sp \forall \; m,n 
\ee
\be
\D_{({\rm t})}^g(m,n) = \D^g_{({\rm t})} (m) + 1 - \d_{m,n} \sp \forall \; m,n 
\ee
\es
which are in agreement with the KZ example in Appendix B.

In what follows, we introduce a unified notation $\r=$ s, t, u for the three
channels and their corresponding blocks $(\F_g^{(\r)})_m{}^n$,
\bs
\be
\bv(\r,g)_m v(\r,g)^n =\d_m^n \;\;\;\; , \;\;\;\;
 v(\r,g)^m_{\a} \bv(\r,g)_m^\b = (I_g)_\a^\b
\label{com2} 
\ee
\be Y_g(y) = \sum_{m,n} d(\r)^m \F^{(\r)}_g(y)_m{}^n  \bv(\r,g)_n \sp 
\r = {\rm s,t,u}  
\ee
\be
\F^{(\r)}_g(y)_m{}^n = \bv(\r,g)_m [ \corrg ] v(\r,g)^n + \ok
\ee
\be
(\F^{(\r)}_g(y)_m{}^n)^* = \F^{(\r)}_g(y^*)_n{}^m
\ee
\label{asbf} \es
where the last relation follows by unitarity, that is, hermiticity of the 
basic matrices in the correlators.

We finally note that the number $B_g$ of affine-Sugawara blocks in 
each of the channels, 
\be 
B_g  =(d_g )^2 
\label{asnb} \ee
is equal to the square of the dimension $d_g$ of the $g$-invariants in
any channel.   

For the special case of the $3 \bar{3} \bar{3} 3$ correlator on 
$SU(3)$, Appendix B provides a check of our high-level blocks against
the finite-level blocks obtained by Knizhnik and Zamolodchikov \cit{kz}
in this case. 
\vs .4cm
\noindent \underline{Crossing relations}
\vs .3cm

Using completeness of the three sets of eigenvectors, one finds 
that the three sets of blocks are related  by the
crossing relations, 
\bs
\be
\F_g^{(\r)}(y)_m {}^n = [ X_g(\r \s) + \ok ]_m {}^p \,  \F_g^{(\s)}(y)_p{}^q \, 
( [ X_g(\r \s) + \ok ]^{-1})_q {}^n 
\ee
\be
\;\;\; \;\;\;\;\;\;\;\;\;\; \;\;\;\;\;\;
  =  X_g(\r \s)_m {}^p  \F_g^{(\s)}(y)_p{}^q 
 X_g^{-1}(\r \s)_q {}^n + \ok 
\;\;\;\;, \;\;\;\; \r ,\s = {\rm s,t,u} 
\label{cro} \ee
\be
 X_g(\r \s)_m {}^n = \bv(\r,g)_m v(\s,g)^n 
\label{xm} \ee
\be
X_g^{-1}(\r \s)_m{}^n = X_g(\s \r)_m{}^n = (X_g(\r \s)_n{}^m)^*
\label{xuni} \ee
\label{ascro} \es
where $\r \neq \s$ and we call $X_g(\r \s)$ in (\ref{xm})
the crossing matrix from channel $\s$ 
to channel $\r$. The last relation (\ref{xuni}) says that the crossing
matrices $X_g(\r \s)_m{}^n$ are unitary $X_g^\dagger = X_g^{-1}$ for each
$\r \neq \s$, and the
crossing matrices explicitly satisfy the consistency relations  
\bs
\be
X_g(\r \s) X_g(\s \t) X_g(\t \r) = X_g(\r \t) X_g(\t \s) X_g(\s \r) = 1   
\ee
\be
(1)_m {}^n  = \d_m^n 
\ee 
\label{ybl} 
\es 
which says that we return to the same blocks when we go around an
s,t,u cycle.

In the special case when $\T^2 \sim \T^3$, the conformal weights exchanged
in the u-channel are the same as in the s-channel. In further detail, we have
\be
L^{ab}_{g,\infty} (\T_a^1 \T_b^3)_\a {}^\b = 
L^{ab}_{g,\infty} (\T_a^1 \T_b^2)_{\a '} {}^{\b '} 
\label{tt} \ee
in this case, where $\a ' = (\a_1 \a_3 \a_2 \a_4) $ and similarly for $\b '$. 
Then we may identify the $g$-invariants of the u-channel in terms of those
of the s-channel  
\be
 v({\rm u},g)^m_\a = v({\rm s},g)^m_{\a '}  \;\;\;\; , \;\;\;\; 
 \bv({\rm u},g)_m^\a = \bv({\rm s},g)_m^{\a '}  
\ee
where $m=(r,\xi,\xi ')$ is the same irrep $\T^r$ in both channels.
It follows from (\ref{asbu}), (\ref{tt}) and 
(\ref{xm}) that
\bs
\be
X_g({\rm su})^{-1} = X_g({\rm su}) \;\;\;\;, \;\;\;\; X_g({\rm su})^2 =1 
\ee
\be
X_g({\rm us})^{-1} = X_g({\rm us}) \;\;\;\;, \;\;\;\; X_g({\rm us})^2 =1 
\label{xip} \ee
\be
\F_g^{({\rm u})}(y)_m{}^n = \F_g^{({\rm s})} (1-y)_m{}^n
\pe
\label{ussy} \ee
\label{tet} \es
Then, using (\ref{tet}) in (\ref{cro}), one finds that the 
s-channel affine-Sugawara 
blocks close under s-u crossing,
\be
\F_g^{({\rm s})}(1-y)_m {}^n = [ X_g({\rm su}) + \ok]_m {}^p 
\,  \F_g^{({\rm s})} (y)_p {}^q \, [ X_g({\rm su}) + \ok]_q {}^n
\ee
as they should in this case.

In the special case when all four representations are the same, one finds
that the unitary crossing matrices are also idempotent $X(\r \s)^2=1$ and 
hence $X(\r \s) = X( \s  \r)$ for all $\r\neq \s$: then,
the Yang-Baxter-like relation
\be
X_g(\r \s) X_g(\s \t) X_g(\t \r) = X_g(\t \r) X_g(\s \t) X_g(\r \s) = 1   
\label{yb}
\ee
follows from the consistency relations (\ref{ybl}).

Using (\ref{cro}) and (\ref{aast}), we finally write down the crossing 
relations among  all three sets 
$\F_g^{({\rm s})}$, $\F_g^{({\rm u})}$, $\Fh_g^{({\rm t})}$
of analytic affine-Sugawara blocks,  
\bs
\be
\F_g^{({\rm s})} = [ X_g({\rm su}) + \ok]\,  \F_g^{({\rm u})} \,  
 [ X_g({\rm su}) + \ok]^{-1} \;\;\; \;\;  
\ee
\be 
\F_g^{({\rm u})} = [ X_g({\rm us}) + \ok]  \, \F_g^{({\rm s})} \,  
[  X_g ({\rm us}) + \ok ]^{-1} \;\;\;\;\;  
\ee
\be
\F_g^{({\rm s})} = [ X_g({\rm st}) + \ok] \,  \Fh_g^{({\rm t})} \,  
[ X_g({\rm st}) U_g^{-1} + \ok ]^{-1} 
\ee
\be  
\Fh_g^{({\rm t})} = [ X_g({\rm ts}) + \ok ]  \, \F_g^{({\rm s})} \,  
[U_g X_g({\rm ts}) + \ok ]^{-1} \;\;\;  
\ee
\be
\F_g^{({\rm u})} = [ X_g({\rm ut}) + \ok ] \, \Fh_g^{({\rm t})} \,  
[  X_g({\rm ut}) U_g^{-1} + \ok ]^{-1} 
\ee
\be 
\Fh_g^{({\rm t})} = [ X_g({\rm tu}) +\ok ]  \, \F_g^{({\rm u})} \,  
[U_g X_g({\rm tu}) + \ok ]^{-1} 
\ee
\label{fcma} \es
where $U_g$ is the non-analytic unitary phase matrix (\ref{pm}) of the
affine-Sugawara constructions. 
It is known \cit{ms} that the crossing matrices of analytic blocks
involve non-analytic factors, and we remark that, according to eqs.
(\ref{pm}) and (\ref{xm}),  the phase matrix  provides the entire $\oko$ 
corrections to the full crossing matrices in (\ref{fcma}).    

\subsection{Non-chiral WZW correlators}
To construct a set of 
high-level  non-chiral WZW correlators from the affine-Sugawara
blocks, we take the diagonal construction in the s-channel blocks (\ref{asbl}),
\bs
\be
Y_g(y^*,y)_\a{}^\b = 
  \sum_{m,n,p}  (\F_g^{({\rm s})}(y)_p{}^m)^* \F_g^{({\rm s})}(y)_p{}^n 
\, v({\rm s},g)_\a^m\bv({\rm s},g)_n^\b + \ok 
\ee
\be
\;\;\;\;\;\;\;\;\;\;\;\;\;\;\;\;\;\;\;\;\;
= \sum_{m,n} 
v({\rm s},g)_\a^m  \; [\F_g^{({\rm s})}(y^*) \F_g^{({\rm s})}(y)]_m{}^n 
\bv({\rm s},g)_n^\b + \ok 
\ee
\label{asc} \es
which shows trivial monodromy around $y=0$. These correlators satisfy 
the high-level forms of the holomorphic and anti-holomorphic KZ equations,
and the corresponding $g$-global conditions
on the left and the right. In the special case of the $ n \bar{n} \bar{n}n$
correlator on $SU(n)$, they also agree with the diagonal construction studied
by Knizhnik and Zamolodchikov in \cit{kz}.

To see that these correlators have trivial monodromy around $y=1$
and $y = \infty$, one uses the crossing relations (\ref{ascro}) of the
affine-Sugawara blocks to rewrite the correlator (\ref{asc}) 
in the two alternate forms
\bs
\be
Y_g(y^*,y)_\a{}^\b = \sum_{m,n}v({\rm u},g)_\a^m \; 
[\F_g^{({\rm u})}(y^*)\F_g^{({\rm u})}(y)]_m{}^n 
\, \bv({\rm u},g)_n^\b + \ok \;\;\;\;\;
\ee
\be 
\;\;\;\;\;\;\;\;\;\;\;\;\;\;\;\; 
= \sum_{m,n} v({\rm t},g)_\a^m \; 
[\F_g^{({\rm t})}(y^*) \F_g^{({\rm t})}(y)]_m{}^n 
\, \bv({\rm t},g)_n^\b + \ok
  \label{nct} \ee
\label{asc2} \es
where the u and t-channel blocks are given in (\ref{asbu}) and (\ref{asbt}).  

We can also express the t-channel form (\ref{nct}) in terms of the analytic
t-channel blocks (\ref{asb2b}),
\bs
\be
Y_g(y^*,y)_\a{}^\b =  
 \sum_{m,n}v({\rm t},g)_\a^m \; [\Fh_g^{({\rm t})}(y^*) U_g(y^*)  
\Fh_g^{({\rm t})}(y) U_g(y)]_m{}^n 
\, \bv({\rm t},g)_n^\b + \ok
 \ee
\be
\;\;\;\;\;\;\;\;\;\;\;\;\;\;\;\;\;\;
 = \sum_{m,n}v({\rm t},g)_\a^m \; [ \Fh_g^{({\rm t})}(y) U_g(y^*)  
U_g(y) \Fh_g^{({\rm t})}(y) ]_m{}^n 
\, \bv({\rm t},g)_n^\b + \ok
 \ee
\be
=\sum_{m,n}v({\rm t},g)_\a^m \;
[\Fh_g^{({\rm t})}(y^*) \Fh_g^{({\rm t})}(y)]_m{}^n 
\, \bv({\rm t},g)_n^\b + \ok
 \ee
\es
where we have used the fact that 
\be
[A,B] = \ok \;\;\;\; {\rm when}\;\;\;\; A,B = \one + \oko 
\label{cid} \ee
and the first property in (\ref{ppm}) of the phase matrix $U_g$. 

Using completeness and 
the form (\ref{asbl}) of the affine-Sugawara blocks, we also find the
summed  form of the non-chiral WZW correlators
\bs
\be
\label{wzw}
\eqalign{
Y_g (y^*,y)_\a {}^\b  =& \left\{ [ \corrgs ]  I_g \right. \cr
 & \left. \ti [ \corrg ] \right\}_\a{}^\b + \ok \cr}
\ee
\be
\;\;\;\;\;\;\;\;\;\;\;\;\;\;\;\;\;\;\;\;\;\;\;\;\;
 = \{[ \one  +
2 L^{ab}_{g,\infty} (\T_a^1 \T_b^2 \ln | y|^2  +
 \T_a^1 \T_b^3 \ln |1-y|^2 ) ]I_g \}_{\a}{}^{\b} + \ok  
\label{ncas2} \ee
\label{ncas}\es
where $I_g$ is the projector (\ref{com}) onto the $G$-invariant subspace
of $\T^1 \oti\cdots \oti \T^4$, and
we have used eq.(\ref{lic}) to obtain the second form, which explicitly
 shows two of the trivial monodromies. 
The third trivial monodromy, around $y=\infty$, also follows immediately
because both terms in (\ref{ncas2}) are proportional to $|y|$ at large $y$.
The correct t-channel singularities are then obtained by an application of the
$g$-global Ward identity (\ref{ggwi}), using $I_g$ in the form (\ref{ocev2}). 
  
Using the $g$-crossing matrices (\ref{xm}),  
Appendix A gives alternate expressions for the $g$-blocks
(\ref{asbf}), the analytic t-channel $g$-blocks (\ref{asb2b})
 and the  $g$-correlators (\ref{asc}) 

\section{The Coset Constructions \label{sec4}}
\subsection{The coset blocks}
The next simplest, and next most symmetric,  set of conformal field theories 
are the $g/h$ coset constructions [1,2,7,8], whose chiral correlators
are defined by (\ref{cor}) with 
\bs
\be
L^{ab}_{g/h,\infty} = {P_{g/h}^{ab} \over 2k}  \sp
P_{g/h} = P_g - P_h 
\label{gcc} \ee
\be
Y_{g/h}(y) \sum_{i=1}^4 \T_a^i =0 \;\;\;\;, \;\;\;\; a=1 \ldots {\rm dim}
\, h
\ee
\es 
where $h \subset g$. These correlators are the high-level solutions of the 
general coset equations of Refs.[15,16,14] on simple $g$, and the results below are
the high-level form of the general coset blocks studied in [22,15,16,14].

We begin by reorganizing the
high-level coset correlators (\ref{cor}) as,
\be
\label{coscor}
\eqalign{
 Y_{g/h}^\a (y)= \left\{  \bv_g \right. & [ \corrg ]  \cr  
&  \left. \ti [ \corrh ] \right\}^{\a} + \ok \cr} 
\ee 
where we have used (\ref{gcc}) and 
moved the terms  of the $h$ theory to the right.

To define the $\r$=s, t and u-channel coset blocks, we need the $g$-invariant
eigenvectors $v(\r,g)^m$, $\bv(\r,g)_m$ of Section \ref{sec3}, 
and also the corresponding
$h$-invariant eigenvectors $\ps(\r,h)$, 
\bs
\be
2 L^{ab}_{h,\infty} \T_a^1 \T_b^2 \ps({\rm s},h)^M = (\D^h_{({\rm s})} (M)  -
\D^h_{M_1} (\T^1)   -\D^h_{M_2} (\T^2) )  \ps({\rm s},h)^M
\label{hev} \ee
\be
 2 L^{ab}_{h,\infty} 
\T_a^1\T_b^3\ps({\rm u},h)^M =(\D^h_{({\rm u})}(M)-\D^h_{M_1}(\T^1)
-\D^h_{M_3}(\T^3))\ps({\rm u},h)^M
\label{hev2} \ee
\be
\label{hev3} 
2 L^{ab}_{h,\infty}
 \T_a^2 \T_b^3 \ps({\rm t},h)^M = (\D^h_{({\rm t})}(M) -\D^h_{M_2}(\T^2) 
-\D^h_{M_3}(\T^3))
\ps({\rm t},h)^M \;  
\ee 
\be
L^{ab}_{h,\infty} \T_a^i \T_b^i \ps(\r,h)^M = \D_{M_i}^h (\T^i)\ps(\r,h)^M 
\;\;\;\;, \;\;\; i=1 \ldots 4 \;\;\;, \;\;\;\; \r ={\rm s,t,u}  
\label{ses} 
\ee
\be
(\sum_{i=1}^4 \T_a^i ) \ps(\r,h)^M = 0 \;\;\;\;, \;\;\;\;a =1 \ldots
{\rm dim }\,h 
\;\;\;\;, \;\;\r ={\rm s,t,u}  
\label{hgwi} \ee
\es
whose properties parallel those of the $g$-invariants.
In particular, the eigenvalue problems 
(\ref{hev}-c)  are compatible with the diagonalization
of the $h$ conformal weights in (\ref{ses}) because the
   matrices $L^{ab}_{h} \T_a^i \T_b^j$ 
and $ L^{ab}_{h} \T_a^i \T_b^i$ commute. The $h$-global Ward identities
(\ref{hgwi}) are also compatible with the eigenvalue problems, whose
matrices are $h$-invariant. It then follows from the high-level form of
the relation
\be
2 L_{h}^{ab}  \T^i_a \T^j_b = 
 L_{h}^{ab}  (\T^i_a + \T^j_a) (\T^i_b + \T^j_b)   
- (L_h^{ab}\T_a^i \T_b^i  + L_h^{ab} \T_a^j \T_b^j) 
\sp 1 \leq i < j \leq 4
\ee 
that the quantities $\D^h_{(\r)}(M)$ in (\ref{hev}-c) are the
high-level forms of  the broken conformal weights of $h$-irreps
in the $\r$-channel (that is, the decomposition of $\T \oti \T'$ into 
$h$-irreps). 

The $h$-invariant eigenvectors also satisfy 
completeness and orthonormality,
\bs
\be
\bps(\r,h)_M \ps(\r,h)^N =
\d_M^N
\;\;\;\;, \;\;\;\;
 \ps(\r,h)^M \bps(\r,h)_M = I_h \;\;\;\;, \;\;\;\; \r={\rm s,t,u}  
\ee
\be
[ L_{h,\infty}^{ab} \T_a^i \T_b^j, I_h] = 0 \sp 1 \leq i,j \leq 4 
 \ee
\es
where $I_h$ is the projection operator onto the $h$-invariant subspace
of $\T^1 \oti \cdots \oti \T^4$. 

As an explicit example, we give the solution for the $U(1)$-invariant 
s-channel eigenvectors of the coset correlator 
\be
(\T^1 , \T^2,\T^3,\T^4)= 
(j_1,j_2,j_3,j_4)
\;\;\;\; {\rm in} \;\;\;\
  {SU(2)\over U(1)}
 \pe
\ee
 In this case we need
\be 
L^{ab}_{U(1),\infty} = {\d^a_3 \d^b_3 \over 2k }
\;\;\;\;, \;\;\;\; 
\T_3^i = \sqrt{\ps_g^2} \left( \matrix{ j_i &  & 0 \cr & \ddots & \cr 0 & & -j_i }
\right) 
\ee
where $\ps_g^2$ is the $SU(2)$ root length squared and
we have taken the usual magnetic quantum number basis for the matrices, with
$\a_i =M_i$, $|M_i| \leq j_i$. 
The solution of the eigenvalue problem (\ref{hev}) is then 
\bs
\be
\ps({\rm s},U(1))^M_{\a} = \d_\a^M \d( \sum_{i=1}^4 M_i=0) \;\;\;\;, \;\;\;\;
M=(M_1, M_2, M_3, M_4)
\ee
\be
\D^{U(1)}_{({\rm s})}(M) = { (M_1 +M_2)^2 \over x } \;\;\;\;, \;\;\;\;
\D^{U(1)}_{M_i} (\T^i) = { M_i^2 \over x} \;\;\;\;, \;\; i=1 \ldots 4
\ee
\es
where $x = 2k / \ps_g^2$ is the invariant level of $g= SU(2)$.
For more general coset correlators 
the eigenvectors $\ps({\rm s},h )$ are squares of products of Clebsch-Gordan
coefficients times Clebsch-Gordan coefficients for branching of
$g$-irreps into
$h$-irreps \cit{swa}.

Using completeness of $v(g), \bv(g)$ and $\ps(h) , \bps (h)$, we have 
[15,16,14] 
\bs
\be
\bv_g^\a = \sum_m d(\r)^m \bv(\r,g)_m^\a
\ee
\be
Y_{g/h}^\a(y) = \sum_{m,M} d(\r)^m \C^{(\r)}(y)_m{}^M \bps(\r,h)_M^\a 
\ee
\es
where $\C^{(\r)}(y)$ are the coset blocks. Further use of
completeness gives the explicit form of the high-level coset blocks
\bs
\be
\C^{(\r)}(y)_m{}^M = 
\F_g^{(\r)}(y)_m{}^n e(\r,g/h)_n{}^N (\F^{(\r)}_h(y)^{-1})_N{}^M
\sp \r = {\rm s,t,u} 
\label{cbl} \ee
\be
 \F^{(\r)}_h(y)_N{}^M = \bps(\r,h)_N [ \corrhp ]  \ps(\r,h)^M 
+ \ok \ee
\be
 (\F^{(\r)}_h(y)^{-1})_N{}^M = \bps(\r,h)_N [\corrh ] \ps(\r,h)^M 
 + \ok \ee
\be
e(\r,g/h)_n{}^N = \bv(\r,g)_n \ps(\r,h)^N
\ee
\label{ccor} \es
where $\F_g^{(\r)}$ are the $\r$-channel $g$-blocks (of the affine-Sugawara
construction on $g$) given in eq.(\ref{asbf}), and 
 $e(\r,g/h)$ is the embedding matrix of the $g$-invariants $v(g)$ 
in the $h$-invariants $\ps (h)$.
The inverse $h$ blocks $\F_h^{-1}$ are the inverse
 of the $h$ blocks $\F_h$. In Ref.\cit{wi}, the exact coset blocks were written
as $(\C)_m{}^M= (\F_g)_m{}^n (\F_h^{-1})_n {}^M$, where 
$(\F_h^{-1})_n{}^M = e(g/h)_n{}^N (\F_h^{-1})_N {}^M$ in the present notation.

The s and u-channel coset blocks in (\ref{ccor}) are high-level forms
of analytic blocks, as above.
To obtain the analytic t-channel coset blocks, we first
 use the continuation formulae
(\ref{rul}) to find the analytic t-channel $h$ blocks $\Fh_h^{({\rm t})}$
and their inverse, 
\bs
\be
\eqalign{
 \Fh^{({\rm t})}_h(y)_N{}^M=\bps({\rm t},h)_N
[\one +  
2 L^{ab}_{h,\infty} (\T_a^1 [\T_b^2 +\T_b^3]\ln  (-y)
& + \T_a^1 \T_b^3 \ln\left(1-\frac{1}{y}\right)) ] 
\ps({\rm t},h)^M \cr 
& + \ok \cr }
\ee
\be
\eqalign{
(\Fh^{({\rm t})}_h(y)^{-1})_N{}^M = \bps({\rm t},h)_N 
[\one  -
2 L^{ab}_{h,\infty} (\T_a^1 [\T_b^2 +\T_b^3]\ln  (-y)
& + \T_a^1 \T_b^3 \ln\left(1-\frac{1}{y}\right)) ] 
 \ps({\rm t},h)^M \cr  
& + \ok \cr}
\ee
\be
\F_h^{({\rm t})}(y) =  \Fh_h^{({\rm t})}(y) U_h (y)  
\sp
\Fh_h^{({\rm t})}(y) =  \F_h^{({\rm t})}(y) U_h (y)^{-1}   
\ee
\be
U_h(y)_M{}^N   = 
\bps({\rm t},h)_M 
\exp [- 2 \p i L_{h,\infty}^{ab} \T_a^1 \T_b^2 \sign (\arg (-y)) ] 
\ps({\rm t},h)^N  
+ \ok 
\ee
\be
U_h (y^*) = U_h(y)^{-1} \sp  U_h(y)^{\dagger} = U_h(y)^{-1}  
\ee 
\es 
whose form closely parallels that of
 the analytic t-channel blocks $\Fh_g^{({\rm t})}$
in (\ref{atcb}). Here $U_h(y)$ is the non-analytic unitary phase matrix
of the $h$ blocks embedded in $g$.  
Then we may rearrange the t-channel coset blocks as follows,
\bs
\be
\C^{({\rm t})}(y) =  
\Fh_g^{({\rm t})}(y) U_g(y) e({\rm t},g/h) U_h(y)^{-1} \Fh^{({\rm t})}_h(y)^{-1}
\;\;\;\;\;\;\;\;\;\;
 \ee
\be
=\Fh_g^{({\rm t})}(y)e({\rm t},g/h)U_g(y) U_h(y)^{-1} \Fh^{({\rm t})}_h(y)^{-1}
\ee
\be
\;\;\;\;\; =[ \Fh_g^{({\rm t})}(y)e({\rm t},g/h) \Fh^{({\rm t})}_h(y)^{-1}] \,
[ U_g(y)U_h(y)^{-1}]  
\label{icb} \ee
\es
where we have used the fact that  
\bs
\be U_g(y)_m{}^p e({\rm t},g/h)_p{}^M= 
e({\rm t},g/h)_m{}^P U_g(y)_P{}^M
\ee
\be
U_g(y)_P{}^M \equiv  
\bps({\rm t},h)_P  
\exp [- 2 \p i L_{g,\infty}^{ab} \T_a^1 \T_b^2 \sign (\arg (-y)) ] 
\ps({\rm t},h)^M   
+ \ok 
\ee
\es
in the second step and the commutation 
identity (\ref{cid})  in the last step.

From (\ref{icb}), we read the form and properties of the analytic t-channel
coset blocks $\Ch^{({\rm t})}(y)$,
\bs
\be
\Ch^{({\rm t})}(y) =  
\Fh_g^{({\rm t})}(y) e({\rm t},g/h) \Fh^{({\rm t})}_h(y)^{-1}
\label{acbt1} \ee
\be
\C^{({\rm t})}(y) =  \Ch^{({\rm t})}(y) U_{g/h} (y)  
\sp
\Ch^{({\rm t})}(y) =  \C^{({\rm t})}(y) U_{g/h} (y)^{-1}  
\label{acbt} \ee
\be
\label{cpm} 
\eqalign{
U_{g/h}(y)_M{}^N & =U_g(y)_M{}^P (U_h(y)^{-1})_P{}^N  \cr
& =  
\bps({\rm t},h)_M 
\exp [- 2 \p i L_{g/h,\infty}^{ab} \T_a^1 \T_b^2 \sign (\arg (-y)) ] 
\ps({\rm t},h)^N + \ok\cr}  
\ee
\be
U_{g/h} (y^*) = U_{g/h}(y)^{-1} \sp  U_{g/h}(y)^{\dagger} = 
U_{g/h}(y)^{-1}  
\ee
\label{sbc2} \es
where $U_{g/h}(y)$ is the non-analytic unitary phase matrix of the
$g/h$ coset constructions. 

The limiting behavior of the analytic coset blocks 
$\C^{({\rm s})}$, $\C^{({\rm u})}$, $\Ch^{({\rm t})}$
follows from their form in (\ref{cbl}) and (\ref{acbt1}), together with
the results above for $g$ and $h$, 
\bs
\be
\label{cb} 
\C^{({\rm s})}(y)_m {}^M 
\mathrel{\mathop \sim_{y \ra 0}}
 \Ga_{g/h}^{({\rm s})} (m,M)  
y^{\D^{g/h}_{({\rm s})}(m,M)-\D_{M_1}^{g/h}(\T^1)
-\D_{M_2}^{g/h}(\T^2) } +\ok \;\;\;\;\;\; \;\;\;  
\ee
\be
\C^{({\rm u})}(y)_m {}^M 
\mathrel{\mathop \sim_{y \ra 1} } 
 \Ga_{g/h}^{({\rm u})} (m,M)  
  (1-y)^{\D^{g/h}_{({\rm u})}(m,M)-\D_{M_1}^{g/h}(\T^1)
-\D_{M_3}^{g/h}(\T^3) } +\ok  
 \ee
\be
\Ch^{({\rm t})}(y)_m {}^M 
\mathrel{\mathop \sim_{y \ra \infty} } 
 \Ga_{g/h}^{({\rm t})} (m,M)  
(- y)^{-\D^{g/h}_{({\rm t})}(m,M)-\D_{M_1}^{g/h}(\T^1)
+ \D_{M_4}^{g/h}(\T^4) } +\ok  
\ee
 \be
\D_{M_i}^{g/h}(\T^i) = \D^g(\T^i) - \D^h_{M_i} (\T^i) \;\;\;\;, 
\;\;i=1 \ldots 4  
\label{cw2} \ee
$$
\;\;\;\;\; \D^{g/h}_{(\r)}(m,M) =  
 \left\{ \matrix{ 
 \D^g_{(\r)}(m)- \D^h_{(\r)}(M) +\ok\phantom{1}   \sp e(\r,g/h)_m{}^M \neq 0 
\hskip 1.55cm  (4.15{\rm e}) \cr       
1+\cO (k^{-1}) \phantom{\D^g_{(\r)}(m)-\D^h_{(\r)}(M)}\sp e(\r,g/h)_m{}^M = 0   
\hskip 1.55cm  (4.15{\rm f}) \cr }
\right.       
$$
$$
\;\;\;\;\Ga_{g/h}^{(\r)}   (m,M) = 
 \left\{ \matrix{ 
e(\r,g/h)_m{}^M + \ok\phantom{-c(\r,g/h)_N{}^M}  & 
\ssp e(\r,g/h)_m{}^M \neq 0 
\hskip 0.9cm  (4.15{\rm g})\cr   
 - e(\r,g/h)_m{}^N c(\r,g/h)_N{}^M + \ok &  \ssp  e(\r,g/h)_m{}^M = 0 
\hskip 0.9cm  (4.15{\rm h})   
 \cr} 
 \right.   
$$
\label{lcb} 
\es
\vskip -.3cm 
\noindent where the matrices $c(\r,g/h)$ 
\bs
\be
c({\rm s},g/h)_N{}^M 
= \bps({\rm s},h)_N 2 L_{g/h,\infty}^{ab} \T_a^1 \T_b^3 \ps({\rm s},h)^M 
\ee
\be
c({\rm u},g/h)_N{}^M 
=\bps({\rm u},h)_N 2 L_{g/h,\infty}^{ab} \T_a^1 \T_b^2 \ps({\rm u},h)^M 
\ee
\be
\;\;\;\;  c({\rm t},g/h)_N{}^M 
= \bps({\rm t},h)_N 2 L_{g/h,\infty}^{ab} \T_a^1 \T_b^3 \ps({\rm t},h)^M 
\ee
\label{ccd} \es
are defined in analogy to those of the $g$ theory. 

The $g/h$  conformal weights in (\ref{cw2}) and
(4.15e) for $e(\r,g/h)_m{}^M \neq 0$ 
are the correct
conformal weights of the external and intermediate  coset-broken affine
primary fields, and the intermediate broken affine primary states contribute
with residue $\cO (k^0)$, in accord with the general OPE (\ref{ope}).

The $(1+\oko)$ conformal weights in (4.15f) are broken
affine secondaries (with residue $\oko$ in (4.15h))
which are not necessarily integer descendants of broken
affine primaries; see for example the exact conformal blocks
\be
n \bar{n} \bar{n} n  
\;\;\;\; {\rm in} \;\;\;\; 
  {SU(n)_{x_1} \ti  SU(n)_{x_2} \over SU(n)_{x_1+x_2} }
\ee
obtained in Ref.\cit{wi}. All the conformal weights in (4.15d-f) check
against the large $x_1=x_2=x$ form of these blocks.

Appendix D studies a coset example on simple $g$
\be
3 \bar{3} \bar{3} 3   
\;\;\;\; {\rm in} \;\;\;\; {SU(3) \over SU(2)_{\rm irr} }  
\ee
in some detail. This case shows a block which begins at  
$\ok$. 

We finally note that the number $B_{g/h}$ of coset blocks in 
each of the channels,
\be 
B_{g/h}  =d_g  \cdot  d_h
\label{cnb} \ee
is the product of the dimensions $d_g$ and $d_h$ of the 
$g$- and $h$-invariants in any channel. In fact $d_h \geq d_g$ 
because $h \subset g$, so that the inequality 
\be
B_{g/h}  \geq B_g 
\label{hgi} \ee
is obtained for comparison of  correlators with fixed external $g$-irreps,
where $B_g $ in (\ref{asnb}) is the number of affine-Sugawara blocks
in each of the channels. The result (\ref{hgi}) is in accord with the intuitive
 expectation that
the number of blocks grows with increased symmetry breaking.  
 
 \vs .4cm
\noindent \underline{Crossing relations}
\vs .3cm

Following the development of the previous section we find the crossing 
relations
for the embedding matrix  and the (inverse) $h$-blocks,
\bs
\be
e(\r,g/h)_m{}^M = X_g(\r \s)_m{}^n e(\s, g/h)_n{}^N X_h^{-1}(\r \s)_N{}^M
\label{cro1} \ee
\be
(\F_h^{(\r)}(y)^{-1}){}_M{}^N = [ X_h(\r \s) + \ok]_M {}^P 
 \, (\F_h^{(\s)}(y)^{-1}){}_P{}^Q \, 
([ X_h(\r \s) + \ok]^{-1})_Q{}^N
\label{cro2} \ee
\label{cro3} \es
where $X_g(\r \s)$ are the $g$-crossing matrices (\ref{xm}) and 
 $X_h(\r \s)$ are the corresponding  $h$-crossing matrices, 
\bs
\be
 X_h(\r \s)_{M}{}^N = \bps(\r,h)_M \ps (\s,h)^N 
\ee
\be
X_h^{-1}(\r \s)_{M}{}^N = X_h(\s \r )_M{}^N =
(X_h(\r \s)_N{}^M)^* 
\ee
\label{ccr} \es
which are also unitary. Using (\ref{cro}) and  (\ref{cro3})   
we obtain the crossing relations  of the coset blocks,
\be
\C^{(\r)} (y)_m{}^M  =[ X_g(\r \s) +\ok]_m{}^n 
 \, \C^{(\s)}(y)_n{}^N \, 
( [ X_h(\r \s ) + \ok]^{-1})_N{}^M  
\label{cocr} \ee
which involve, as expected, the crossing matrices $X_g$ and $X_h$ 
of $g$ and of $h$. 

The $h$-crossing matrices satisfy the same consistency relations, 
\bs
\be
X_h(\r \s) X_h (\s \t) X_h (\t \r) =
X_h(\r \t ) X_h (\t  \s ) X_h (\s  \r ) = 1   
\ee
\be
(1)_M{}^N = \d_M^N
\ee
\label{ybl2}  \es 
which were seen for the $g$-crossing matrices  in (\ref{ybl}).

When the external $g$-irreps satisfy
$\T^2 \sim \T^3$, we find that 
$X_h({\rm us})^2 =1 $ and  $\F_h^{({\rm u})} (y) =\F_h^{({\rm s})} (1-y) $, as for
the $g$-blocks. Together with the
corresponding relations for the $g$-quantities in this case, this implies
\be
 e({\rm u},g/h) =e({\rm s},g/h)
\sp 
 \C^{({\rm u})} (y) =\C^{({\rm s})} (1-y)
\ee
and then,
\be
\C^{({\rm s})} (1-y)_m{}^M  =[ X_g ({\rm su}) +\ok]_m{}^n 
\,  \C^{({\rm s})}(y)_n{}^N \, 
[ X_h ({\rm su}) +\ok]_N{}^M 
\ee
so that the s-channel
 coset blocks are closed under crossing in this case, as expected.

Using (\ref{cocr}) and (\ref{acbt}), we finally write down the crossing 
relations among all three sets
$\C^{({\rm s})}$, $\C^{({\rm u})}$, $\Ch^{({\rm t})}$
 of analytic coset blocks, 
\bs
\be
\C^{({\rm s})} = [ X_g({\rm su}) + \ok]  \, \C^{({\rm u})} \,  
 [ X_h({\rm su}) +\ok]^{-1} \;\;\;\;\;  
\ee
\be
\C^{({\rm u})} = [ X_g({\rm us}) + \ok ] \, \C^{({\rm s})} \,  
 [ X_h ({\rm us}) +\ok]^{-1} \;\;\;\;\;  
\ee
\be
\C^{({\rm s})} = [ X_g({\rm st}) + \ok] \, \Ch^{({\rm t})} \,  
[  X_h({\rm st}) U_{g/h}^{-1} + \ok]^{-1} 
\ee
\be
\Ch^{({\rm t})} = [ X_g({\rm ts}) + \ok] \, \C^{({\rm s})} \,  
[U_{g/h} X_h({\rm ts}) + \ok ]^{-1} 
\ee
\be
\C^{({\rm u})} = [ X_g({\rm ut}) + \ok] \, \Ch^{({\rm t})} \,  
[  X_h({\rm ut}) U_{g/h}^{-1} +\ok]^{-1}  
\ee
\be
\Ch^{({\rm t})} = [ X_g({\rm tu}) +\ok] \,  \C^{({\rm u})} \,  
[U_{g/h} X_h({\rm tu}) +\ok ]^{-1} 
\ee
\label{fcmc} 
\es
where $U_{g/h}$ is the non-analytic unitary phase matrix (\ref{cpm}) of the
$g/h$ coset constructions. As seen above for the affine-Sugawara 
constructions, the phase matrix provides the entire $\oko$ corrections
to the full coset crossing matrices in (\ref{fcmc}).  

\vs .4cm
\noindent \underline{Fixed external $h$ representations}
\vs .3cm 

The crossing relations (\ref{cocr}) of the coset blocks mix the
internal $h$-irreps ($M$) which arise from different external irreps of
$h$ (that is, the $h$-irreps 
which arise from the $h$-decomposition of the $g$-irreps $\T^i$).

To obtain blocks characterized by
fixed external irreps of $h$, we introduce a hermitean projection operator
$\P_h=\P(T^{h1},T^{h2},T^{h3},T^{h4})$ to select any four external $h$-irreps
of interest, 
\bs
\be
 \ps(\r,h)_a^{\tM} \bps  (\r,h)
^\b_{\tM} = (\P_h)_\a{}^\b 
\ee
\be
\P_h \ps(\r,h)^M =  \ps(\r,h)^{\tM} \d_{\tM}^M 
\label{hpo} \ee
\be
[L^{ab}_{h,\infty}\T_a^i \T_b^j,\P_h] = 
[L^{ab}_{g/h,\infty}\T_a^i \T_b^j,\P_h] = 0 \sp 1 \leq i,j \leq 4  
\label{lpc} \ee
\es 
where $\tM$ runs over the eigenvectors  associated to the fixed external set of
$h$-irreps. The inverse $h$ blocks are block diagonal under this decomposition
\be
\eqalign{
(\F_h^{(\r)}& (y)^{-1})_M{}^{\tN } \cr  
& = \bps(\r,h)_M  
[\corrh ] \ps(\r,h)^{\tN}  
 + \ok \cr
& = \bps(\r,h)_M  
[\corrh ] \P_h \ps(\r,h)^{\tN}  
 + \ok \cr
& = \bps(\r,h)_M \P_h   
[\corrh ]  \ps(\r,h)^{\tN}  
 + \ok \cr
& = \d_M^{\tM} (\F_h^{(\r)}(y)^{-1})_{\tM}{}^{\tN }\cr}  
\ee
where we have used the first relation in (\ref{lpc}). Then the corresponding
subset of  coset blocks is
\be
(\C)_m {}^{\tM} =( \F_g)_m {}^n e(g/h) _n{}^N (\F_h^{-1})_N{}^{\tM} 
=( \F_g)_m {}^n e(g/h) _n{}^{\tilde{ N} }  (\F_h^{-1})_{\tilde{ N}}{}^{\tM} 
\label{cbr} \pe \ee
Similarly, the $h$-crossing matrices are block diagonal under this 
decomposition,
\be
 \label{bdc} 
X_h^{-1}(\r \s )_M{}^{\tilde{N}}  = 
\bps (\s ,h)_M \ps(\r,h)^{\tilde{N}}
= \bps (\s ,h)_M \P_h \ps(\r,h)^{\tilde{N}} 
= \d_M^{\tM} X_h^{-1}(\r \s )_{\tM}{}^{\tilde{N}} \pe 
\ee
Then, it follows from (\ref{cocr}) and (\ref{bdc}) that 
\be
\C^{(\r)}(y)_m{}^{\tM} = [ X_g (\r \s ) + \ok]_m{}^n \, 
 \C^{(\s )}(y)_{n }{}^{\tilde{N}} \,  
( [ X_h(\r \s ) + \ok]^{-1})_{\tilde{N}}{}^{\tM} 
\label{rcr} \ee
which shows that the selected subset of coset blocks is closed under crossing

The selected subset of analytic coset blocks $\C^{({\rm s})}(y)_m{}^{\tM}$, 
$\C^{({\rm u})}(y)_m{}^{\tM}$  and $\Ch^{({\rm t})}(y)_m{}^{\tM}$ 
is also closed under crossing.
To see this we need the fact the non-analytic coset phase matrix (\ref{cpm})
is also block diagonal, 
\bs
\be
U_{g/h}(y)_M{}^{\tN} = \d_M^{\tM}  U_{g/h}(y)_{\tM}{}^{\tN}    
\label{cpmd} \ee
\be
\C^{({\rm t})}(y)_m{}^{\tM}  =  \Ch^{({\rm t})}(y)_m{}^{\tN} 
U_{g/h} (y)_{\tN}{}^{\tM}  
\sp
\Ch^{({\rm t})}(y)_m{}^{\tM}  =  \C^{({\rm t})}(y)_m{}^{\tN} 
(U_{g/h} (y)^{-1})_{\tN}{}^{\tM}  
\label{acbts}
\ee
\es
which follows from (\ref{cpm}) and the second relation in 
(\ref{lpc}).  The restricted phase matrix $U_{g/h}(y)_{\tM}{}^{\tN}$
is unitary in each subspace. Then, we have for example that
\bs
\be
\C^{({\rm s})}(y)_m{}^{\tM} = X_g ({\rm st})_m{}^n \Ch^{({\rm t} )}(y)_n{}^R  
U_{g/h} (y)_R{}^{\tN}   X_h^{-1}({\rm st})_{\tN}{}^{\tM} +\ok  
\ee
\be
\;\;\;\;\;\;\;\;\; \;\;\;\;\;\;\;\;\;\;\;  
= X_g ({\rm st})_m{}^n \, [ \, \Ch^{({\rm t})}(y)_n{}^{\tR}  ] \,  
U_{g/h}(y)_{\tR}{}^{\tN}   X_h^{-1}({\rm st})_{\tN}{}^{\tM} + \ok  
\ee
\es
where the last step follows from (\ref{cpmd}). 

The explicit form of these projection operators can be quite complicated
in the general case, but there are some simple, highly symmetric cases
where the form of $\P_h$ is very simple. As an example, consider the situation 
when each of the four external $g$-irreps branches into a single $h$-irrep,
so that the   
 $g/h$-broken conformal weights of $g$-irrep $\T^i$ are degenerate,
\be
(L^{ab}_{g/h,\infty} \T_a^i  \T_b^i)_{\a}{}^{\b} = \D^{g/h} (\T^i) \d_{\a}^{\b}
\;\;\;\;, \;\;\;\; i =1 \ldots 4
\pe
\ee
In this case, all the coset-broken components of the $g$-irrep $\T^i$ are
on an equal footing, and one may choose the trivial projector  
\be
\P_h = \one 
\pe \ee 
This is the situation, e.g., in 
\be
 \T = (\T_1,1)
\;\;\;\; {\rm in} \;\;\;\; {g_{x_1} \times g_{x_2} \over g_{x_1 +x_2} } 
\label{ee1}  \ee
examples of which were studied in Ref.\cit{wi}. Examples on simple $g$ 
include 
\bs 
 \be
\T  = n \;\,{\rm or}\;\, \bar{n} 
\;\;\;\; {\rm in } \;\;\;\; 
 {SU(n)_x \over SO(n)_{2x}} =
\left\{ \matrix {SU(3)_x \over SU(2)_{4x} & \sp n=3 \cr 
{SU(n)_x \over SO(n)_{2x}} & \sp n \geq 4  \cr} \right. 
\label{ex1} \ee
\be
\T  = 2 n 
\;\;\;\; {\rm in } \;\;\;\; 
{SO  (2n )_x \over SO(n)_{x} \ti SO(n)_x }  
\label{ex1b} \ee
\label{e1} 
 \es 
and the case $n=3$ of (\ref{ex1}) will be considered in detail in 
Appendix D.  
In (\ref{ex1})  the $n$ of $SU(n)$ is the $n$
 of $SO(n) \subset SU(n) $, while in (\ref{ex1b}) the $2n$  of $SO(2n)$ is 
 the $(n, n)$ of $(SO(n) \ti SO(n)) \subset SO(2n) $.
As we will discuss below, these simple cases are examples of a more general
situation in ICFT (see Section \ref{sec5}). 
\subsection{Non-chiral coset correlators}

To construct a set of high-level non-chiral correlators for the coset
constructions, we take
the s-channel diagonal construction,  
\be
Y_{g/h} (\P_h| y^*,y) 
=\sum_{m, \tM } |\,  \C^{({\rm s})} (y)_m{}^{\tM} |^2 + \ok 
\label{nccc} \ee
which shows trivial monodromy around $y=0$. 
To see that (\ref{nccc}) has trivial monodromy
around $y=1$ and $y= \infty$, one uses the crossing relations (\ref{rcr})
of the coset blocks to rewrite the coset correlator (\ref{nccc}) in
the two alternate forms,  
\bs
 \be
Y_{g/h} (\P_h| y^*,y)  =
\sum_{m, \tM } | \, \C^{({\rm u})} (y)_m{}^{\tM} |^2 + \ok 
\ee
\be
\;\;\;\;\;\;  \;\; \;\;\;\;\;\;\;\; \;\;\;\;\;\;\;\;\;
  =\sum_{m, \tM } | \, \C^{({\rm t})} (y)_m{}^{\tM} |^2  +\ok 
\pe \label{tcfc} \ee
\es
 We can also use (\ref{acbts}) to express the t-channel form (\ref{tcfc})
 of the correlator
in terms of the analytic t-channel  coset blocks,  
\be
Y_{g/h} (\P_h| y^*,y)  =
\sum_{m,\tN ,\tM } |\,\Ch^{({\rm t})} (y)_m{}^{\tN} U_{g/h}(y)_{\tN}{}^{\tM} |^2
+ \ok =\sum_{m, \tN } | \,  \Ch^{({\rm t})} (y)_m{}^{\tN} |^2  +\ok 
\ee
where the last step follows from the unitarity 
of the restricted coset phase matrix.

Using completeness and the explicit form (\ref{ccor}) of the coset blocks, 
the summed form of these coset correlators is
\bs
\be
\eqalign{Y_{g/h} (\P_h|y^*,y )
= & {\rm Tr} \left\{ [ \corrghs ] I_g \right.  \cr
& \;\;\left. \ti [\corrgh ] \P_h \right\} +\ok \;\;\;\;\;\;\;\;\;\;\;\;\;\; \cr}
\ee
\be
\;\;\;\;\;\;\;\;\;\;\;\;\;\;\;\;\;\;\;\;\;\;\;\;\;\;  
= {\rm Tr}   [(\one +  2 L^{ab}_{g/h,\infty} (\T_a^1 \T_b^2 \ln | y|^2 + 
\T_a^1 \T_b^3 \ln |1-y|^2 ))  I_g \P_h ] +\ok     
\ee
\label{sncc} \es
where $I_g$ is the projector onto the $G$-invariant subspace of
$\T^1 \oti \cdots \oti \T^4$ and
$ \P_h $ is the projector onto the desired subset of external $h$ 
representations. To obtain the second form, which explicitly shows two of
the trivial monodromies, we used the second relation in (\ref{lpc}).
Trivial monodromy around $y=\infty$ is also easily seen following the  
discussion below eq.(\ref{ncas}).

\boldmath 
\section{A Simple Class of Correlators in ICFT \label{sec5}}
\unboldmath 

\boldmath 
\subsection{$L(g;H)$-degenerate states and correlators}
\unboldmath 

In this section, we use the intuition gained in our discussion of the
affine-Sugawara and coset constructions above to identify what we believe 
to be the simplest, most 
highly symmetric processes in ICFT.

In the first place , we restrict our attention to the ICFTs with a 
symmetry, that is, to the $H$-invariant CFTs on $g$, whose inverse inertia 
tensors $L_H$ satisfy 
\be
\o (H) L_H \, \o(H)^{-1} = L_H
\sp \o(H) \in H
\label{lhi} \ee
where
$ H \subset G$ is any  subgroup of $G$, including finite groups and the Lie
groups. The matrix $\o(H)_a{}^b $ is in the adjoint of $g$. For the $H$-invariant
CFTs, the conformal weight matrix of 
irrep $\T$ of $g$ and hence the broken conformal weights 
$\D_\a^H (\T) $ are $H$-invariant,  
\bs
\be
\O (H,\T) L_H^{ab}\, \T_a \T_b \; \O^{-1 }(H,\T) = L_H^{ab} \, \T_a \T_b  
\sp \O(H,\T)  \in H 
\ee
\be
\O (H,\T)_{\a}{}^{\b} [\D_\a^H (\T) - \D_\b^H (\T) ] = 0 
\label{cwi} \ee
\es
where $\O (H,\T)_{\a}{}^{\b} $ is in irrep $\T$ and we have used (\ref{cwm})
to obtain (\ref{cwi}).  

In the $H$-invariant CFTs, we further restrict ourselves to the
 most symmetric broken affine primary fields, that is, to the irreps
$\T$ of $g$ whose $L^{ab}$-broken
 conformal weights $\D_\a^H(\T) = \D^H (\T)$, $\a = 1 \ldots {\rm dim}\,\T$
 are completely
degenerate
\be
(L_H^{ab} \T_a \T_b)_{\a}{}^{\b} = \D^H(\T) \d_{\a}^{\b}
\label{cwd}
\ee
at all levels.
In what follows, such irreps of $g$ are called the $L(g;H)$-degenerate
states because, in these cases, the irrep of $g$ decomposes into a unique
irrep of $H$. Finally, we restrict the discussion to the 
 $L(g;H)$-degenerate
processes, which are those correlators all of whose external states are 
$L(g;H)$-degenerate. In this sense, the $L(g;H)$-degenerate processes are 
the most symmetric correlators in ICFT. 

Although they are by no means generic, it is easy to find examples of
$L(g;H)$-degenerate states in the $H$-invariant CFTs. The simplest cases
of $L(g;H)$-degenerate states are all the affine primary states of all
the affine-Sugawara constructions, which are in fact $L(g;G)$-degenerate.
 
Examples of $L(g;h)$-degenerate states in the 
 $g/h$ coset constructions include those
 mentioned in 
(\ref{ee1}) and  (\ref{e1}).
These are RCFT examples  in the Lie $h$-invariant CFTs, and in principle
many  irrational examples, beyond the coset constructions,
 can be found among the
 Lie $h$-invariant
CFTs. 

Irrational examples in the much larger set of
$H$-invariant CFTs,
beyond the Lie $h$-invariant CFTs, are already known, including 
 the irrational cases  \cit{nuc} 
\bs 
 \be
\T = n \;\,{\rm or}\;\, \bar{n}
\;\;\;\; {\rm in} \;\;\;\;
(SU(n)_x)^\#_M 
\label{ex2} \ee
 \be
\T = 2 n 
\;\;\;\; {\rm in} \;\;\;\;
(SO  (2n )_x)^\#_M 
\label{ex2b} \ee
\label{e2} \es
where $H$ is a finite subgroup of $SO(n)\subset SU(n)$ and $(SO(n) \ti SO(n)) 
\subset SO(2n)$ in (\ref{ex2}) and (\ref{ex2b}) respectively.  
The case $n=3$ in (\ref{ex2}) will be considered in detail in 
Section \ref{sec6}. 

We should also remark that the $L(g;H)$-degenerate conformal weights 
of the coset  examples in (\ref{e1})
and the irrational examples in (\ref{e2}) all obey the unified
 conformal weight formula,  
\be
\D^H_\a (\T) =  \D^H (\T)  = {c \over 2 x n  } 
\ee
where $x$ is the invariant level of $g$ and
$c$ is the central charge, which is rational for the coset
constructions and  irrational for $SU(n)^\#_M$ and
$SO(2n)^\#_M$.
The  occurence of \newline
a) $L(g;H)$-degenerate states \newline
b) a  unified form of the
conformal weights \newline
for these rational and irrational families is  not totally surprising, since 
both families of
constructions are contained in the same (maximally-symmetric) ansatz
\cit{nuc} of the Virasoro master equation.

In what follows, we will find 
uniform formulae for the high-level conformal blocks and correlators
of all possible $L(g;H)$-degenerate processes in ICFT.

\subsection{Conformal blocks in ICFT} 
We study only the class of
$L(g;H)$-degenerate correlators in ICFT. 
 Fig.2 shows these correlators
generically, with one degenerate conformal weight $\D_i^H \equiv \D^H (\T^i)$, 
$ i =1 \ldots 4$ for 
each external state. 
\begin{center}
\begin{picture}(200,130)
\put (10,40){\line(4,3){40}}
\put (10,100){\line(4,-3){40}}
\put (50,70){\line(1,0){90}}
\put (140,70){\line(4,3){40}} 
\put (140,70){\line(4,-3){40}} 
\put(-10,35){$\D_1^H$}
\put(-10,100){$\D_2^H$}
\put(185,100){$\D_3^H$}
\put(185,35){$\D_4^H$}
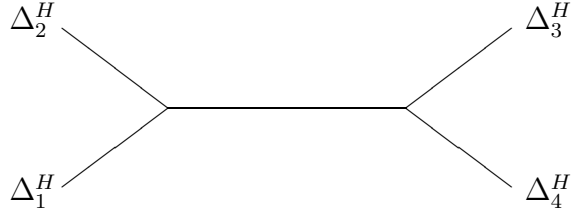
\put(-20,0){Fig. 2. The $L(g;H)$-degenerate correlators.}
\end{picture} 
\end{center} 

In this case, the chiral correlators (\ref{cor}) take the form,
\bs
\be
Y_H^\a (y) = \bv_g^\b \L_H (y)_\b{}^\a  + \ok 
\;\;\;\;\;\;\;\;\;\;
\ee
\be
\label{la1} 
\L_H (y) 
 = \one + 2 L^{ab}_{H,\infty} [ \T_a^1 \T_b^2 \ln y +  \T_a^1\T_b^3  \ln (1-y)]
\;\;\; \;\;\;\;\;\;\;\;\;\;\;\;\; \;\;\;\;\;\;\;\;\;  
\ee
\be
\label{la2} \eqalign{
\;\;\;\;\;\;\;\;\;\;\;\;\;\;\;\;\;\;\;\; 
= \one & + [L^{ab}_{H,\infty} (\T_a^1+\T_a^2) (\T_b^1+\T_b^2)-
 (\D_1^H +\D_2^H )\one] \ln y \cr 
& +[L^{ab}_{H,\infty} (\T_a^1+\T_a^3) (\T_b^1+\T_b^3) - (\D_1^H + \D_3^H)\one] 
\ln (1-y)  \cr} 
\ee
\be
Y_H \O  (H)  =Y_H  \sp \O(H)= \prod_{i=1}^4 \O(H,\T^i)  
\pe
\label{lh2}
\ee
\label{lhc} \es 
Here we have used the high-level forms of the
identities
\bs
\be
L_{H}^{ab} \T_a^i \T_b^i =  \D_i^H \one \sp   i =1 \ldots 4
\label{lghd} \ee
\be
2 L_{H}^{ab}  \T^i_a \T^j_b = 
 L_{H}^{ab}  (\T^i_a + \T^j_a) (\T^i_b + \T^j_b)   
-( \D^H_i +\D^H_j) \one 
\sp 1 \leq i < j \leq 4  
\label{ttr} \ee
\es
to obtain the alternate form in (\ref{la2}). The statement in (\ref{lghd}) is
the $L(g;H)$-degeneracy of each external state.  
The condition (\ref{lh2}), which enforces the $H$-symmetry of the system,
follows from the $H$-invariance of the relevant matrices  
\be
[\L_H ,\O  (H)]  =0
\ee
and the fact that $\bv_g$, being $g$-invariant, is also invariant under $\O(H)$.

To find  $\r=$ s, t  and u-channel block bases for the conformal blocks, we
 introduce the $H $-invariant
  eigenvectors  $\ps(\r,H)$ of the $\r$-channel, 
\bs
\be
 2 L^{ab}_{H,\infty} \T_a^1\T_b^2 \ps({\rm s},H)^M = (\D_{({\rm s})}^H(M)  
-\D_1^H -\D_2^H)  \ps({\rm s},H)^M
\label{evga} \ee
\be
2  L^{ab}_{H,\infty} \T_a^1\T_b^3 \ps({\rm u},H)^M = 
(\D_{({\rm u})}^H(M)  -\D_1^H -\D_3^H)  \ps ({\rm u},H)^M
\ee
\be
 2 L^{ab}_{H,\infty} \T_a^2\T_b^3 \ps({\rm t},H)^M = 
(\D_{({\rm t})}^H(M)  -\D_2^H -\D_3^H)  \ps({\rm t},H)^M
\ee
\be
\bps(\r,H)_M \ps(\r,H)^N = \d_M^N \;\;\;\;, \;\;\;\;
 \ps(\r,H)_\a^M \bps(\r,H)_M^\b  = (I_H )_\a^\b 
\label{gcl} \ee
\be
\O^{-1} (H) \ps(\r,H)^M = \ps(\r,H)^M 
\sp  \bps(\r,H)_M \O (H) = \bps(\r,H)_M     
\ee
\label{evg} \es
where $(I_H)_\a^\b$ is the  projector onto the $H$-invariant subspace of
$\T^1 \otimes \cdots \otimes \T^4$. 
According to the identity (\ref{ttr}), the quantities  
$\D_{(\r)}^H(M)$ are  the $L^{ab}$-broken
high-level
conformal weights  of the broken affine primary states  in the $\r $-channel.

We remind the reader that the correlators  (\ref{lhc}) include all the
correlators in $H$-invariant CFTs with $L(g;H)$-degenerate external 
states. This includes
in particular all the correlators of all the affine-Sugawara constructions, 
in which case the eigenvectors $\ps(\r,H)$ may be
taken as the $g$-invariants $v(\r,g)$ of Section \ref{sec3}, and all
the coset correlators whose external states are
$L( g;h)$-degenerate, in which case  the eigenvectors $\ps(\r,H)$
may be identified as the $h$-invariants $\ps(\r,h)$ of Section \ref{sec4}.

The $\r= $ s, t  and u-channel conformal blocks $\B^{(\r)}$ 
are then obtained by inserting completeness sums in 
(\ref{lhc}), according to 
\be
\L_H =\L_H I_H = \L_H  \ps(\r,H)^M  \bps(\r,H)_M\sp\forall\; \r  
\pe
\ee 
In this way, we obtain the three expansions,
\bs
\be
Y_H^\a (y) = \sum_{m,M} d({\rm s})^m \B^{({\rm s})} (y)_m{}^{M} \bps({\rm s},H)_M^\a
\ee
\be
\;\;\;\;\;\;\;\;\; = \sum_{m,M} d({\rm u})^m \B^{({\rm u})} (y)_m{}^{M} \bps({\rm u},H)_M^\a
\ee
\be
\;\;\;\;\;\;\;\; = \sum_{m,M} d({\rm t})^m \B^{({\rm t})} (y)_m{}^{M} \bps({\rm t},H)_M^\a
\ee
\es
where the $\r$-channel blocks $\B^{(\r)} (y)$ are
\bs
\be
\label{gsub}
\B^{(\r)} (y)_m{}^{M} = \bv(\r,g )_m  \L_H (y) \ps(\r,H)^M + \ok 
\;\;\;\;\;\;\;\;\;\;\;\;\;\;\;\;\;\;\;\;\;\;\;\;\;\; \;\;\;\;\;\;\;\;\;\;\;\;\; 
\ee
\be
\;\;\;\;\;\;\;\;\;\;\;\;\;\;\;\;\;\;  
= e (\r,H)_m{}^N \bps(\r,H)_N \L_H (y) \ps(\r,H)^M +\ok   
\sp \r = {\rm s,t,u} 
\label{efb} \ee
\vskip -.3cm
\be
\L_H (y) 
= \one+2 L^{ab}_{H,\infty} [ \T_a^1 \T_b^2 \ln y +  \T_a^1\T_b^3  \ln (1-y) ]  
\ee
\vskip -.3cm
\be
e(\r,H)_m{}^M = \bv(\r,g)_m \ps(\r,H)^M  \pe  
\label{gem} \ee
\label{gb} 
\es
Here $e(\r,H)$ is the embedding matrix of the $g$-invariants in the
$H$-invariants.  

The s and u-channel blocks  $\B^{({\rm s})}$ and $\B^{({\rm u})}$  
are analytic blocks, as above, and 
the  analytic t-channel blocks $\Bh^{({\rm t})}$,  
\bs
\be  
\label{atcbh} 
\Bh^{({\rm t})} (y)_m{}^{M} = \bv({\rm t},g )_m  \hat{\L}_H (y) 
 \ps({\rm t},H)^M + \ok 
\;\;\;\;\;\;\;\;\;\;\;\;\;\;\;\;\;\;\;\;\;\;\;\;\;\; \;\;\;\;\;\;\;\;\;\;\;\;\; 
\ee
\be
\;\;\;\;\;\;\;\;\;\;\;\;\;\;\;\;\;\;  
= e ({\rm t},H)_m{}^N \bps({\rm t},H)_N \hat{\L}_H (y) \ps({\rm t},H)^M +\ok   
\phantom{\sp \r = {\rm s,t,u}}  
\ee 
\be 
\hat{\L}_H(y) =
\one + 2 L^{ab}_{H,\infty} (\T_a^1 [\T_b^2 +\T_b^3]  \ln  (-y)    
  + \T_a^1 \T_b^3 \ln\left(1-\frac{1}{y}\right) )  
\ee 
\be
\B^{({\rm t})} (y) = 
\Bh^{({\rm t})} (y) U_H (y) 
\sp \Bh^{({\rm t})} (y) = 
\B^{({\rm t})} (y) U_H (y)^{-1}  
\label{agtb} \ee
\be
\label{hpm} 
U_H(y)_M{}^N =  
\bps({\rm t},H)_M \exp [- 2 \p i L_{H,\infty}^{ab} \T_a^1 \T_b^2 
\sign (\arg (-y)) ] \ps({\rm t},H)^N  
 + \ok  
\ee
\be
U_{H}(y^*)= U_H(y)^{-1} \sp  U_{H}(y)^\dagger   = U_{H}(y)^{-1}
\label{phpm} \ee
\label{sb2} \es
are also obtained by now-familiar steps, including the
continuation rules (\ref{rul}). 
The quantity $U_H(y)$ in (\ref{hpm}) is the non-analytic unitary phase matrix of
the $L(g;H)$-degenerate correlators in ICFT.
 
The expressions (\ref{gsub},b) and (\ref{atcbh},b) for the high-level analytic
blocks  $ \B^{({\rm s})}$,  $ \B^{({\rm u})}$, $\Bh^{({\rm t})}$
of the $L(g;H)$-degenerate correlators in ICFT are among the central
results of this paper.  

To study the limiting behavior of the analytic blocks, we use
the eigenvalue problems (\ref{evg}) to rearrange the blocks in each of
the channels as follows,
\bs
\be
\eqalign{
 \B^{({\rm s})}(y)_m {}^M  
= e({\rm s},H)_m{}^N 
\bps({\rm s},H)_N [ \one + & 2  L^{ab}_{H,\infty} \T_a^1 \T_b^3 \ln (1-y) ] 
\ps({\rm s},H)^M  \cr  
&\ti  [ 1 +  (\D^H_{({\rm s})}(M) -  \D_1^H - \D_2^H ) \ln y ] + \ok \;\;\;\;\;
\;\;\;\; \cr}
\ee
\be
\;\;\;\;\;\;\;\;\;=  e({\rm s},H)_m{}^N 
\left[ \d_N^M - c({\rm s},H)_N{}^M \sum_{p=1}^{\infty} {y^p \over p} \right] 
y^{\D^H_{({\rm s})}(M) - \D_1^H - \D_2^H } 
+ \ok   
\ee  
\be
c({\rm s},H)_N{}^M  = \bps({\rm s},H)_N 2 L^{ab}_{H,\infty} \T_a^1 \T_b^3 
\ps({\rm s},H)^M 
\ee
\be
 \B^{({\rm u})}(y)_m {}^M  
=  e({\rm u},H)_m{}^N 
\left[ \d_N^M - c({\rm u},H)_N{}^M \sum_{p=1}^{\infty} {(1-y)^p \over p} \right]
(1-y)^{\D^H_{({\rm u})}(M) - \D_1^H - \D_3^H } 
+ \ok   
\ee  
\be
c({\rm u},H)_N{}^M  = \bps({\rm u},H)_N 2 L^{ab}_{H,\infty} \T_a^1 \T_b^2  
\ps({\rm u},H)^M 
\ee
\be
 \eqalign{
 \Bh^{({\rm t})}(y)_m {}^M  
= \bv({\rm t},g)_m  [\one +  & 2L^{ab}_{H,\infty} \T_a^1 [\T_b^2 +\T_b^3] 
\ln (-y) ] \cr 
& \ti [ \one 
+ 2 L^{ab}_{H,\infty} \T_a^1 \T_b^3  \ln \left( 1 -\frac{1}{y} \right) ] 
\ps({\rm t},H)^M +\ok  \;\;\;\;\;\;\;\;\; \;\;\;\;\;\;\;\;\;\;\;  \cr}  
\ee 
\be
\label{bba} \eqalign{
\;\;\;\;\;\; \; \;\; \;\;\;  = e({\rm t},& H)_m{}^N  \bps({\rm t},H)_N 
   [\one +  
 2  L^{ab}_{H,\infty} \T_a^1 \T_b^3 
\ln \left( 1- \frac{1}{y}  \right) ] \cr  
 & \ti [\one  -  (2  L^{ab}_{H,\infty} \T_a^2 \T_b^3  +\sum_{i=1}^3 
\D^H_i  - \D^H_4 ) \ln (-y) ]  \ps({\rm t},H)^M  + \ok \cr}  
\ee
\be  
\;\;\;\;\;\;\;\;\;\;\;\;\;\;
= e({\rm t},H)_m{}^N 
\left[ \d_N^M  - c({\rm t},H)_N{}^M   \sum_{p=1}^{\infty} 
\left(\frac{1} {y}\right)^p { 1 \over p}  \right] 
(-y)^{-\D^H_{({\rm t})}(M) - \D^H_1 + \D^H_4 } 
+ \ok 
\ee  
\be
c({\rm t},H)_N{}^M   = \bps({\rm t},H)_N 2 L^{ab}_{H,\infty} \T_a^1 \T_b^3 
\ps({\rm t},H)^M   
\pe
\ee
\label{ebb} \es
To obtain the form  (\ref{bba}) of the analytic  
t-channel blocks we also used
the  $g$-global Ward identity on the $g$-invariants $\bv(\r,g)_m $,  
\bs
\be
\bv(\r,g)_m [ 2 L_{H}^{ab}( 
\T_a^1\T_b^2+  
\T_a^2\T_b^3 +
\T_a^3\T_b^1) - \g_H \one    ] 
= 0 
\ee
\be
\g_H =\D^H_4 - \D^H_1  - \D^H_2 - \D^H_3  
\ee 
 \label{higw} 
\es
applied here at high level for the case $\r$=t. 

Using the expressions in (\ref{ebb}), we find the limiting behavior of the 
conformal blocks 
\bs
\be
\label{bss} 
\B^{({\rm s})}(y)_m{}^{M} 
\mathrel{\mathop\sim_{y \ra 0} }
\Ga_H^{({\rm s})} (m,M)  
y^{\D_{({\rm s})}^H (m,M)-\D_1^H-\D_2^H}+\ok \;\;\;\;\;\;\;\;\;\;\; \;\;\;\;\; 
 \ee
\be
\label{bus} 
\B^{({\rm u})}(y)_m{}^{M}
\mathrel{\mathop\sim_{y \ra 1} }
\Ga_H^{({\rm u})} (m,M)  
 (1-y)^{\D_{({\rm u})}^H(m,M) -  \D_1^H-\D_3^H}+\ok \;\;\;\;\;\;\;\;  
\ee
\be
\label{bts} 
\Bh^{({\rm t})}(y)_m{}^{M} 
\mathrel{\mathop\sim_{y \ra \infty} }
\Ga_H^{({\rm t})} (m,M)  
( -y)^{-\D_{({\rm t})}^H (m,M)- \D_1^H+\D_4^H}
 +\ok \;\;\;\;\;\;\;\;\;  
\ee
$$
\;\;\;\;\;\;\;\;\;\;\;\;\; \D^{H}_{(\r)}(m,M) =  
 \left\{ \matrix{ 
 \D^H_{(\r)}(M)  +\ok\phantom{1}   \sp e(\r,H)_m{}^M \neq 0 
\hskip 2.9cm  (5.16{\rm d}) \cr       
1+\cO (k^{-1}) \phantom{\D^H_{(\r)}(M)}\sp e(\r,H)_m{}^M = 0   
\hskip 2.9cm  (5.16{\rm e}) \cr }
\right.       
$$
\be
\setcounter{equation}{6}
\label{lre} 
\Ga_H^{(\r)}  (m,M)  
= 
 \left\{ 
\matrix{e(\r,H)_m{}^M +\ok &     \sp  e(\r,H)_m{}^M \neq 0 \cr
 -e(\r,H)_m{}^N c(\r,H)_N{}^M   + \ok 
 &  \sp e(\r,H)_m{}^M = 0 \cr} \right. 
\ee
\es
followed by integer-spaced secondaries.
The blocks with $e(\r,H)_m{}^M \neq 0$ begin at $\cO (k^0)$ and
 exhibit leading singularities (with $\cO (k^0)$ residues) whose 
 high-level conformal weights  $\D_{(\r)}^H(M)$ in (\ref{evg})
are those    of the correct
broken affine-primary states in each of the three channels. 
The remaining blocks, which begin at  $\oko $, show leading singularities 
 which are
broken affine secondaries. As noted for the affine-Sugawara and coset
constructions in Sections \ref{sec3} and \ref{sec4}, 
this pattern is in agreement with the general OPE in (\ref{ope}).  

Further discussion of these conformal weights follows that given for the 
coset constructions below (\ref{ccd}). 
In particular, as noted for the cosets,
the $(1 +\oko)$  $\r$-channel conformal weights in (5.16e) are broken affine
secondaries which need not be
integer descendants of broken affine primary states.  
Identification of these states is therefore an important open problem in ICFT.

\vs .4cm
\noindent \underline{Number of blocks}
\vs .3cm 

We finally note that, for an $L(g;H)$-degenerate process,
the number $B_{H}$ of blocks in each of the channels 
\be 
B_{H}  =d_g \cdot d_H
\label{hnb} \ee
is the product of the dimension $d_g $ of $g$-invariants and the dimension
$d_H$ of $H$-invariants in any channel. We know that 
$d_H \geq d_h  \geq d_g $ 
when $H$ is a finite subgroup of the Lie group
generated by $h \subset g$, and hence we obtain the double inequality 
\be
B_H  \geq B_{g/h}  \geq B_g 
\label{die} \ee
for comparison of correlators with fixed external $g$-irreps,
where $B_{g/h} $ and $B_g$ in (\ref{cnb}) and (\ref{asnb}) are the
number of coset and affine-Sugawara blocks respectively in any channel. 
This double
 inequality summarizes the symmetry hierarchy within the
$L(g;H)$-degenerate processes, and is in accord with the expectation that the
number of blocks increases with increased symmetry breakdown in ICFT.  

In Appendices B and  D and Section \ref{sec6}, we study the $L(g;H)$-degenerate
correlator $3 \bar{3} \bar{3} 3$ under the three constructions,
\ben
\item[$\bullet$] the affine-Sugawara construction on $SU(3)$
\item[$\bullet$] the coset construction $SU(3)/SU(2)_{\rm irr} $
\item[$\bullet$] the irrational construction $\sumh$ 
\een
to illustrate the double inequality (\ref{die}). As discussed below, the
symmetry hierarchy for these three constructions is
 $SU(3) \supset SU(2)_{\rm irr} \supset O$, where $SU(2)_{\rm irr}$ is the
irregular embedding of $SU(2)$ in $SU(3)$ and $O$ is the
octohedral group symmetry of the irrational construction. 

\subsection{Crossing relations} 
Using the completeness relations (\ref{com2}) and (\ref{gcl})
 of the $g$-invariant and 
$H$-invariant eigenvectors  respectively, we verify 
the crossing relations among the blocks, 
\bs
\be
\B^{(\r)} (y)_m{}^{M} = [ X_g (\r \s) +\ok]_m{}^n 
\, \B^{(\s)} (y)_n{}^{N} \, ([ X_H(\r \s) + \ok]^{-1})_N{}^M
\label{gcro} \ee
\be
X_H(\r \s)_M{}^N = \bps(\r,H)_M \ps(\s,H)^N 
\label{gxm} \ee 
\be
X_H^{-1}(\r \s)_M{}^N 
=X_H(\s \r)_M{}^N 
= (X_H(\r \s)_N{}^M)^*
\ee
\label{gc} \es
where $X_g (\r \s )$ is the affine-Sugawara crossing matrix defined  in 
(\ref{ascro}) and
$X_H(\r \s )$ in (\ref{gxm})  
 is another set of unitary crossing matrices, called the
$H$-crossing matrices, from the $\s $-channel to the
$\r $-channel. 

 The $H$-crossing matrices satisfy the same consistency relations
\bs
 \be
X_H(\r \s) X_H(\s \t) X_H(\t \r) = X_H(\r \t) X_H(\t \s) X_H(\s \r) =  1 
\ee
\be
(1)_M{}^N  = \d_M^N 
\ee
\label{ybl3} \es 
found for $g$ and $h$ in (\ref{ybl}) and (\ref{ybl2}).

When the external $g$-irreps satisfy  $\T^2 \sim \T^3$, we may take
\be
\ps({\rm u},H)^M_\a = \ps({\rm s},H)^M_{\a '} \;\;\;\;, \;\;\;\;
\bps({\rm u},H)_M^\a = \bps({\rm s},H)_M^{\a '} 
\label{csc} \ee
and then one finds that,
\bs
\be
\L_H(y)_{\a '}{}^{\b '} = 
\L_H(1-y)_{\a }{}^{\b } 
\ee
\be
\B^{({\rm u})} (y) = \B^{({\rm s})} (1-y) 
\ee
\be
X_H({\rm su}) = X_H^{-1}({\rm su}) = X_H({\rm us})
\label{zip} \ee
\es
where $\a ' =(\a_1 \a_3 \a_2 \a_4)$. It follows that the 
set of s-channel blocks
 is closed under crossing 
\be
\B^{({\rm s})} (1-y)_m {}^{M } = [ X_g({\rm us}) + \ok]_m{}^n 
\, \B^{({\rm s})} (y)_n  {}^{N } \, [ X_H({\rm us}) + \ok]_N{}^M  
\label{bsi} \ee
as it should be in this case. Similar relations
hold when any two external states are the same.

Using (\ref{gcro}) and (\ref{agtb}), 
we finally write down the crossing relations of the three sets 
$\B^{({\rm s})}$, $\B^{({\rm u})}$,   $\Bh^{({\rm t})}$  of  
analytic blocks,  
\bs
\be
\B^{({\rm s})} = [ X_g ({\rm su}) +\ok] \,  \B^{({\rm u})} \,  
[ X_H({\rm su}) +\ok]^{-1} \;\;\;\;\;  
\ee
\be 
\B^{({\rm u})} = [ X_g  ({\rm us}) +\ok]  \, \B^{({\rm s})} \,  
 [ X_H ({\rm us}) +\ok]^{-1} \;\;\;\;\;  
\ee
\be
\B^{({\rm s})} = [X_g  ({\rm st}) + \ok] \, \Bh^{({\rm t})} \,  
[  X_H({\rm st}) U_H^{-1}  +\ok]^{-1}
\ee
\be 
\Bh^{({\rm t})} = [ X_g ({\rm ts}) +\ok]  \, \B^{({\rm s})} \, 
[U_H X_H({\rm ts}) +\ok ]^{-1} \;\;\;  
\ee
\be
\B^{({\rm u})} = [ X_g  ({\rm ut}) + \ok ] \,  \Bh^{({\rm t})} \,  
[  X_H({\rm ut}) U_H^{-1} + \ok]^{-1}
\ee
\be 
\Bh^{({\rm t})} = [ X_g ({\rm tu}) +\ok]  \, \B^{({\rm u})} \,  
[U_H X_H({\rm tu}) + \ok ]^{-1} 
\label{satc} \ee
\label{fcmh} \es
where $U_H$ is the non-analytic unitary phase matrix (\ref{hpm}) of the
$L(g;H)$-degenerate correlators. As seen above for the affine-Sugawara and
coset constructions, the phase matrix provides the entire $\oko$ 
corrections to the full crossing matrices in (\ref{fcmh}).  

For the special case of the  $L(g;h)$-degenerate coset correlators 
(with $L_H = L_{g/h}$ and $\ps (H) = \ps(h)$)
the general high-level blocks (\ref{gb}), (\ref{sb2}) reduce  
precisely to the $L(g;H)$-degenerate subset of
 high-level coset blocks computed in (\ref{ccor}), (\ref{sbc2}).
In the same way, the crossing relations (\ref{fcmh}) reduce in this case
to the coset crossing relations in (\ref{fcmc}).

\subsection{Non-chiral correlators in ICFT}

For the general $L(g;H)$-degenerate process, we construct a set of high-level
 non-chiral correlators using the diagonal construction in the
s-channel blocks (\ref{gb}),   
\be
Y_H(y^*,y)  
=\sum_{m,M} | \, \B^{({\rm s})} (y)_m{}^{M} |^2 + \ok 
\label{ncc} \ee
which shows trivial monodromy around $y=0$. Using the crossing relations  
(\ref{gc}) we can also express this correlator in terms of u or t-channel
blocks
\bs 
 \be
Y_H (y^*,y) 
=\sum_{m,M} | \, \B^{({\rm u})} (y)_m{}^{M} |^2 + \ok  
\ee
\be
 \;\;\;\;\;\;\;\;\;\;\;\;\;\;\;\;
=\sum_{m,M} | \, \B^{({\rm t})} (y)_m{}^{M} |^2  + \ok  
\ee
\es 
which show trivial monodromy around $y=1 $ and $y=\infty$ respectively. 

The non-chiral correlator can also be expressed in terms
of the analytic t-channel blocks (\ref{atcbh}), 
 \be
Y_H(y^*,y) =  
\sum_{m,N,M } | \, \Bh^{({\rm t})} (y)_m{}^{N} U_H(y)_N{}^M  |^2 + \ok 
 =\sum_{m,N} | \, \Bh^{({\rm t})} (y)_m{}^{N}   |^2  + \ok  
\ee
where the last step  uses the unitarity (\ref{phpm}) of
the phase matrix $U_H$. 

Using completeness and
the explicit form of the conformal blocks, we also obtain the
summed  form of the non-chiral correlators 
\bs
\be
\label{c1}
\eqalign{ Y_H (y^*,y) = & {\rm Tr} \{ [ \corrHs ] I_g  \cr
  &\ti [\corrH ] \} + \ok \cr}
\ee
\be
\;\;\;\;\;\;\;\;\;\;\;\;\;\;\;\;\;\;\;\;\;\;\;\;\;\;\;\;\;\;
 = {\rm Tr}[
(\one  +
2 L^{ab}_{H,\infty} (\T_a^1 \T_b^2 \ln  |y|^2  +\T_a^1 \T_b^3 \ln |1-y|^2 ))I_g]
+\ok
\ee 
\label{ncsf} \es  
where $I_g$ is the projector onto the $G$-invariant subspace of
$\T^1 \oti \cdots \oti \T^4$.
The last form explicitly shows two of the trivial monodromies, and trivial 
monodromy around $y=\infty$ is easily seen following the  
discussion below eq.(\ref{ncas}).  
One also sees the expected crossing symmetry
\be
 Y_H(1-y^*,1-y) = Y_H(y^*,y)
\label{snc} \ee
when $\T^2 \sim \T^3$. We finally note that the general 
$L(g;H)$-degenerate 
 correlators (\ref{ncsf}) correctly include the
$L(g;h)$-degenerate  coset
correlators obtained from (\ref{sncc}) when \linebreak $\P_h=1$.  

Using the embedding matrices (\ref{gem}) and the $H$-crossing matrices 
(\ref{gxm}),
Appendix A gives alternate expressions for the blocks and correlators of
the $L(g;H)$-degenerate processes in ICFT.

\boldmath 
\section{Blocks and Correlators in ${\bf SU(3)_M^\#}$ \label{sec6}} 
\unboldmath 

As an explicit example in irrational conformal field theory, we work out 
here the high-level conformal blocks and non-chiral correlators for a 
particular
$L(g;H)$-degenerate process in the unitary irrational level-family \cit{nuc}
\be
(SU (3)_x)_M^\#
 \ee
where $x$ is the invariant level of $SU(3)$. For simplicity below,
this construction is often called $\sumh $. The construction is included in 
the larger maximally-symmetric ansatz for
all simply-laced $g$, which was in fact the first set of ICFTs found in the
Virasoro master equation. The closely related coset construction 
$SU(3)_x/SU(2)_{4x}$, which also resides in the maximally-symmetric ansatz,
is studied in Appendix D.

The exact forms of the central charge and the conformal weights of the
3 and $\bar{3}$ representations under  $\sux$ are
\bs
\be
c [(SU (3)_x)_M^\#] ={2 x \over x+3} \left[ 2   - 
{ x^2 - 8x + 17 
\over 
\sqrt{ 4 x^4 - 28 x^3 + 17x^2 + 160x -128} }\right]   
\label{cc} \ee
\be
\D( \T_{(3)}) =\D(\T_{(\bar{3})} ) = {c \over 6x } 
\label{sucw} \ee  
\es
where the 3-fold degenerate conformal weights in (\ref{sucw}) strongly suggest
that the 3 and $\bar{3}$ are $L(g;H)$-degenerate representations.

As discussed further in Appendix C, the level-family $\sux$ has a
finite group symmetry 
\be
H (\sumh) = O \subset SU(2)_{\rm irr}
\ee
where $O$ is the octohedral group and $SU(2)_{\rm irr}$ is the
irregularly embedded $SU(2) \subset SU(3)$. The degeneracy of the 
3 and $\bar{3}$ is due to the octohedral symmetry of the construction,
which mixes all three components of each representation. Thus the
3 and $\bar{3}$ are $L(SU(3);O)$-degenerate representations in
$\sux$, as desired. 

For the high-level computations in $\sux$ below, we need only the
high-level forms of the inverse inertia tensor (in the  Gell-Mann basis)
and the degenerate conformal weights,
\bs
 \be
L^{ab}_{O,\infty} = {1 \over x  \psi_g^2  } \theta_a \d_{ab}   
\sp  
\th_a = \left\{ \matrix{ 1 & a =1,4,6 \cr 0 &  a = 3,8, 2,5,7  \cr} \right. 
\label{ihl}\ee
\be
c = 3 + \cO (x^{-1})
\ee
\be
\D^O(\T_{( 3)}) =\D^O(\T_{(\bar{3})} ) =  {1 \over 2 x} + \cO (x^{-2})   
\ee
\es
which identifies $P^{ab} = \th_a \d_{ab}$ as the high-level projector of
$\sumh$. Moreover, we will consider only the $L(SU(3);O)$-degenerate 
process $3 \bar{3} \bar{3} 3$ in $\sumh$, 
\be
\T^1 = \T^4 = \T_{(3)} \sp \T^2 = \T^3 =\T_{(\bar{3})}
\ee
shown schematically in Fig.3. The matrix irrep of the 3 and $\bar{3}$ in the
Gell-Mann basis are given by, 
\bs
\be
\T_{(3)} = {\sqrt{\ps_g^2} \over 2} \l_a \sp
\T_{(\bar{3})} ={\sqrt{\ps_g^2} \over 2} \bar{\l}_a 
\label{fur} \ee
\be
 \bar{\l}_a =- \l_a^T = \left\{ \matrix{-  \l_{a} & a=3,8,1,4,6 \cr
\l_a & a=2,5,7 \cr} \right.
\label{gmb} \ee
\label{gmb2} \es
where $\l_a$, $a=1 \ldots 8$  are the Gell-Mann matrices.

\begin{center}
\begin{picture}(200,130)
\put (10,40){\line(4,3){40}}
\put (10,100){\line(4,-3){40}}
\put (50,70){\line(1,0){90}}
\put (140,70){\line(4,3){40}} 
\put (140,70){\line(4,-3){40}} 
\put(0,35){$3$}
\put(0,100){$\bar{3}$}
\put(185,100){$\bar{3}$}
\put(185,35){$3$}
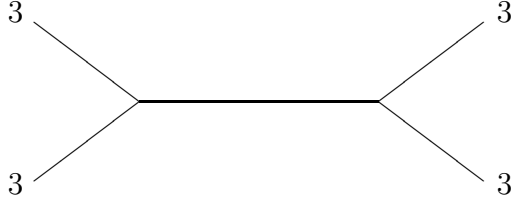
\put(-55,0){Fig. 3.  An $L(SU(3);O)$-degenerate correlator in $\sumh$.}
\end{picture} 
\end{center} 

To compute the high-level blocks in the s-channel, we need to solve 
the eigenvalue problem (\ref{evga}) for the s-channel 
$O$-invariant  eigenvectors $\ps({\rm s},O)$, 
which reads in this case,
\bs 
\be
  [ - \frac{1}{2x } \sum_{a=1,4,6}  \l_a^1  
\l_a^2  +\frac{1 }{x} \one ]_{\a}{}^{\b}  \; \ps({\rm s},O)^M_{\b }  = \D_{({\rm s})}^O(M) 
        \ps({\rm s},O)^M_{\a }   
\label{iev} \ee
\be 
\prod_{i=1}^4 (\o_l^i)_{\a_i}{}^{\b_i} \ps({\rm s},O)_{\b}^M =
 \ps({\rm s},O)_{\a}^M     
 \sp l =1,2 
\label{ihi} \ee
\be  
\o_1 = \exp ( i \pi \l_2/2 ) =   
 \left(  \matrix{ 0 & 1 &  0 \cr  
-1 & 0 &  0 \cr  0 & 0 &  1 \cr}
\right)
\ee
\be 
 \o_2 = \exp (i \pi\l_5/2 ) \exp( i \pi \l_7/2 ) =  
 \left(  \matrix{ 0 & -1 &  0  \cr  
0  & 0  &  1  \cr  -1 & 0 &  0 \cr}
\right)
\pe
\ee
\es
\vskip -.3cm
\noindent
The matrices $\o_1$ and $\o_2$ which appear in the $O$-invariance
condition (\ref{ihi}) may be taken as the generators of $O$.

After some algebra, one finds the following orthonormal set of s-channel 
eigenvectors $\ps({\rm s},O)^M$ and their eigenvalues $\D_{({\rm s})}^O(M)$,
\bs
\be
 \ps({\rm s},O)_\a^1   
= \frac{1}{3} 
\d_{ \a_1 \a_2} \d_{\a_3 \a_4} 
\sp  \D_{({\rm s})}^O(1) = 0
\label{cwv}\ee
\be
\ps({\rm s},O)_\a^2    
=  {1 \over 2 \sqrt{3 }} [ 
\d_{ \a_1 \a_3 } \d_{\a_2  \a_4} 
+ \d_{ \a_1 \a_4  } \d_{\a_2  \a_3 } 
-2 \d_{ \a_1 \a_2  } \d_{\a_3   \a_4} 
\d_{\a_1 \a_3  } ]
 \sp  \D_{({\rm s})}^O(2) = \frac{1}{2 x } 
\label{cwa1} \ee
\be
\ps({\rm s},O)_\a^3     
= \frac{1}{3 \sqrt{2} } [  
\d_{ \a_1 \a_2} \d_{\a_3 \a_4}    
- 3  \d_{ \a_1 \a_2  } \d_{\a_3   \a_4} \d_{\a_1 \a_3} ]  
\sp \D_{({\rm s})}^O(3) =\frac{3}{2x }  
\label{cwa2} \ee
\be
\ps({\rm s},O)_\a^4      
=  {1 \over 2 \sqrt{3 }} [ 
\d_{ \a_1 \a_3 } \d_{\a_2  \a_4} 
-  \d_{ \a_1 \a_4  } \d_{\a_2  \a_3 }]  
\sp  \D_{({\rm s})}^O(4) = \frac{3}{2x } 
\label{cwa3} \ee
\be
\bps({\rm s},O)^\a_M  = ( \ps({\rm s},O)^M_\b)^* \eta^{\b \a} = \ps({\rm s},O)^M_\a  
\ee 
\label{ies} \es 
where the last relation says that the left and right eigenvectors 
coincide in this case.

In ICFT,  the high-level fusion rules [16,14] of the broken affine primaries
follow the Clebsch-Gordan coefficients of their corresponding matrix irreps,
so  the s-channel
should show the exchange of  broken affine primary states corresponding
to the vacuum and the adjoint representation,
\be
 3 \oti \bar{3} = 1 \oplus 8 + \oko  
\pe
\label{fr1} \ee
Indeed, the first conformal weight in (\ref{cwv}) is the conformal weight
of the vacuum, and the other three high-level conformal weights in 
(\ref{cwa1}-d) are precisely the high-level form
of the  three
degenerate subsets of broken conformal weights of the adjoint (see
Appendix C). 

Similarly, we can solve for the u and t-channel eigenvectors, which are
given by 
\bs 
\be
 \ps({\rm u},O)^M = \ps({\rm s},O)^M \vert_{2 \lra 3}  \sp \D_{({\rm u})}^O(M) = \D_{({\rm s})}^O(M) 
\label{icuv} \ee
\be 
\ps({\rm t},O)^M = \ps({\rm s},O)^M \vert_{2 \lra 4 }  \sp \D_{({\rm t})}^O(M) =\frac{2}{x} -
  \D_{({\rm s})}^O(M) 
\ee
\es
 where $2 \lra 3$ and $2 \lra 4$ mean respectively $\a_2 \lra \a_3$ and
$ \a_2 \lra \a_4$ in the explicit expressions of the s-channel eigenvectors
(\ref{ies}).   
The result in (\ref{icuv}) is in accord with 
 (\ref{csc}) since  
$\T^2 \sim \T^3$, so that  the u-channel conformal weights are identical to
the ones in the s-channel. 
The conformal weights found in the t-channel, 
\be
 \D_{({\rm t})}^O(M) = (\frac{2}{x},\frac{3}{2x},\frac{1}{2x},\frac{1}{2x})
\label{tiw} \ee
are also in agreement with the conformal weights of broken affine primaries
 in the known high-level fusion rule
\be
 3 \otimes 3 = 6 \oplus \bar{3} + \oko  
\pe
\label{fr2} \ee
In particular, the last value in (\ref{tiw}) is
the completely degenerate conformal weight of the $\bar{3}$ and the first three 
coincide with the three degenerate subsets (\ref{cw6}) of the 6, according
to  the high-level form (\ref{hli}).

Using eq.(\ref{gxm}), the high-level s-u and s-t $O$-crossing matrices 
are computed from the eigenvectors  as
\bs 
 \be
X_O ({\rm us})_M{}^N = \ps({\rm u},O)^M  \ps({\rm s},O)^N  =
\frac{1}{6} \left( 
\matrix{ 2 & 2 \sqrt{3} &- 2 \sqrt{2} &  2 \sqrt{3}   \cr
2 \sqrt{3}  & 3   &\sqrt{6 } & - 3  \cr
- 2 \sqrt{2}  & \sqrt{6}  & 4 &  \sqrt{6 } \cr
 2 \sqrt{3}  &  -3  &  \sqrt{6 } & 3   \cr}
\right) 
\ee 
\be
X_O ({\rm  ts})_M{}^N = \ps({\rm t},O)^M  \ps({\rm s},O)^N  =
\frac{1}{6} \left( 
\matrix{ 2 & 2 \sqrt{3} & - 2 \sqrt{2} & - 2 \sqrt{3} \cr
2 \sqrt{3}  & 3   &\sqrt{6 } &   3   \cr
- 2 \sqrt{2}  &  \sqrt{6}  & 4 & -  \sqrt{6 } \cr
- 2 \sqrt{3}  &  3  & -  \sqrt{6 } & 3  \cr}
\right)  
\ee 
\label{icr} \es 
\vskip -.3cm 
\noindent which are orthogonal and idempotent  matrices in this case. The third 
$O$-crossing
matrix 
\be
X_O({\rm ut}) =X_O({\rm us}) X_O({\rm ts})
\ee
 follows from the consistency relation  (\ref{ybl3}).  

For the crossing of the blocks one also needs the high-level affine-Sugawara
crossing matrices (\ref{xm}) for $g=SU(3)$.
 The $\r$-channel $SU(3)$-invariant
eigenvectors and the corresponding crossing matrices are 
\bs
\be
v({\rm s},SU(3))^V_\a = \frac{1}{3 } \d_{\a_1 \a_2} \d_{\a_3 \a_4} 
\label{asca} \ee
\be
v({\rm s},SU(3))^A_\a = \frac{1}{2 \sqrt{2 } }[  \d_{\a_1 \a_3 }\d_{\a_2  \a_4} 
- \frac{1}{3}   \d_{\a_1 \a_2} \d_{\a_3 \a_4}] 
\label{ascb}\ee
\be 
 v({\rm u},SU(3)) = v({\rm s},SU(3))\vert_{2 \lra 3}
\label{ascc} \ee
\be
v({\rm t},SU(3))^{6}_\a = \frac{1}{2 \sqrt{6}}  
[  \d_{\a_1 \a_2  } \d_{\a_3   \a_4} 
+   \d_{\a_1 \a_3 } \d_{\a_2  \a_4}] 
\label{ascd} \ee
\be
 v({\rm t},SU(3))^{\bar{3}}_\a = \frac{1}{2 \sqrt{3 }}  
[  \d_{\a_1 \a_2  } \d_{\a_3   \a_4} 
-    \d_{\a_1 \a_3 } \d_{\a_2  \a_4}] 
\label{asce} \ee
 \be
X_{SU(3)} ({\rm us})_m{}^n   = v({\rm u},SU(3))^m  v({\rm s},SU(3))^n= 
\frac{1}{3} \left( \matrix{  1 & 2 \sqrt{2} \cr 
                          2 \sqrt{2} & -1 \cr }\right) 
\label{ascf} \ee       
\be 
X_{SU(3)} ({\rm ts})_m{}^n   = v({\rm t},SU(3))^m  v({\rm s},SU(3))^n= 
\frac{1}{3} \left( \matrix{  \sqrt{6}  &  \sqrt{3} \cr 
                           \sqrt{3} & - \sqrt{6}  \cr }\right) 
\label{ascg} \ee 
\label{ascr2} \es 
\vskip -.3cm
\noindent
where the labels $V,A$ stand for vacuum and adjoint irrep, and $6,\bar{3}$ for 
 symmetric and antisymmetric irrep. The third $g$-crossing matrix is given by
$X_{SU(3)}({\rm ut}) = X_{SU(3)}({\rm us})$ $X_{SU(3)}^{-1}({\rm ts})$. 

Finally, we write down the 8 
high-level s-channel conformal blocks (\ref{gb}) of the $3 \bar{3} \bar{3} 3$
correlator in $\sumh$,  
\be
\Bo^{({\rm s})}(y)_m{}^M  = e({\rm s},O)_m {}^N  [1    +  
(\D_{({\rm s})}^O- \frac{1}{x } \on  )\ln y + (Q_{ ({\rm su})}^O- \frac{1}{x }
 \on  ) 
 \ln (1-y)]_N{}^M 
   +\cO (x^{-2})     
\label{suhb} \ee
where $(1)_N{}^M = \d_N^M$ and   
\bs
 \be
e({\rm s},O) _m{}^M = v({\rm s},SU(3))^m \ps({\rm s},O) ^M =
\left(
\matrix{ 1 & 0 & 0 & 0 \cr
 0 & \frac{1}{4}\sqrt{6} & -\frac{1}{2} & \frac{1}{4}\sqrt{6}  \cr }
\right)
\label{hem} \ee
\be
(\D_{({\rm s})}^O) _N{}^M = \D_{({\rm s})}^O(M)  \d_N^M  
 \sp 
\D_{({\rm s})}^O(M) = 
 \D_{({\rm u})}^O(M)  = (0,\frac{1}{2x},\frac{3 }{ 2 x}, \frac{3 }{2x} ) 
\label{sumd} 
\ee 
\be
(Q_{({\rm su})}^O)_N{}^M  
= \sum_{L }       X_O({\rm us})_{N}{}^L  \D_{({\rm u})}^O(L )    X_O ({\rm us})_L  {}^M 
= \frac{1}{12x} \left( \matrix{ 
12 & -4 \sqrt{3}  & 0 & 0 \cr 
-4 \sqrt{3}  & 9  & \sqrt{6}  & - 3  \cr 
0 &  \sqrt{6}  & 12  & 3 \sqrt{6}  \cr 
0 &  -3   & 3 \sqrt{6} & 9   \cr} \right)  
\pe
\ee
\label{sicb} \es
Here we have used the alternate expression (\ref{gba}) for the 
$L(g;H)$-degenerate blocks in Appendix A.
The u and t-channel blocks can be computed from the s-channel blocks above 
using the crossing relation (\ref{gc}) and the explicit forms of the
 crossing matrices $X_{SU(3)}({\rm us}), X_{SU(3)}({\rm ts})$ in (\ref{ascr2}), 
and $ X_O({\rm us}), X_O({\rm ts})$ in (\ref{icr}).   
Moreover, using the explicit form of the non-analytic phase matrix 
(\ref{hpm}) for this process,  
\be
U_O(y)_M{}^N  =\sum_L  
 X_O({\rm ts})_{M}{}^L  \exp\left(-\pi i[ \D_{({\rm s})}^O(L )-\frac{1}{x} ] 
\sign(\arg(-y)) \right)  X_O ({\rm ts})_L {}^N  
+ \ox 
\ee
the analytic t-channel blocks follow from the crossing relation (\ref{satc}) 

Using (\ref{bss},d-f) we obtain the following limiting behavior as
$y \ra 0$ for the 8  s-channel blocks (\ref{suhb}) of this
correlator,  
\bs 
\be
\Bo^{({\rm s})}(y)_m{}^{M} 
\mathrel{\mathop\sim_{y \ra 0} }
\Ga_O^{({\rm s})} (m,M)  
y^{\D_{({\rm s})}^O (m,M)-1/x} +\cO (x^{-2}) \sp m=V,A \sp M = 1,2,3,4
 \ee
\be
\D_{({\rm s})}^O (V,1) = 0 + \cO (x^{-2}) 
\sp 
\D_{({\rm s})}^O (A,2) = \frac{1}{2x}  + \cO (x^{-2}) 
\label{cwp} \ee
\be
\D_{({\rm s})}^O (A,3) = \frac{3}{2x}  + \cO (x^{-2}) 
\sp 
\D_{({\rm s})}^O (A,4) = \frac{3}{2x}  + \cO (x^{-2}) 
\ee
\be
\D_{({\rm s})}^O (V,2) = 1 + \cO (x^{-1}) 
\sp 
\D_{({\rm s})}^O (A,1) = 1 + \cO (x^{-1}) 
\label{cws} \ee
\be
\D_{({\rm s})}^O (V,3) =  \cO (x^{0}) 
\sp 
\D_{({\rm s})}^O (V,4) =  \cO (x^{0}) 
\label{cwu} \ee
\be
\Ga_0^{({\rm s})} (m,M)  
= \left( \matrix{ 1  & \frac{1}{3x} \sqrt{3} & 0 & 0 \cr
\frac{1}{4x} \sqrt{2} & \frac{1}{4} \sqrt{6} &  - \frac{1}{2} &
\frac{1}{4} \sqrt{6} \cr} 
\right) +  \cO (x^{-2}) 
\pe \ee
\es
The explicit form of these residues was obtained using
(\ref{lre}), the embedding matrix (\ref{hem}) and the relation
 $c({\rm s},O) = Q^O_{({\rm su})} -\frac{1}{x}\on $.  

The four conformal weights in (\ref{cwp},c) are the broken affine primary
states in (\ref{fr1}), whose residues are $\cO (x^0)$ in accord with
the general OPE in (\ref{ope}). The two conformal weights in (\ref{cws}) are 
 broken affine secondary states (with residues which are $\oxo$) which
are not necessarily integer descendants of broken affine primary 
states. The conformal weights in (\ref{cwu}) cannot be determined through this 
order because their residues $\Ga_O^{({\rm s})}$ are zero through $\oxo$,
and indeed these entire blocks begin at order $\ox $,  
\be
\Bo(y)_V{}^3 \;\;, \;\;  
\Bo(y)_V{}^4
\;\; = \;\; \ox  
\label{oktb} \ee
a phenomenon also encountered in the coset example of Appendix D. 
To see (\ref{oktb}) directly 
 from (\ref{suhb}) note that, for these blocks,  
 $e({\rm s},O)_m{}^M=0$ and 
 $e({\rm s},O)_m{}^N (Q^O_{({\rm su})})_N{}^M=0$.  
 In the u-channel we also find two blocks which begin at $\ox$, while in the
 t-channel there is one such block. 

In agreement with (\ref{hnb}), the number of blocks for this 
$L(SU(3);O)$-degenerate
process is
\be
B_O  = 2 \cdot 4 =8 
\pe 
\label{inb} \ee
Because of the increasing symmetry breakdown,
\be 
O \subset SU(2)_{\rm irr} \subset SU(3) 
\ee
the number (\ref{inb}) is larger than the number of blocks
\be
 B_{SU(3)}  = 2 \cdot 2 = 4 \sp 
B_{SU(3)/SU(2)} = 2 \cdot 3 = 6 
\label{nbs} \ee
for the same correlator under the affine-Sugawara construction  (see
Appendix B) and the
closely related coset construction studied in Appendix D.    
Taken together, (\ref{inb}) and (\ref{nbs}) are an illustration of the
double inequality (\ref{die}). 

Using eqs.(\ref{agc}), (\ref{gesm}) we also find  the following expression 
for the high-level non-chiral correlators of $\sumh$,
\bs 
 \be
Y_O(y^*,y)  
\! \! = \! \! \sum_{M } E({\rm s},O)_{M}{}^M  
[ 1   +
 (\D_{({\rm s})}^O- \frac{1}{x }\on )\ln | y|^2  + (\D_{({\rm s})}^O- \frac{1}{x}\on )
 \ln | 1-y|^2 ]_M{}^M 
   +\cO (x^{-2})     
\ee
\be 
 E({\rm s},O) _M {}^N =  \sum_m ( e({\rm s},O)_m {}^M)^*  e({\rm s},O)_m{}^N  
= \left(
\matrix{ 1 & 0 & 0 & 0   \cr
 0 & \frac{3}{8 }  &-\frac{1}{8} \sqrt{6} & \frac{3}{8 }   \cr 
 0 &-\frac{1}{8} \sqrt{6 } &  \frac{1}{4}  &  -\frac{1}{8} \sqrt{6 } &  \cr 
 0 & \frac{3}{8 }  & -\frac{1}{8 }\sqrt{6} & \frac{3}{8} &  \cr}
\right)
\ee
\es 
\vskip -.3cm 
\noindent where we have used $ X_O({\rm us}) E({\rm s},O) X_O({\rm us}) = E({\rm s},O)$ 
and the diagonal s-channel conformal weight matrix $\D_{({\rm s})}^O$ is
given in (\ref{sumd}). This result explicitly shows the crossing symmetry
(\ref{snc}), as it should since $\T^2 \sim \T^3$ in this case.

We finally remark that
 the high-level blocks and correlators
of the K-conjugate theory 
\be
SU(3)/SU(3)^\#_M \sp 
\tL = L_{SU(3)}-L
\ee
can be easily obtained from the results above, by substituting everywhere the
K-conjugate conformal weights $\tilde{\D}(\T) = \D^g(\T) - \D (\T)$ for 
the conformal weights $\D(\T)$. Moreover,
the results above can easily be extended to the $L(g;H)$-degenerate correlators
$ n \bar{n} \bar{n} n$ in the larger family of ICFTs called $SU(n)^\#_M $ 
\cit{nuc};
in this case, the number of $H$-invariant tensors stays the same, with 
closely analogous forms for all the more general results.

\section{Conclusions}

The generalized KZ equations of ICFT provide a uniform description of the 
chiral correlators of rational and irrational conformal field theory, and the 
solution 
of these equations is known at high level on simple $g$. The apparent simplicity
of this result is deceptive, however, because the solution describes a vast
variety of generically irrational conformal field theories ranging from the
most symmetric (the RCFTs) to totally asymmetric (the generic ICFT).

In this paper, we have begun the resolution of the high-level chiral
correlators into high-level conformal blocks and non-chiral correlators,
beginning with the simplest and most symmetric classes. 

In particular,
we began by working out the high-level blocks and correlators of all the 
\ben
\item[$\bullet$] affine-Sugawara constructions on simple $g$
\item[$\bullet$] coset constructions on simple $g$.
\een
Both results are new, and the results for the cosets are apparently
inaccessible by other methods.

Based on this analysis, we then identified what we believe to be the
simplest and most symmetric class of correlators in ICFT. These are the
\ben
\item[$\bullet$] $L(g;H)$-degenerate processes in $H$-invariant CFTs on simple $g$
\een
which are those correlators whose external states have entirely degenerate
conformal weights $\D_\a = \D $. This class of correlators includes all the 
affine-Sugawara
correlators, a highly-symmetric subset of  coset correlators and
a presumably large set of irrational correlators, examples of which are
known.

For this simple class of correlators we were able to find the general 
expression for the high-level blocks and non-chiral correlators, and we 
worked out an irrational example with octohedral symmetry on $SU(3)$.

Our results emphasize that the $L(g;H)$-degenerate correlators are a very
special class of correlators indeed, since they have a finite number of 
conformal blocks (at least in the semi-classical approximation), whereas
the generic correlator in ICFT is expected to involve an infinite number
of blocks. We are intrigued to find that ICFT resembles RCFT in this simple
domain, and we are optimistic that the simplicity of the $L(g;H)$-degenerate
correlators can provide a foothold for further exploration. 

Additional information is needed, however, 
 to go beyond the leading orders of the
$L(g;\!H)$-degenerate processes in ICFT. The central question here is whether
the number of conformal blocks remains finite, as we found in the 
semi-classical approximation, or increases with the order of $k^{-1}$.
At finite values of the level, one will also need to consider the roles
of the affine cutoff [4,15] and fixed-point resolution \cit{fss}.

The more immediate open direction is to find the high-level conformal 
blocks of irrational correlators beyond the set of
 $L(g;H)$-degenerate processes. 
An ever-increasing number of blocks is expected here   as one confronts 
the progressively larger 
symmetry breakdown of ICFT, signalled by the $L^{ab}$-broken conformal 
weights $\D_{\a}$.

In this direction, we remind the reader of the known singularities of the 
invariant 
flat connections $W$ which govern the exact (finite level) correlators of
ICFT. For example, it is known that \cit{rev} 
$$
\eqalignno{ \;\;\;\; 
W(\tu,u)_\a{}^\b & \limit{=}{\tu,u\ra 0}
   \left({u\over \tu}\right)^{\D_{\a_1}(\T^1) + \D_{\a_2}(\T^2) -
    \D_{\b_1}(\T^1) - \D_{\b_2}(\T^2) }
  { (2L^{ab}\T_a^1\T_b^2)_\a{}^\b \over u }
&(7.1{\rm a}) \cr & \cr  
& = \left\{ \matrix{ 
              {(2L_\infty^{ab} \T_a^1 \T_b^2)_\a{}^\b \over u}  
+\cO (k^{-2})   
& \;\;\;\;(\mbox{high}\;k) \phantom{(L(g;H)\mbox{-degenerate})} 
\hskip 1.6cm  (7.1{\rm b}) \cr       
              {(2L^{ab} \T_a^1 \T_b^2)_\a{}^\b \over u}  
\phantom{+\cO (k^{-2})}   
  & \;\;\;\;(L(g;H)\mbox{-degenerate}) \phantom{(\mbox{high}\;k)}
 \hskip 1.6cm  (7.1{\rm c}) \cr }
\right.       
} 
$$

\noindent where $u$ and $\tilde{u}$ are the 
variables of the theory and its K-conjugate 
theory
respectively. The result (7.1a) shows
 the apparently 
non-Fuchsian $\alpha,\beta$ dependent shielding factor, which is hidden in
the high-level limit (7.1b), and which simplifies to unity at all levels,
shown in (7.1c), for the $L(g;H)$-degenerate processes. 
 We believe that this phenomenon underlies the  
 simplicity of the class of $L(g;H)$-degenerate processes
in ICFT, and it may be necessary to consider this factor in the
physical interpretation of the high-level logarithmic
singularities of correlators
beyond the simple class we have considered here.

\addcontentsline{toc}{section}{Acknowledgements}
\section*{Acknowledgements}
We thank L. Alvarez-Gaum\'e, D. Gepner, E. Kiritsis and 
C. Schweigert for helpful discussions.
We also thank the theory groups at Niels Bohr Institute and CERN for their
hospitality and support during the course of this work. 

The work of MBH was supported in part by the Director, Office of 
Energy Research, Office of High Energy and Nuclear Physics, Division of 
High Energy Physics of the U.S. Department of Energy under Contract 
DE-AC03-76SF00098 and in part by the National Science Foundation under 
grant PHY90-21139. The work of NO was supported in part by the European 
Community, Project ERBCHBG CT93 0273.
MBH and NO also acknowledge financial support
from the Danish Science Research Council while visitors at NBI.

\vskip 1.0cm
\setcounter{equation}{0}
\def\theequation{A.\arabic{equation}}
 \boldmath 
\addcontentsline{toc}{section}{Appendix A: 
Alternate expressions for blocks and correlators}
\centerline{\bf Appendix A: Alternate expressions for blocks and correlators}
\unboldmath 
\vskip 0.5cm
In this appendix, we use the relevant crossing matrices to give alternate 
expressions for the conformal
blocks and correlators of any set of external states in the
affine-Sugawara constructions 
(see Section \ref{sec3})
and of any $L(g;H)$-degenerate process  in the more 
general $H$-invariant CFTs (see Section \ref{sec5}). 

\vs .4cm
\noindent \underline{Affine-Sugawara constructions}
\vs .3cm 

We begin with the $\r$-channel affine-Sugawara blocks in (\ref{asbf}),
\be
\F_g^{(\r)}(y)_m {}^n 
= \bv(\r,g)_m [ \corrg ] v(\r,g)^n + \ok 
\pe
\ee
Using the definitions (\ref{asev2}), (\ref{ocev}) 
of the $g$-invariant $\r$-channel 
eigenvectors $v(\r,g)$ and the $g$-crossing matrices $X_g$ in (\ref{xm}),
we have the $g$-crossing relations,
\be
\bv(\r,g)_m = X_g (\r\s)_m{}^n \bv(\s,g)_n 
\sp 
v(\r,g)^m = v(\s,g)^n X_g (\s\r)_n{}^m
\pe \ee
Using these relations, we obtain 
 the alternate form of the affine-Sugawara blocks, 
\be
\label{eas} 
\eqalign{
\F_g^{(\r)}(y)_m{}^n  
=  [ 1   & +  
(Q^g_{(\r {\rm s})} - (\D^g(\T^1) + \D^g(\T^2))  \on    )\ln y \cr
& +  ( Q^g_{(\r {\rm u})}- (\D^g(\T^1) + \D^g(\T^3)) \on     ) \ln (1-y)]_m{}^n 
   +\ok \cr}    
\ee
where
\be
(1)_m{}^n = \d_m^n \sp 
 (Q^g_{(\r \s )})_m{}^n = \left\{ 
\matrix{  (\D^g_{(\r)})_m{}^n =  \D^g_{(\r)}(m)  \d_m^n & \sp \r = \s \cr
 \sum_{l }  X_g ( \r \s)_{m}{}^l \D^g_{(\s)}(l ) X_g  ( \s \r)_l{}^n 
&  \sp \r \neq \s \cr}\right.  
\label{qde}\ee

\vskip -.1cm 
\noindent and $\D^g_{(\r)}(m)$ are the $\r$-channel affine-Sugawara conformal 
 weights in (\ref{asev2}) and (\ref{ocev}).

We also give the corresponding
alternate form of the analytic t-channel affine-Sugawara 
blocks (\ref{atcb}), 
\bs
\be
\label{f2t} 
\eqalign{
\Fh_g^{({\rm t})}(y)_m{}^n  =  [1  & +    
(Q^g_{({\rm ts})} +  Q^g_{({\rm tu})} - (2\D^g(\T^1) + \D^g(\T^2) + \D^g(\T^3)) 
\on    )\ln(-y) ) \cr
& + (Q^g_{({\rm tu})} - (\D^g(\T^1) + \D^g(\T^3))  \on   ) \ln 
\left(1-\frac{1}{y} \right)]_m{}^n 
   +\ok      \cr}
\ee
\be
\label{f2tb}
\eqalign{ 
\phantom{(\Fh_g^{({\rm t})}(y)_m{}^n}  =  [1   & -    
(Q^g_{({\rm tt})} +  (\D^g(\T^1) - \D^g(\T^4)) \on    )\ln(-y) ) \cr
& + (Q^g_{({\rm tu})} - (\D^g(\T^1) + \D^g(\T^3))  \on    ) \ln 
\left(1-\frac{1}{y} \right)]_m{}^n 
   +\ok      \cr}
\ee
\es
where $Q^g_{({\rm t} \s)}$, $\s={\rm s,t,u}$ is given in (\ref{qde}). 
Here, the second form (\ref{f2tb}) follows from (\ref{f2t}) 
 using the $\r$=t form of the
 conformal weight sum rule,
\be
\D^g_{(\r)} + X_g ( \r \s  ) \D^g_{(\s)} X_g(\s \r ) 
+ X_g (\r \t ) \D^g_{(\t)} X_g(\t \r ) = \left( \sum_{i=1}^4 \D^g (\T^i)  
\right) \on \;\;,\;\; \r \neq \s \neq \t \neq \r
\ee
which is itself a direct consequence of the $g$-global Ward identity
(\ref{ggwi}). 

Substitution of the alternate forms (\ref{eas}) of the affine-Sugawara blocks
 in 
the expression (\ref{asc}) for the affine-Sugawara correlators then gives
the corresponding alternate form for the non-chiral correlators, 
\be
\eqalign{
Y_g (y^*,y)_\a {}^\b   =& 
 \sum_{m,n} [1    +  
(Q^g_{(\r {\rm s})} - (\D^g(\T^1) + \D^g(\T^2))  \on    )\ln |y|^2   
\cr
&   +  ( Q^g_{(\r {\rm u})}- (\D^g(\T^1) + \D^g(\T^3))\on )\ln |1-y|^2 ]_m{}^n 
\, v(\r,g)_\a^m\bv(\r,g)_n^\b + \ok 
\cr}
\ee
which explicitly shows the  trivial monodromies   around $y=0$,  1 and
$\infty$. 

\vs .4cm
\noindent \underline{$L(g;H)$-degenerate processes}
\vs .3cm 

Following the development for the affine-Sugawara constructions above,
we may find similar alternate forms 
 for the blocks and correlators of the general $L(g;H)$-degenerate process.

Using the definitions
(\ref{evg}) of the $H$-invariant eigenvectors $\ps(\r,H)$ and  the 
$H$-crossing matrices $X_H$ in (\ref{gc}), we have the $H$-crossing 
relations,
\be
\bps(\r,H)_M = X_H (\r\s)_M{}^N \bps(\s,H)_N  
\sp 
\ps(\r,H)^M = \ps (\s,H)^N X_H (\s\r)_N{}^M 
\pe \ee
Using these relations in the expressions (\ref{efb}) for the blocks, we obtain 
 the following alternate form of the $L(g;H)$-degenerate blocks,  
\be
\label{gba} 
\eqalign{ \B^{(\r)}(y)_m{}^M  = e(\r,H)_m{}{}^N 
 [1   & +  
(Q_{(\r {\rm s})}^H - (\D_1^H + \D_2^H)  \on )\ln y \cr  
& +  ( Q_{(\r {\rm u})}^H- (\D_1^H + \D_3^H)  \on  ) \ln (1-y)]_N{}^M  
   +\ok \cr}  
\ee
where $e(\r,H)$ are the $\r$-channel embedding matrices (\ref{gem}) and 
\be
(1)_M{}^N = \d_M^N  \sp  (Q_{(\r \s )}^H)_M{}^N = \left\{ 
\matrix{ (\D_{(\r)}^H)_M{}^N =  \D_{(\r)}^H(M)  \d_M^N & \sp \r = \s \cr
 \sum_{L } X_H ( \r \s)_{M}{}^L \D_{(\s)}^H(L ) X_H  ( \s \r)_L{}^N 
&  \sp \r \neq \s \cr}\right.  
\label{qde2}\ee

\vskip -.1cm 
\noindent 
with  $\D_{(\r)}^H (M)$ the $\r$-channel conformal weights in (\ref{evg}).  
These results include all the correlators of the
 affine-Sugawara constructions and all the 
 $L(g;h)$-degenerate processes of the
$g/h$ coset constructions.
 
We also give the corresponding
alternate form for the analytic t-channel blocks (\ref{atcbh}),
\bs
\be
\label{h2t} 
\eqalign{
\Bh^{({\rm t})}(y)_m{}^M  =  e({\rm t},H)_m{}^N  [1   & +    
(Q^H_{({\rm ts})} +  Q^H_{({\rm tu})} - (2\D^H_1+ \D^H_2 + \D^H_3) 
\on  )\ln(-y) ) \cr
& + (Q^H_{({\rm tu})} - (\D^H_1 + \D^H_3)  \on  ) \ln 
\left(1-\frac{1}{y} \right)]_N{}^M  
   +\ok      \cr}
\ee
\be
\label{h2tb}
\eqalign{ 
\phantom{(\Bh^{({\rm t})}(y)_m{}^M}  =  
  e({\rm t},H)_m{}^N  [1   & -     
(Q^H_{({\rm tt})} +  (\D^H_1 -  \D^H_4) \on  )\ln(-y) ) \cr
& + (Q^H_{({\rm tu})} - (\D^H_1 + \D^H_3)  \on  ) \ln 
\left(1-\frac{1}{y} \right)]_N{}^M  
   +\ok      \cr}
\ee
\es
where $Q_{({\rm t} \s)}^H$, $\s={\rm s,t,u}$ is given in (\ref{qde2}). 
Here, the second form (\ref{h2tb}) follows from (\ref{h2t}) 
using the $\r=$t form of the conformal weight sum rule
in $L(g;H)$-degenerate processes, 
\be
\eqalign{
e(\r,H) [\D_{(\r)}^H + &  X_H ( \r \s  ) \D_{(\s)}^H X_H (\s \r ) 
+ X_H (\r \t ) \D_{(\t)}^H X_H(\t \r )] \cr
& =e(\r,H) \sum_{i=1}^4 \D_i^H 
\;\;\; \sp \;\;\;  \r  \neq \s  \neq \t  \neq \r \cr} 
\ee 
which is itself a direct consequence of the $g$-global Ward identity 
(\ref{higw}).
 
Finally, we give the corresponding alternate form of the non-chiral
correlators (\ref{ncc}), using
the expression (\ref{gba}) for the blocks,
\be
\label{agc}
\eqalign{
Y(y^*,y) =  \sum_{M,N} E(\r,H) _M{}^N  
 [ 1    & +  
(Q_{(\r {\rm s})}^H - (\D_1^H + \D_2^H)  \on  )\ln |y|^2   \cr
& +  ( Q_{(\r {\rm u})}^H- (\D_1^H + \D_3^H)  \on  ) \ln |1-y|^2 ]_N {}^M  
+\ok \cr}   
\ee
where
\be
E(\r,H)_M{}^N = \sum_m (e(\r,H)_m{}^M)^* e(\r,H)_m{}^N 
\label{gesm} \pe \ee
This form of the correlator explicitly shows the trivial monodromies
around $y=0$, 1  and $\infty$.

\vskip 1.0cm
\setcounter{equation}{0}
\def\theequation{B.\arabic{equation}}
 \boldmath 
\addcontentsline{toc}{section}{Appendix B: Comparison with the blocks of
Knizhnik and Zamolodchikov}
\centerline{\bf Appendix B: Comparison with the blocks of
Knizhnik and Zamolodchikov}
\unboldmath 
\vskip 0.5cm

In this appendix, we check our high-level affine-Sugawara blocks 
(\ref{iss}), (\ref{asbu3}) and 
(\ref{asb2b}) against the exact blocks obtained in Ref.\cit{kz}  
for the $3 \bar{3} \bar{3} 3$ correlator on 
$SU(3)$.  

To find the explicit form of our high-level blocks in this case, 
we need first the high-level form of the affine-Sugawara construction
on $g=SU(3)$, 
\be
L_{g,\infty}^{ab}= \frac{1}{x\ps_g^2} \d_{ab}  
\ee
where $\ps_g$ is the highest root of $SU(3)$, $x$ is the invariant level
of affine $SU(3)$  and 
we have used the  Gell-Mann basis. The matrix irreps
 of the 3 and $\bar{3}$ are given in (\ref{gmb2})
and the corresponding high-level conformal weights are
\be
\D^g (\T_{(3)} )=  
\D^g (\T_{(\bar{3})} ) =\frac{4}{3x} + \cO (x^{-2} ) 
\pe \ee
Using the 
 $\r$-channel invariants in (\ref{asca}-e) in the eigenvalue problems
(\ref{asev})  and (\ref{asevu},b),  
we also obtain the high-level intermediate $\r$-channel affine-primary 
conformal weights,
\bs
\be
\D^g_{(\r)} (m) = \left\{ \matrix{   
\D^g(\T_{(1)}) =  0 + \cO (x^{-2} ) & \sp m = V \cr
\D^g(\T_{(8)}) = \frac{3}{x} +\cO (x^{-2} ) & \sp m= A \cr}
\right.   
\sp \r ={\rm s,u} \ee
\be
\D^g_{({\rm t})} (m) =  
\left\{ \matrix{
\D^g(\T_{(6)}) =  \frac{10}{3x} +   \cO (x^{-2} )  & \sp m = 6   \cr  
\D^g(\T_{(\bar{3})}) = \frac{4}{3x} +\cO (x^{-2} ) & \sp m = \bar{3} \cr } 
\right.   
\ee
\es
where $m=(V,A)$ labels the vacuum and adjoint representations in the 
 s and u-channels and $m=(6,\bar{3})$ labels the symmetric and 
antisymmetric representations in the t-channel.

Finally, we use the corresponding crossing matrices (\ref{ascf},g) to compute
the matrices $c(\r,g)$ in (\ref{ascm}), (\ref{ascmu}) and (\ref{ascmt}),     
\bs
\be 
c({\rm s},g) = c({\rm u},g) = X_g ({\rm su}) [\D^g_{({\rm u})} - 
  \frac{8}{3x} \on ]
 X_g({\rm us}) 
= \frac{1}{3x} \left(
\matrix{0 &  - 2\sqrt {2} \cr
 - 2\sqrt {2} & - 7 \cr } 
\right) 
\ee
\be
c({\rm t},g) = X_g ({\rm tu}) [\D^g_{({\rm u})} - \frac{8}{3x} \on ]
 X_g({\rm ut} ) 
= \frac{1}{3x} \left(
\matrix{-5  &  3 \sqrt {2} \cr
 3 \sqrt {2} & -2 \cr } 
\right) 
\ee
\es 
where 1 and $\D^g_{({\rm u})}$ are defined in (\ref{qde}).
 With these data we obtain the explicit form of our high-level blocks
for the $3 \bar{3} \bar{3} 3 $ correlator, 
\bs
\be
\F_g^{({\rm s})}(y)_m{}^n 
=\left( \matrix{ 1 & 0  \cr 0  & 1 \cr} \right) 
+\frac{1}{3x} \left( \matrix{ -8 & 0  \cr 0  & 1 \cr} \right) \ln y  
+\frac{1}{3x} \left(\matrix{ 0 & -2\sqrt{2}\cr -2\sqrt{2}   & -7 \cr}\right) 
\ln (1-y) 
+ \ox  
\label{hls}\ee
 \be
\F_g^{({\rm u})}(y)_m{}^n 
= \F_g^{({\rm s})}(1-y)_m{}^n 
\label{hlu} \ee
\vskip -.3cm
\be
\label{hlt} 
\Fh_g^{({\rm t})}(y)_m{}^n 
=\left( \matrix{ 1 & 0  \cr 0  & 1 \cr} \right) 
-\frac{1}{3x} \left( \matrix{ 10 & 0  \cr 0  & 4 \cr} \right) 
\ln (- y)   
+ \frac{1}{3x} \left( \matrix{-5  &  3 \sqrt {2} \cr
 3 \sqrt {2} & -2 \cr } \right) 
\ln \left( 1- \frac{1}{y} \right) 
+ \ox 
\ee
\label{ashl} \es 
where $m=(V,A)$ in (\ref{hls},b) and $m=(6,\bar{3})$ in (\ref{hlt}).
The relation (\ref{hlu}) is in accord with (\ref{ussy}), since $\T^2 \sim \T^3$
in this case.

We wish to check our high-level blocks (\ref{ashl}) 
against the exact results obtained by 
KZ, who,
however, use a different basis for the $SU(3)$-invariant tensors, 
\be
\bv^{\rm KZ}_1(g) = \d_{\a_1 \a_2} \d_{\a_3 \a_4}  
\sp \bv^{\rm KZ}_2(g) = \d_{\a_1 \a_3} \d_{\a_2 \a_4}  
\pe
\label{kzi} 
\ee
Comparing these invariants with our Clebsch  basis (\ref{asca}-e)
of $SU(3)$-invariants, we learn that the basis transformation of our 
$\r$-channel blocks (\ref{ashl})
to the KZ basis is,
\bs
\be
d_m \F_g^{(\r)}(y)_m{}^{\m} = \F_g^{(\r)}(y)_m{}^n L_n{}^\m    
\sp 
L_n{}^\m  = \frac{1}{12} \left( \matrix{ 4 & 0 \cr - \sqrt{2}
& 3 \sqrt{2}  \cr}\right)   
\sp m=V,A \sp  \r = {\rm s,u} 
\label{btk} \ee
\be
d_m \Fh_g^{({\rm t})}(y)_m{}^{\m} = \Fh_g^{({\rm t})}(y)_m{}^n M_n{}^\m    
\sp 
M_n{}^\m  = \frac{1}{12} \left( \matrix{ \sqrt{6} & \sqrt{6} \cr 2 \sqrt{3}
& -2 \sqrt{3}  \cr}\right)   
\sp m=6,\bar{3}   
\ee
\es
where $\m=1,2$ labels the KZ invariants, the normalization constants
$d_m$ are arbitrary and the blocks  $\F_g^{({\rm s,u})}(y)_m{}^\m$, 
$\Fh_g^{({\rm t})}(y)_m{}^\m $ are
the KZ blocks. More explicitly, our prediction for the high-level analytic
KZ blocks is then,
\bs
\be
\F_g^{({\rm s})}(y)_m{}^\m 
=\left( \matrix{ 1 & 0  \cr 1  & -3 \cr} \right) 
+\frac{1}{3x} \left( \matrix{ -8 & 0  \cr 1  & -3 \cr} \right) \ln y  
+\frac{1}{3x} \left( \matrix{ 1 & -3  \cr 1   & 21  \cr} \right) \ln (1-y) 
+ \ox  
\label{kzs} 
\ee
\be
\F_g^{({\rm u})}(y)_m{}^\m=  
\F_g^{({\rm s})}(1-y)_m{}^\m 
\label{kzu} \ee
\vskip -.3cm
\be
\Fh_g^{({\rm t})}(y)_m{}^\m 
=\left( \matrix{ 1 & 1  \cr 1  & -1 \cr} \right) 
-\frac{1}{3x} \left( \matrix{ 10  & 10  \cr 4  & -4  \cr} \right) \ln (-y)  
+\frac{1}{3x} \left( \matrix{ 1 & -11  \cr 1   & 5   \cr} \right) 
\ln \left(1-\frac{1}{y}\right) 
+ \ox  
\label{kzt} 
\ee
\label{dva} 
\es
where we have chosen the particular values of the normalization constants 
\be
d_V = \frac{1}{3} \sp
d_A = -{1 \over 6 \sqrt{2} } \sp  
d_6 = \frac{1}{2 \sqrt{6} } \sp
d_{\bar{3}} = {1 \over 2 \sqrt{3} } 
\ee
with some pedagogical foresight.

To check that these blocks are precisely the high-level limit of the 
 KZ blocks on $SU(3)$,  we recall the exact form of 
the s-channel KZ blocks 
$\F_g^{({\rm s})}(y)_m{}^\m$ on  $SU(3)$ \cit{kz}, 
\bs
\be
\F_g^{({\rm s})}(y)_V{}^1 = 
y^{-2 \D^g(\T_{(3)})} (1-y)^{\D^g(A)-2 \D^g(\T_{(3)})}  
F(\l,-\l, 1- 3 \l ;y)  
\ee
\be
\F_g^{({\rm s})}(y)_V{}^2 = 
\frac{1}{x} y^{1 -2 \D^g(\T_{(3)})} (1-y)^{\D^g(A)-2 \D^g(\T_{(3)})}  
F(1+ \l,1- \l, 2- 3 \l ;y)  
\ee
\be
\F_g^{({\rm s})}(y)_A{}^1 = 
 y^{ \D^g(A) -2 \D^g(\T_{(3)})} (1-y)^{\D^g(A)-2 \D^g(\T_{(3)})}  
F(2  \l,4  \l, 1+  3 \l ;y)  
\ee
\be
\F_g^{({\rm s})}(y)_A{}^2  = 
 - 3 y^{ \D^g(A) -2 \D^g(\T_{(3)})} (1-y)^{\D^g(A)-2 \D^g(\T_{(3)})}  
F(2  \l,4  \l,  3 \l ;y)  
\ee
\be
\D^g(\T_{(3)}) = {4 \over 3 (x+3)} \sp 
\D^g(A)= \D^g(\T_{(8)})  = {3 \over  x+3}  
\sp
\l = { 1 \over x+3} 
\ee
\label{ekz} \es 
where $m=V,A$ label the vacuum and adjoint blocks, $\D^g(\T_{(3)})$ is
the conformal weight of the 3, $\D^g(A)$ is the conformal
weight of the adjoint representation and $F(a,b,c;y)$ is the hypergeometric
function. Using the high-level expansions,
\bs
\be
F\left({a \over x + d},{b \over x + e},{c \over x + f};y\right) = 1 
- {a b \over c x} \ln (1-y) + \ox 
\ee
\be
F\left({a \over x + d},{b \over x + e},1 + {c \over x + f};y\right) = 1 + \ox  
\ee
\be
F\left({a \over x + d},1+ {b \over x + e},1 + {c \over x + f};y\right) = 1 
- {a  \over  x} \ln (1-y) + \ox 
\ee
\be
F\left(1 + {a \over x + d},1+ {b \over x + e},2 + {c \over x + f};y\right) = 
- { \ln (1-y) \over y } + \oxo   
\ee
\label{hlr} \es 
one finds that the high-level limit of
(\ref{ekz}) agrees precisely with the predicted form in (\ref{kzs}).
For the u-channel the check follows the same steps, 
with the replacement $y \ra 1-y$ everywhere. 

To continue the s-channel KZ blocks (\ref{ekz}) to the t-channel, 
one uses the standard
continuation formula \cit{bat}
\be
\eqalign{
F(a,b,c;y) = & {\Ga (c) \Ga (b-a)\over \Ga (b) \Ga(c-a) } (-y)^{-a}
F(a,1-c+a,1-b+a;\frac{1}{y})  \cr  
& + {\Ga (c) \Ga (a-b)\over \Ga (a) \Ga(c-b) }(-y)^{-b}
F(b,1-c+b,1-a+b;\frac{1}{y})  \sp |\arg(-y)| < \p \pe \cr}  
\ee
According to the expansions (\ref{hlr}),
the high-level limit of this formula in the cases of all 
four blocks in (\ref{ekz}) is  
\be 
\ln(1-y) = \ln(-y) + \ln\left( 1-\frac{1}{y} \right) \sp | \arg(-y) | < \p 
\ee
and this limit is identical to the first continuation rule (\ref{ru1}) 
used in the text. To continue the factors in front of the hypergeometric 
functions, we use the relations
\bs
\be
(1-y)^\a = (-y)^\a \left( 1-\frac{1}{y} \right)^\a 
\ee
\be
y^\b = (-y)^\b \exp[- i \p \b \sign(\arg(-y))]   
\label{flr} \ee
\es
which are equivalent, finite forms of the continuation rules (\ref{ru1},b).

Factoring out the non-analytic phases generated by (\ref{flr}) (which then
appear in the crossing matrices to the t-channel), we 
find the finite-level form of the analytic t-channel KZ blocks,   
\bs
\be
\Fh_g^{({\rm t})}(y)_6{}^1 = 
(-y)^{- \D^g (\T_{(6)} )} 
\left( 1- {1 \over y}  \right)^{\D^g(A)-2 \D^g(\T_{(3)})}  
F( 4 \l,  \l, 1 + 2 \l ;\frac{1}{y})  
\ee
\be
\Fh_g^{({\rm t})}(y)_6{}^2   = 
(-y)^{ -\D^g (\T_{(6)} )} 
\left( 1- {1 \over y}  \right)^{\D^g(A)-2 \D^g(\T_{(3)})}  
F(4 \l, 1+  \l, 1 + 2 \l ;\frac{1}{y})  
\ee
\be
\Fh_g^{({\rm t})}(y)_{\bar{3}}{}^1 = 
(-y)^{ -\D^g (\T_{( \bar{3})} ) }  
\left( 1- {1 \over y} \right)^{\D^g(A)-2 \D^g(\T_{(3)})}  
F(2 \l,- \l, 1- 2 \l ;\frac{1}{y})  
\ee
\be
\Fh_g^{({\rm t})}(y)_{\bar{3}}{}^2  = -  
(-y)^{ -\D^g (\T_{(\bar{3})}) }  
\left( 1-{1 \over y}  \right)^{\D^g(A)-2 \D^g(\T_{(3)})}  
F(2 \l, 1-  \l,  1- 2 \l ;\frac{1}{y})  
\ee
\be
\D^g(\T_{(6)} )=    {10  \over  3( x+3)}  
\sp \D^g(\T_{(3)} ) = \D^g (\T_{(\bar{3} ) } )    = {4  \over  3( x+3)}  
\sp \l = {1 \over x +3} 
\pe \ee
\es
The high-level forms of these blocks agree precisely with our prediction 
(\ref{hlt}).

For completeness, we finally give the finite-level forms of
the s-channel blocks
\bs$\F_g^{({\rm s})}(y)_m{}^n$ in our Clebsch basis,
\be
\F_g^{({\rm s})}(y)_V{}^V  =
y^{-2 \D^g(\T_{(3)})} (1-y)^{\D^g(A)-2 \D^g(\T_{(3)})}  F(\l,-\l, - 3 \l ;y)
\ee
\be
\F_g^{({\rm s})}(y)_V{}^A =
\frac{2\sqrt{2} \l}{3(1-3\l)} y^{1 -2 \D^g(\T_{(3)})} (1-y)^{\D^g(A)-2 \D^g(\T_{(3)})}
F(1+ \l,1- \l, 2- 3 \l ;y)
\ee
\be
\F_g^{({\rm s})}(y)_A{}^A  =
y^{\D^g(A) -2 \D^g(\T_{(3)})} (1-y)^{\D^g(A)-2 \D^g(\T_{(3)})}
F(2  \l,4  \l, 3 \l ;y)
\ee
\be
\F_g^{({\rm s})}(y)_A{}^V =
\frac{2 \sqrt{2}  \l }{3 (1+ 3 \l)}  
y^{1+ \D^g(A) -2 \D^g(\T_{(3)})} (1-y)^{\D^g(A )-2 \D^g(\T_{(3)})}
F(1+2  \l,1+4  \l, 2+  3 \l ;y)
\ee
\be
\D^g(\T_{(3)}) = {4 \over 3 (x+3)} \sp 
\D^g(A)= \D^g(\T_{(8)})  = {3 \over  x+3}  
\sp
\l = { 1 \over x+3}
\ee 
\label{okz} \es
which are easily obtained from the s-channel KZ blocks (\ref{ekz}) and the 
basis transformation (\ref{btk}). The basis transformation gives two of these
blocks as linear combinations of two hypergeometric functions, 
which we have then combined into 
a single hypergeometric function using Gauss' contiguous relations.

The exact blocks (\ref{okz}) also show quite explicitly the high-level
pattern discussed in the text for the general affine-Sugawara blocks in 
our Clebsch basis: the diagonal blocks begin at  
order $\cO (k^0)$ with leading singularities which are
affine primary states, while  the off-diagonal blocks begin at
 $\oko $ with leading singularities which are affine secondary
states. Moreover, one sees that the conjectured result (\ref{exr}) for the
exact conformal weights of the general blocks is indeed correct in this case.

\vskip 1.0cm
\setcounter{equation}{0}
\def\theequation{C.\arabic{equation}}
 \boldmath 
\addcontentsline{toc}{section}{Appendix C: The level-families 
$ {\bf SU(3)^\#_M } $ and ${\bf   SU(3)/SU(2)_{\rm irr} } $}   
\centerline{\bf Appendix C: The level-families $ {\bf SU(3)^\#_M } $
and $ {\bf  SU(3)/SU(2)_{\rm irr} } $}   
\unboldmath 
\vskip 0.5cm
In this appendix we review [24,17] 
 various results for the unitary irrational 
level-family 
\be (SU(3)_x)^\#_M
\ee 
and the closely-related level-family of the coset construction 
\be
{SU(3) \over SU(2)_{\rm irr}} = {SU(3)_x \over SU(2)_{4x} }
\ee
both of which  occur in the maximally-symmetric ansatz on $SU(3)$.
$SU(2)_{\rm irr}$ denotes the 
irregularly embedded $SU(2)$ subgroup of $SU(3)$ generated by $J_{2,5,7}$.
The results given here are used in Section \ref{sec6} and Appendix D.

In the (Cartesian) Gell-Mann basis (\ref{gmb2}), the maximally-symmetric 
 construction $( SU(3)_x)_M^\# $  has the form \cit{nuc} 
\bs
\be
L^{ab} = \frac{1}{\ps_g^2} \ell_a \d_{ab} \sp \ell_a = \left\{ 
\matrix{ \ell_c & a =3,8 \cr
  \ell_h  & a =2,5,7  \cr
  \ell_r   & a =1,4,6   \cr} \right. 
\ee
\be
T = \frac{1}{\ps_g^2} \xx [ \ell_c (J_3^2 +J_8^2) + \ell_h (J_2^2 + J_5^2 +J_7^2) 
  +  \ell_r  (J_1^2 + J_4^2 +J_6^2)] \xx  
\ee
\be
 c = x ( 2 \ell_c + 3 \ell_h + 3 \ell_r) 
\ee
\es 
where $T$ is the stress tensor, 
 $\ps_g$ is the highest root of $SU(3)$, 
$ x $ is the affine level and $c$ is the central charge.
The exact 
form of $c$ is given in (\ref{cc}), but we refer to \cit{nuc} 
for the exact${}^{\rm d}$\footnotetext{${}^{\rm d}$The relation to the notation 
of Ref.\cit{nuc} is
$\ell_c = 3 \l$, $\ell_h = (L_- - L_+)/2$ and $\ell_r = (L_- + L_+)/2$.}  
forms of the coefficients $\ell_{c,h,r} $.
 The construction above  includes the  coset construction 
 $SU(3)/SU(2)_{\rm irr} $ as a special case when the further symmetry relation
\be
\ell_c = \ell_r \equiv \ell_{g/h}
 \label{esy}\ee
is obeyed.

$SU(3)^\#_M$ is an $H$-invariant CFT with symmetry group  \cit{hl} 
\be
H ( SU(3)_M^\#) = O = \mbox{ octohedral group} \subset SU(2)_{\rm irr}  
\ee
where $O$ is the octohedral group (rotational symmetry group of
the cube, with order 24) and $SU(2)_{\rm irr}$ is the irregular embedding
of $SU(2)$ in $SU(3)$. The octohedral group includes  the elements
\be
\O_{(2)} = \exp ( i { \pi \over \sqrt{\ps_g^2} } J_2(0) )  
\;\;\;, \;\;\;
\O_{(5)} = \exp ( i { \pi \over \sqrt{\ps_g^2} } J_5(0) )
\;\;\;, \;\;\;
\O_{(7 )} = \exp ( i { \pi \over \sqrt{\ps_g^2} } J_7(0) )
\label{omm} \ee
where $J_a(0)$ are the zero modes of the currents $J_a$,  
and in particular we may take the two elements
$\o_1$ and $\o_2$,
\be 
\o_1 = \O_{(2)} \sp \o_2 = \O_{(5)} \O_{(7)} 
\label{ohge} \ee
which satisfy
\bs
\be
\o_1^4 = 1 \sp \o_2^3 = 1  
\ee 
\be
\o_1 \o_2^2 \o_1 = \o_2 \sp \o_1 \o_2 \o_1 = \o_2 \o_1^2 \o_2 
\ee
\label{ohg} \es
as the generators of the octohedral group.

The coset construction $SU(3)/SU(2)_{\rm irr} $ has the 
larger Lie group symmetry 
\be
H ( SU(3)/SU(2)_{\rm irr} ) = SU(2)_{\rm irr}   
\ee 
because of  the symmetry relation (\ref{esy}).

The 3 and $\bar{3}$ are $L(g;H)$-degenerate irreps
of $\sumh$ and $\suc$ with completely degenerate conformal weights, 
\be
 \D (\T_{(3)}) = \D (\T_{(\bar{3})}) = \frac{c}{6x} \; \;\;\;\;\;\;\;(3)  
\label{cw3} \ee
where the number in parentheses denotes the degeneracy. 

For $(  SU(3)_x ) ^\#_M$ one also finds the $L^{ab}$-broken conformal 
weights of the  8 (adjoint) and  6 (symmetric),   
\bs
\be 
\D(\T_{(8)}) = \left\{ \matrix{ 
\ell_c + \frac{1}{2} ( 3 \ell_h +  \ell_r) &  \;\;\;\;(3 ) \cr
\frac{3}{2} (\ell_h +\ell_r) &  \;\;\;\;(2) \cr
\ell_c + \frac{1}{2} (\ell_h + 3 \ell_r) &  \;\;\;\;(3 ) \cr }\right.  
\label{cw8} \ee
\be
\D(\T_{(6)}) =\D(\T_{(\bar{6})})  = 
\left\{ \matrix{ 
\frac{2}{3 } (2 \ell_c  +3 \ell_r) & \;\;\;\; (1) \cr
\frac{1}{3} \ell_c + \frac{3}{2} ( \ell_h +  \ell_r) &  \;\;\;\;(3 ) \cr
\frac{4}{3} \ell_c + \frac{1}{2} (3 \ell_h +  \ell_r) & \;\;\;\; (2) \cr
} \right.  
\pe
\label{cw6} \ee
\label{cw86} \es 
\vskip -.3cm 
\noindent 
These forms show that the 8 and the 6 each split into three subsets of 
degenerate
weights, in agreement with the block analysis of Section \ref{sec6}.

For  $SU(3)_x /SU(2)_{4x} $ 
the splitting is reduced to two subsets,
\bs
 \be 
\D_{g/h} (\T_{(8)}) = 
\left\{ \matrix{ \frac{3}{2} (\ell_h +\ell_{g/h} ) & \;\;\;\; (5) \cr
     \frac{1}{2} (\ell_h + 5  \ell_{g/h} ) &  \;\;\;\; (3 ) \cr} \right. 
\label{cw8c}  \ee
\be
\D_{g/h} (\T_{(6)}) =\D_{g/h} (\T_{(\bar{6})})  = 
 \left\{ \matrix{ 
\frac{10}{3 } \ell_{g/h}   &  \;\;\;\; (1) \cr
\frac{3 }{2} \ell_h + \frac{11}{6}  \ell_{g/h}  &  \;\;\;\; (5 ) \cr  
} \right. 
\label{cw6c} \ee
\label{cw86c} \es 
\vskip -.3cm 
\noindent 
according to  (\ref{esy}) and (\ref{cw86}). These forms are in agreement with 
the coset 
block analysis of Appendix D. 

For the computations of Section \ref{sec6} and Appendix D, we need the
high-level forms of the two constructions,
\bs
\be  
 ( SU(3)_x)^\#_M \;\;: \ell_r = \frac{1}{x} +\cO (x^{-2})  
 \sp \ell_c = \ell_h = \cO (x^{-2} ) \sp c = 3 + \cO (x^{-1})  
\label{hli} \ee
\be
{ SU(3)_x \over SU(2)_{4x} }\;\; : \ell_{g/h} =   \frac{1}{x}
+ \cO (x^{-2})  \sp \ell_h=\cO (x^{-2})  
\sp c = 5  + \cO (x^{-1})
\label{hlc} \ee
\label{suc} \es  
which can be used with (\ref{cw3}), (\ref{cw86}) and (\ref{cw86c}) to
obtain the high-level forms of all the quantities discussed in this
appendix.  

\vskip 1.0cm
\setcounter{equation}{0}
\def\theequation{D.\arabic{equation}}
\boldmath 
\addcontentsline{toc}{section}{Appendix D: Blocks and correlators in    
$ {\bf SU(3) /SU(2)_{\rm irr } } $} 
\centerline{\bf Appendix D: Blocks and correlators in    
${\bf  SU(3) /SU(2)_{\rm irr }}  $} 
\unboldmath 
\vskip 0.5cm
 As an explicit example in rational conformal field theory, we work out in
this appendix  the high-level conformal blocks and correlators of
a particular  $L(g;h)$-degenerate process in the level-family of 
 the coset construction
\be
{g \over h} = { SU(3)_x \over SU(2)_{4x} }= {SU(3) \over SU(2)_{\rm irr}}
\ee
which is included in the family of coset examples  (\ref{ex1}).

This  level-family has the Lie symmetry $SU(2)_{\rm irr}$, and 
 the 3 and $\bar{3}$ representations
 are  $L(SU(3);SU(2)_{\rm irr})$-degenerate.

For the high-level computations in $\suc$  below, we need the high-level 
form of the inverse inertia tensor  (in the Gell-Mann basis)
and the degenerate conformal weights,
\bs
\be
L^{ab}_{g/h,\infty} = {1 \over x  \psi_g^2  } \theta_a \d_{ab}   
\sp  
\th_a = \left\{ \matrix{ 1 & a =3,8,1,4,6 \cr 0 &  a = 2,5,7  \cr} \right. 
\label{chl} \ee
\be
\D^{g/h}(\T_{( 3)}) =\D^{g/h}(\T_{(\bar{3})} )  = {5\over 6x} + \cO (x^{-2})   
\ee
\es
and we will consider here the same process, that is $3 \bar{3} \bar{3} 3$, 
which we studied for the affine-Sugawara construction
and the irrational construction $\sumh$ in Appendix B and Section 6
 respectively.
 
To compute the high-level blocks in the s-channel, we need to determine
the s-channel eigenvectors $ \ps({\rm s},SU(2) )$ 
from the eigenvalue problem (\ref{evg}),
which reads in this case, 
\bs 
\be
  [ - \frac{1}{2x } \sum_{a={3,8 \atop 1,4,6}}  \l_a^1  
  \l_a^2  +\frac{5}{3x} \one ]_{\a}{}^{\b}  
      \;    \ps({\rm s},SU(2))^M_{\b }  = \D_{({\rm s})}^{g/h} (M)  
        \ps({\rm s},SU(2))^M_{\a }   
\label{cev} \ee
\be
\sum_{i=1}^4 (\l_a^i)_{\a}{}^{\b} \ps({\rm s},SU(2))_{\b}^M =0 \sp a = 2,5,7 
\label{cic}  \pe    \ee  
\es
Here we have used the properties (\ref{gmb}) of the Gell-Mann matrices,
and the global condition (\ref{cic}) enforces 
the $SU(2)_{\rm irr}$-invariance of the coset construction.

After some algebra, the following orthonormal
 set of s-channel eigenvectors is found
\bs
\be
 \ps({\rm s},SU(2) )_\a^1  
= \frac{1}{3} 
\d_{ \a_1 \a_2} \d_{\a_3 \a_4} 
\sp  \D^{g/h}_{({\rm s})}(1) = 0
\label{cwcv} \ee
\be
\ps({\rm s},SU(2))_\a^2 
=  {1 \over 2 \sqrt{5  }} 
[  \d_{ \a_1 \a_3   } \d_{\a_2  \a_4  } 
+  \d_{ \a_1 \a_4   } \d_{\a_2    \a_3 } - \frac{2}{3} 
\d_{ \a_1 \a_2  } \d_{\a_3   \a_4} ]  
\sp    \D^{g/h}_{({\rm s})}(2 ) = \frac{3 }{2 x } 
\label{cwca1}  \ee
\be
\ps({\rm s},SU(2))_\a^3 
=  {1 \over 2 \sqrt{ 3   }}  
[  \d_{ \a_1 \a_3   } \d_{\a_2  \a_4  } 
-  \d_{ \a_1 \a_4   } \d_{\a_2    \a_3 }]  
  \sp \D^{g/h}_{({\rm s})}(3  ) = \frac{5}{2x } 
\label{cwca2} \ee
\be
\bps({\rm s},SU(2))^\a_M  = ( \ps({\rm s},SU(2))^M_\b)^* \eta^{\b \a} 
= \ps({\rm s},SU(2))^M_\a  
\ee 
\label{ces}  \es 
where the last relation says that the left and right eigenvectors 
coincide in this case. 

The results (\ref{ces}) are in agreement with the high-level fusion
rule (\ref{fr1}); in particular, the conformal weight in (\ref{cwcv}) 
corresponds to the
vacuum, while the remaining two weights in (\ref{cwca1},c) are 
the high-level form of the two degenerate subsets of the coset-broken 
conformal weights of the
adjoint (see eqs.(\ref{cw8c})  and  (\ref{hlc})). 
 
Similarly, we can solve for the u and t-channel eigenvectors, which are
given by   
\bs 
\be
\ps({\rm u},SU(2))^M = \ps({\rm s},SU(2))^M \vert_{2 \lra 3}  
\sp \D^{g/h}_{({\rm u})}(M) 
= \D^{g/h}_{({\rm s})}(M) 
\label{cuv} \ee
\be
\ps({\rm t},SU(2))^M = \ps({\rm s},SU(2))^M \vert_{2 \lra 4 }  \sp 
\D^{g/h}_{({\rm t})}(M) = {10 \over 3x}
 - \D^{g/h}_{({\rm s})}(M) 
\ee
\es
where $2 \lra 3$ and $ 2\lra 4$ mean $\a_2 \lra \a_3$ and 
$\a_2 \lra \a_4$ in the explicit expressions (\ref{ces}) for the
s-channel eigenvectors. Since $\T^2 \sim \T^3$, 
the result in (\ref{cuv}) is a special case of 
 (\ref{csc})  
and  the u-channel conformal weights are identical to
those in the s-channel. 
The conformal weights of the t-channel, 
\be
 \D^{g/h}_{({\rm t})}(M) = (\frac{10}{3x},\frac{11}{6x},\frac{5}{6x})
\label{tcw} \ee
are also in agreement with the coset-broken conformal weights of the
 high-level fusion rule (\ref{fr2}). In particular, the last value is
the completely degenerate conformal weight of the $\bar{3}$ and the first two
coincide with the two degenerate subsets (\ref{cw6c}) of the 6, 
according to the  high-level form (\ref{hlc}). 

Using (\ref{gxm}), the $SU(2)$-crossing matrices are computed
from the eigenvectors as
\bs
\be 
X_{SU(2)}({\rm us})_M{}^N  = 
\ps({\rm u},SU(2))^M \ps({\rm s},SU(2))^N= \frac{1}{6}  \left( 
\matrix{ 2 & 2 \sqrt{5} & 2 \sqrt{3} \cr  
         2 \sqrt{5} & 1 & -  \sqrt{15 } \cr  
         2 \sqrt{3}  & -  \sqrt{15} & 3 \cr} \right)
\label{cxu} \ee
\be 
X_{SU(2)} ({\rm ts})_M{}^N  = 
\ps({\rm t},SU(2))^M \ps({\rm s},SU(2))^N= \frac{1}{6}  \left( 
\matrix{ 2 & 2 \sqrt{5} & - 2 \sqrt{3} \cr  
         2 \sqrt{5} & 1 &   \sqrt{15 } \cr  
         - 2 \sqrt{3}  &   \sqrt{15} & 3 \cr} \right)
 \ee
\label{ccr2} \es  
\vskip -.3cm 
\noindent 
which are orthogonal and idempotent matrices in this case. The third crossing matrix  
$X_{SU(2)}({\rm ut}) = X_{SU(2)}({\rm us} ) X_{SU(2)} ({\rm ts} )$ follows 
from the consistency  relations (\ref{ybl3}).  

Finally, we use the $SU(3)$ eigenvectors (\ref{ascr2}) and the  
alternate expression (\ref{gba}) for the general $L(g;H)$-degenerate blocks
 to write down the 6 s-channel  coset blocks of the $3\bar{3}\bar{3}3$ 
correlator in $SU(3)/SU(2)_{\rm irr}$, 
\be
\C^{({\rm s})}(y)_m{}^M  = e({\rm s}, g/h )_m{}^N  [1     +  
(\D_{({\rm s})}^{g/h} - \frac{5}{3x} \on  )\ln y 
 + (Q_{ ({\rm su})}^{g/h}- \frac{5}{3x }\on  ) 
 \ln (1-y)]_N{}^M 
   +\cO (x^{-2})      
\label{sucb} \ee
where $(1)_N{}^M = \d_N^M$ and 
\bs
 \be
e({\rm s}, g/h ) _m{}^M = v({\rm s},SU(3))^m \ps({\rm s},SU(2) )^M =
\left(
\matrix{ 1 & 0 & 0  \cr
 0 & \frac{1}{4}\sqrt{10  } & \frac{1}{4}\sqrt{6} &  \cr }
\right)
\label{cem} \ee
\be
(\D_{  (s)}^{g/h}) _N{}^M = \D_{({\rm s})}^{g/h}(M)  \d_N^M 
   \sp \D_{({\rm s})}^{g/h}(M) 
=  \D_{({\rm u})}^{g/h}(M )  = ( 0, \frac{3}{2x} , \frac{5}{2x} ) 
\label{dsw} \ee 
\be
(Q_{({\rm su})}^{g/h})_N{}^M  
= \sum_{L }  X_{SU(2)} ({\rm us})_{N}{}^L  \D_{({\rm u})}^{g/h}(L) X_{SU(2)}({\rm us})_L  {}^M 
= \frac{1}{12x} \left( \matrix{ 
20  & -4 \sqrt{5}  & 0  \cr 
-4 \sqrt{5}  & 13  & -3\sqrt{15}    \cr 
0 &  -3\sqrt{15}    &  15     \cr} \right)  
\;.  \ee
\label{scb} \es
 The u and t-channel blocks can be computed from the s-channel blocks above 
using the crossing relation (\ref{gcro}) and the explicit forms of the
 crossing matrices $X_{SU(3)}({\rm us})$, $X_{SU(3)}({\rm ts})$ in (\ref{ascr2})  and 
$X_{SU(2)}({\rm us}), X_{SU(2)}({\rm ts})$ in (\ref{ccr2}). 
Moreover, using the explicit form of the non-analytic phase matrix 
(\ref{hpm}) for this process,  
\be
U_{g/h}(y)_M{}^N  =\sum_L  
X_{SU(2)}({\rm ts})_{M}{}^L  
\exp\left(-\pi i[ \D_{({\rm s})}^{g/h}(L )-\frac{5}{3x} ] 
\sign(\arg(-y)) \right)  X_{SU(2)} ({\rm ts})_L {}^N  
\ee
$$
+ \ox  
$$
the analytic t-channel blocks follow from the crossing relation (\ref{satc}) 

Using (\ref{bss},d) we obtain the following limiting behavior as
$y \ra 0$ for the 6 s-channel blocks (\ref{sucb}) of this correlator, 
\bs 
\be
\C^{({\rm s})}(y)_m{}^{M} 
\mathrel{\mathop\sim_{y \ra 0} }
\Ga_{g/h}^{({\rm s})} (m,M)  
y^{\D_{({\rm s})}^{g/h} (m,M)-5/3x} +\cO (x^{-2}) \sp m=V,A \sp M = 1,2,3
 \ee
\be
\D_{({\rm s})}^{g/h} (V,1) = 0 + \cO (x^{-2}) 
\sp 
\D_{({\rm s})}^{g/h} (A,2) = \frac{3}{2x}  + \cO (x^{-2}) 
\sp 
\D_{({\rm s})}^{g/h} (A,3) = \frac{5}{2x}  + \cO (x^{-2}) 
\label{cwpb} \ee
\be
\D_{({\rm s})}^{g/h} (V,2) = 1 + \cO (x^{-1}) 
\sp 
\D_{({\rm s})}^O (A,1) = 1 + \cO (x^{-1}) 
\label{cwsb} \ee
\be
\D_{({\rm s})}^O (V,3) = \cO (x^{0}) 
\label{cwub} \ee
\be
\Ga_{g/h}^{({\rm s})} (m,M)  
= \left( \matrix{ 1  & \frac{1}{3x} \sqrt{5} & 0 \cr
\frac{5}{12x} \sqrt{2} & \frac{1}{4} \sqrt{10} &  \frac{1}{4} \sqrt{6} \cr}  
\right) +  \cO (x^{-2}) 
\pe \ee
\es
The explicit form of the residues was obtained using
(\ref{lre}), the embedding matrix (\ref{cem}) and the relation 
$c({\rm s},h) = Q^{g/h}_{({\rm su})} -\frac{5}{3x} \on $.  

The three conformal weights in (\ref{cwpb}) are broken affine primary
states and the two in (\ref{cwsb}) are broken affine secondary states
which are not necessarily integer descendants of broken affine primary
states. 
The conformal weight in (\ref{cwub}) cannot be determined through this order
because the residue of the corresponding block $\C (y)_V{}^3 $, and the
block itself,  is zero through order $\oxo$, so this block begins at
$\ox$.
To see this directly from (\ref{sucb})
note that for this block $e({\rm s},g/h)_m{}^M=0$ and 
$e({\rm s},g/h)_m{}^N (Q^{g/h}_{({\rm su})} )_N{}^M =0$.
We also find  one block which begins at $\ox$ in the u-channel and in the
t-channel. 

In accord with (\ref{cnb}), the number of blocks in this process is
\be
 B_{SU(3)/SU(2)}   = 2 \cdot 3 = 6
\ee 
while
the same process under the affine-Sugawara construction on
$SU(3)$ and the irrational construction  $\sumh$ showed 4 and  8  blocks
respectively. This is in accord with the double inequality (\ref{die}) and 
the increasing symmetry breakdown 
$ O   \subset SU(2)_{\rm irr} \subset SU(3)$ of the three constructions.  

Using eqs.(\ref{agc}), (\ref{gesm}), we also find  the following expression 
 for the high-level non-chiral correlators of $\suc$,
\bs 
 \be
Y_{g/h} (y^*,y)  
\! = \! \sum_{M} E({\rm s},g/h)_{M}{}^M  
 [ 1  +
 (\D_{({\rm s})}^{g/h} - \frac{5}{3x } \on  )\ln | y|^2  
 + (\D_{({\rm s})}^{g/h}- \frac{5}{3x }\on  )
 \ln | 1-y|^2 ]_M{}^M 
   +\cO (x^{-2})    
\ee
\be 
 E({\rm s},g/h )_M {}^N  =  
\sum_m ( e({\rm s},g/h )_m {}^M)^*  e({\rm s},g/h)_m{}^N 
 = \left(
\matrix{ 1 & 0 & 0  \cr
 0 & \frac{5}{8 }  & \frac{1}{8 }\sqrt{15 } &  \cr 
 0 & \frac{1 }{8 } \sqrt{15}   & \frac{3 }{8 } &  \cr }
\right) 
\ee
\es 
\vskip -.3cm 
\noindent 
where we have used $ X_{SU(2)}({\rm us}) E({\rm s},g/h) X_{SU(2)}({\rm us}) = 
E({\rm s},g/h)$ and
where the diagonal s-channel conformal weight matrix $\D_{({\rm s})}^{g/h}$ is
given in (\ref{dsw}). This result explicitly shows the crossing 
symmetry (\ref{snc}), as it should  since $\T^2 \sim \T^3$ in this case.

\end{document}